\documentclass[structabstract]{aa}  
\usepackage{graphicx}
\usepackage{ gensymb }
\usepackage{ tipa }
\usepackage{ textcomp }
\usepackage{ wasysym }
\usepackage{float}

\usepackage{rotating}
\usepackage{txfonts}
\usepackage{lscape}
\begin{document}
   \title{A proposed new diagnostic for Herbig disc geometry
\thanks{Based on observations made with ESO Telescopes at the La Silla Paranal Observatory under programme ID 088.C-0898A and programme ID 084.C-1002(A).}}
   \subtitle{FWHM versus $J$ of CO ro-vibrational lines}

 \author{R. P. Hein Bertelsen
	\inst{1}
	\and
  	I. Kamp\inst{1}
	\and
	G. van der Plas\inst{2}
 	\and
	M. E. van den Ancker\inst{3}
	\and
	L. B. F. M. Waters\inst{4}\inst{,5}
	\and
	W.-F. Thi\inst{6}
	\and
	P. Woitke\inst{7}
	}

                 \institute{Kapteyn Astronomical Institute Rijks-universiteit Groningen (RuG),
              Landleven 12, Groningen 9747 Netherlands\\
\email{bertelsen@astro.rug.nl}
     		\and
Departamento de Astronom\'{\i}a, Universidad de Chile, Casilla 36-D, Santiago, Chile
		\and
European Southern Observatory, Karl-Schwarzschild-Str.2, D 85748 Garching bei M\"unchen, Germany
		\and
Anton Pannekoek Astronomical Institute, University of Amsterdam, PO Box 94249, 1090 GE Amsterdam, The Netherlands 
		\and
SRON Netherlands Institute for Space Research, Sorbonnelaan 2, 3584 CA Utrecht, The Netherlands 
  		\and
UJF-Grenoble 1 / CNRS-INSU, Institut de Plan\'{e}tologie et d'Astrophysique (IPAG) UMR 5274, Grenoble, F-38041, France
		\and
SUPA, School of Physics \& Astronomy, University of St. Andrews, North Haugh, St. Andrews KY16 9SS, UK             		
       			}
 

 
  \abstract
 {}
  {CO ro-vibrational lines observed from Herbig group II discs are seen to be often broad while the same lines observed from group I discs are often narrow. This difference is not well understood. In this paper we explore the underlying cause for this difference and provide a pathway for a better understanding of the geometry and structure of the inner discs around Herbig Ae/Be stars.}
   {High-spectral-resolution infrared spectra of CO ro-vibrational emission from six Herbig Ae/Be candidate stars were taken with CRIRES (CRyogenic high-resolution InfraRed Echelle Spectrograph) at the VLT (Very Large Telescope). From these spectra, we produce individual and co-added CO ro-vibrational line profiles. We investigate line profile shape differences, and we explore the FWHM (Full Width Half Maximum) variations with $J$ quantum number in the context of disc geometry. Furthermore, we put our new sources into the context of earlier observed sources to study a large sample. For comparison, we also investigate the FWHM variations with $J$ of modelled CO ro-vibrational lines from two typical disc geometries produced with the thermo-chemical disc modelling code ProDiMo. }
   {For our new observations of CO ro-vibrational lines, we find that the FWHM of individual lines range from 10-60 km/s. We find both narrow and broad single peaked emission lines, but only Hen~2-80 displays double peaked emission lines. For HD~250550, the FWHM of the CO lines is increasing with $J$ value, indicating a radially extended emitting region, while Hen~2-80 shows a constant FWHM versus $J$ behaviour, indicating a narrow emitting region. This qualitatively agrees with the two different modelled disc geometries.  Comparing dust and gas inner disc geometries (inferred by SED and CO ro-vibrational emission) for the expanded sample of observed Herbig discs, we find no clear correspondence between the SED groups of the sources and their inner CO radius.}
   {The FWHM versus $J$ is a potential new gas diagnostic for the inner disc with e.g. a constant FWHM versus $J$ indicating the presence of a large gas hole or gap. Both models and observations indicate the potential of this new diagnostic. Our extended sample does not fully support the previous trend where group I discs have CO ro-vibrational emission lines with small FWHM. Instead, our CO ro-vibrational data from a handful of sources indicates different inner disc geometries for gas and dust of these sources. }

   \maketitle
%

\section{Introduction}
Infrared CO ro-vibrational emission lines are promising tracers of the geometry and structure of the inner protoplanetary discs. They are routinely detected in ground based observations of discs around YSOs (Young Stellar Objects) and their line profiles reveal that they originate within a few tens of au from the central star \citep{carr2001,najita2003,blake2004,brittain2007,plas2014}, coinciding with the region of planet formation.  

It has been found empirically that Herbig Ae/Be stars show two types of SEDs: Group I, displaying a rise in the mid-infrared, and group II with no rise \citep{meeus2001}. Group I disks seem to be dominated by large gaps and possibly depleted inner disks. E.g. \citet{maaskant2013} showed, using Q- and N-band imaging and radiative transfer modelling for four HAeBe discs, classified as group I from their SEDs, that all four required solutions with a large dust gap separating the inner and outer discs. They finally conclude that many, if not all, group I discs seem to have gaps. Group II disks have more compact disks, with growing evidence that gaps in the few to 10 au range are present \citep[see e.g.][]{menu2015}.

CO ro-vibrational emission lines from Herbig discs have been studied on several occasions. {For example, a recent re-analysis of CRIRES archival data by \citet{banzatti2015} find that Herbig star spectra show single component line profiles, contrary to T Tauri spectra that often show two distinct components. Observations} do not always show symmetric double peaked line profiles, instead flat topped profiles, symmetric or asymmetric single peaked profiles, profiles with shoulders or other variations hereof have been observed \citep{blake2004, brittain2007, salyk2011,hein2014, plas2014}. These shapes can occur if the profile is a composite of CO emission coming from both the inner and outer disc, if a disc wind is present \citep{bast2011, pontoppidan2011} or if eccentricities are present in the disc. However, flat topped profiles can also be a result of low spectral resolution, asymmetries can arise randomly in noisy profiles, and non-Keplerian profile shapes can arise from instrumental effects like slit loss \citep{hein2014}. Keeping these challenges in mind, the interpretation of CO ro-vibrational line profiles from the HAeBe discs provide a powerful tool to trace the disc geometry in planet forming environments.

{\citet{plas2014}} were the first to attempt to perform a statistical analysis of CO ro-vibrational emission lines from a larger sample of Herbig Ae/Be discs, including both group I and group II sources. They found that the CO emission in the discs, classified as group II from the SED, originates much closer to the star than in the discs classified as group I from the SED. Their sample included only four group I discs, limiting general conclusions about the radial origin of CO emission in these type of discs. 
Additionally, {\citet{plas2014}} identify a correlation between the excitation mechanism for the CO lines and group I/II classification. The group I discs in the sample had a dominant fluorescent excitation mechanism, indicated by $T_{\rm vib}>T_{\rm rot}$, and the group II discs had a dominant thermal {(LTE)} excitation mechanism, indicated by $T_{\rm vib}\le T_{\rm rot}$. 
From the study by \citet{brittain2007}, UV fluorescence is an important excitation mechanism in dust-depleted discs, and \citet{maaskant2013} showed that all group I discs could in fact be discs with gaps. The presence of a gap would 'force' the various CO transitions to be emitted in a narrow region, similar for all $J$ levels, at the inner rim of the outer disc. This might cause the UV fluorescence to become more important. An example of this is the disc around HD~100546, where {the CO ro-vibrational emission starts at} the inner wall of the outer disc and UV fluorescence is a dominant excitation mechanism \citep{brittain2013,hein2014}. If we consider instead a continuous disc without gaps, the CO transitions can be emitted from a broader region (higher $J$ closer to the star resulting in wider lines and lower $J$ lines dominated by emission from the outer regions resulting in narrow lines) and {allow for the possibility that thermal excitation at small radii becomes more important}. The sources from {\citet{plas2014}}, that seem to require fluorescence from $T_{\rm vib}$ > $T_{\rm rot}$, are all sources with FWHM versus $J$ constant. 

Our new observations of CO ro-vibrational emission will expand the current sample of Herbig Ae/Be discs (especially group I) and provide the basis for a better understanding of the geometry and structure in the two types of HAeBe discs. This is key in establishing the CO ro-vibrational lines as a tracer of planet forming processes across stars of different masses and evolutionary states.
In Sect. \ref{sec:sample} and \ref{sec:obs} we present our sample and our new observations. In Sect. \ref{sec:datared} we describe the details of our data reduction and analysis methods and present our results. 
In Sect. \ref{sec:discus} we discuss the overall trend for the CO emitting region and disc geometry that can be derived from a large sample of observed HAeBes \citep[this paper,][]{salyk2011,plas2014}. 
In Sect. \ref{sec:mod} {we present CO ro-vibrational emission from two modelled disc geometries for the purpose of comparison}, and in Sect. \ref{sec:modeldiscus} we discuss, for both models and observations, the implications of the FWHM versus $J$ behaviour on the disc geometry and the effects of disc inclination. Finally in Sect. \ref{conclusion} we present our conclusions.

\section{Sample} \label{sec:sample}
\begin{figure*}[!htbp]
\begin{center}$
\begin{array}{cc}
   \includegraphics[width=.75\textwidth]{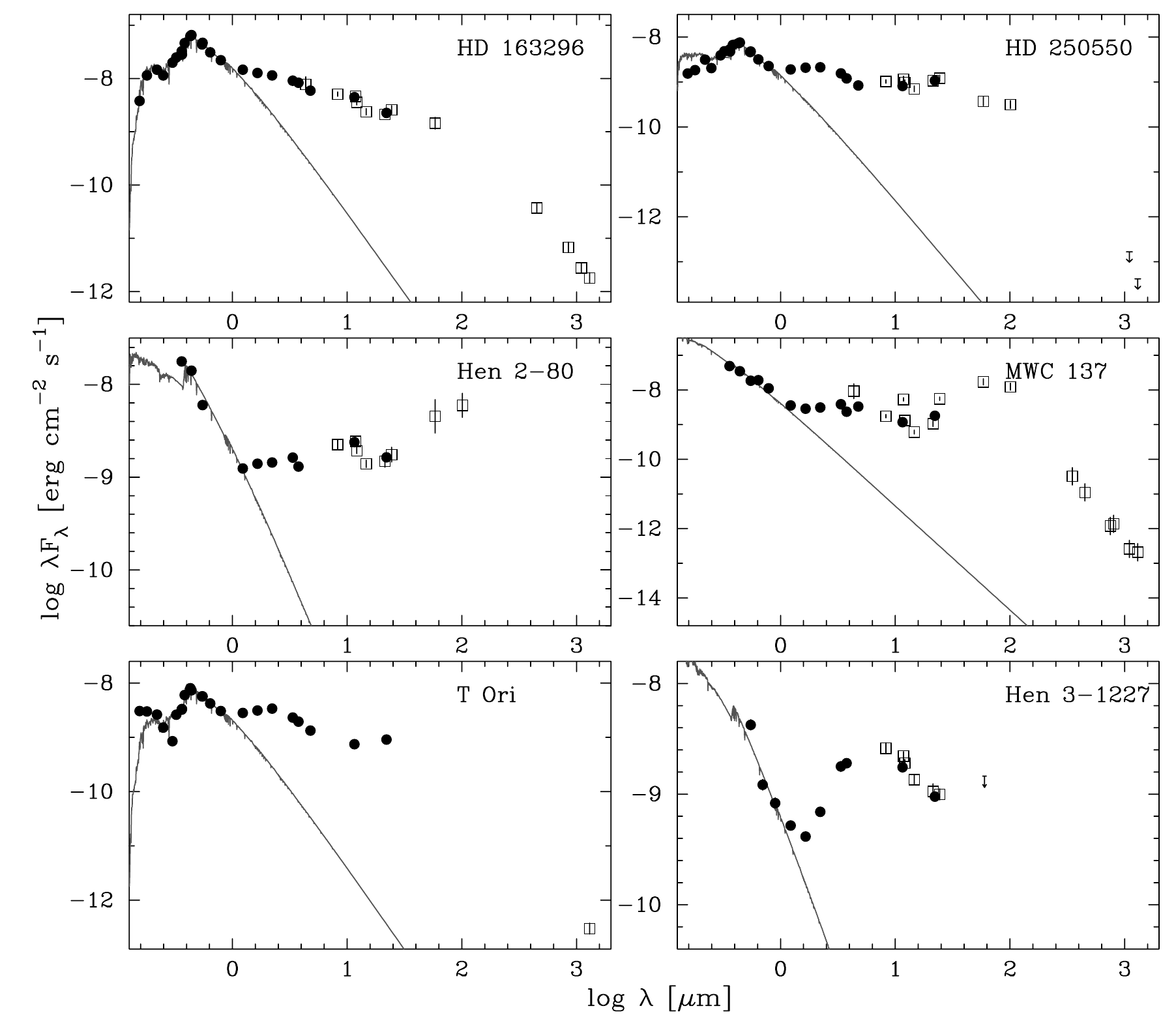}
 \end{array}$
\end{center}
\caption{Extinction corrected SEDs of all sources in our sample. Unreddened \citet{kurucz1991} photospheric models corresponding to $T_{\rm eff}$ and $\log g$ from the literature, are fitted to the extinction-corrected observed photometry. Measurements with beam sizes \textless 3" are shown as solid dots and measurements with beam sizes \textgreater 3" are shown as open squares.}
         \label{fig:sed}
\end{figure*}

Our sample of sources was selected from the list of HAeBe candidates by \citet{the1994}.  Preference in the target selection was given to Meeus et al. group I targets, as these stars were under represented in previous studies of CO emission from HAeBe discs.
An overview of our sample is shown in Table \ref{table:sam} and background details of the individual sources are discussed in the appendix \ref{sample_app}. 
Extinction-corrected SEDs of all stars were constructed from the literature and are shown in Fig. \ref{fig:sed}. The sources have been given a rough classification based on their position in a diagram of near-infrared over total infrared luminosity versus IRAS [12]-[60] colour similar to that in \citet{boekel2005} (Fig. \ref{fig:nir_miras} in the appendix). With the exception of HD~163296, which is close to the boundary line between the two groups, and T~Ori, (due to a lack of IRAS data for this source) all stars in our sample fall within the area occupied by group I discs.

\begin{sidewaystable}
\caption{Stellar parameters of the observed sources. }             
\label{table:sam}      
\centering                          
\begin{tabular}{l c c cc c ccccccc }        

\hline            
Star Name &  Alt. Name &   R.A.  (2000.0)&  Dec.  &    Region     &     d  &Sp.T.   &  $T_{\rm eff}$ & A$_{\rm v}$& log($L_*$)&	 $v_{\rm{rad}}$ & incl.\\
 &  & &  &    &  [pc]   &  & [K] &  [mag] &&  [km/s]&	& 	\\
\hline            
HD~163296 &  MWC 275  &    17 56 21.29 & -21 57 21.8  &Sco OB2-1    &     119 $^{\rm HIP}$  &   A3Ve $^{\rm G98}$&     8720  &   0.1&     1.4 &	-4.0 $^{\rm B00}$&46$\degree$ $^{\rm I07}$	 \\
HD~250550 &  MWC 789   &   06 01 59.99  &+16 30 56.7&L1586&       280 $^{\rm K98}$ &   B8--A0IVe $^{\rm G98}$   & 10750  &    0.4 &    1.3&	+16 $^{\rm F84}$&10$\degree$	$^{\rm F11}$	  \\
Hen~2-80  &  PK 299-00.1 & 12 22 23.18  &-63 17 16.8 &             &    $>$ 750 $^{\rm B15}$&    B6Ve $^{\rm C10}$&    14000   &   4.2   & $>$2.7 &–22 $^{\rm C10}$ \\
MWC~137  &   V1308 Ori   & 06 18 45.50 & +15 16 52.4 &      &  > 1000 $^{\rm B15}$&    B0ep $^{\rm S81}$ &  30000 &4.0 & $>$4.1 &	+43 $^{\rm Z03}$& \\
T~Ori     &  MWC 763    &  05 35 50.44 & -05 28 34.9 & Orion OB1c   &     510 $^{\rm D99}$   &  A3IVe $^{\rm M01}$  &   8660   &  1.2    & 1.79 &+52 $^{\rm T09}$		 \\
Hen 3-1227&  PK 338+01.1 & 16 38 28.62 & -45 23 41.5&           &       $>$ 400 $^{\rm B15}$  &  Be $^{\rm P03}$  &  20000  &  4.0 &   $>$2.0  \\
\hline
\end{tabular}
\tablefoot{The effective temperatures, T$_{\rm eff}$, were derived based on the spectral type and the calibrations 
by \citet{schmidt1982}. The A$_{\rm v}$ and the luminosity were derived using the distances and the SEDs shown in Fig. \ref{fig:sed}. For sources where no distance is known, the spectral type was used to infer a stellar luminosity using the \citet{schmidt1982} calibrations, and a minimum distance was then computed based on this. References: B15 = This paper (minimum distance based on luminosity), C10 = \citet{carmona2010}, D99 = \citet{zeeuw1999}, F84 = \citet{finkenzeller1984b} ,F11 = \citet{fedele2011} G98 = \citet{gray1998}, HIP = Hipparcos distance, K98 =\citet{kawamura1998}, M01 = \citet{mora2001}, P03 = \citet{pereira2003}, S81 =  \citet{sabaddin1981}, I07=\citet{Isella2007}., Z03=\citet{zickgraf2003},  T09=\citet{tobin2009}, B00=\citet{barbier2000}.}

\end{sidewaystable}

\section{Observations} \label{sec:obs}
High resolution spectra of 10 sources were obtained on March 5 and March 6, 2012, with the VLT cryogenic high-resolution infrared echelle spectrograph \citep[CRIRES,][]{kaeufl2004}. {Four of these sources (Hen 2-14, Hen 3-1386, Th 35-101, V883 Ori) have been rejected as Herbig Ae/Be stars on the basis of a careful literature review and are therefore not discussed further in this paper. A more detailed discussion of these sources can be found in Hein Bertelsen (Ph.D. thesis 2015).}
For each target, the wavelength region from 4.5 $\mu$m to 5 $\mu$m was covered using 3 or 6 different grating settings (4.6575, 4.7363 and 4.9948 $\mu$m and the additional three settings 4.6376, 4.8219, and 5.0087 $\mu$m). The observations were made through a slit with a width of 0.2" and the telescope was nodded by 10" along the slit for every other exposure to subtract the sky emission. Since the CO ro-vibrational lines also arise in the earth's atmosphere, we chose the observations to be made in a period of time where the velocity shift of our individual targets would be as large as possible for most of the targets. Furthermore, we observed nearby telluric standard stars directly after or before each target to remove the telluric absorption lines. Flat fields were collected during the daytime (within 12 hours of the science observations).
See Table \ref{table:log} for the summary of the observations.

\begin{table}
\caption{Log of our CRIRES observations}             
\label{table:log}      
\centering                          
\begin{tabular}{l cc cl}        
\hline            
 \multicolumn{2}{c}{March 5, 2012}&  \\   
 \hline

Object &$\lambda_{\rm ref}$ &$t_{\rm exp}$ & airmass			\\   
 		&		[$\mu$m]  &[s] & 			\\   
MWC~137 &4.9948/4.6575/4.7363	& 60.	&	1.30-1.35	\\   
STD$_{\rm MWC137}$ &4.9948/4.6575/4.7363	& 30.	&	1.30-1.38	\\   
T~Ori 	&4.9948/4.6575/4.7363&120.	&1.24-1.75	\\   
STD$_{\rm T~Ori}$ 	&4.9948/4.6575/4.7363&30.	&1.30-1.38	\\   
Hen~2-80 &4.9948/4.6575/4.7363&150.&1.28-1.30\\
	&5.0087/4.6376/4.8219&150.&1.29-1.33	\\   
STD$_{\rm Hen~2-80}$ &4.9948/4.6575/4.7363/&40.&1.22-1.29\\
	&5.0087/4.6376/4.8219&	\\   
Hen~3-1227 &4.9948/4.6575/4.7363&60.	&1.22-1.29	\\   
STD$_{\rm Hen~3-1227}$ &4.9948/4.6575/4.7363&40.	&1.28	\\   
\hline 
 \multicolumn{2}{c}{March 6, 2012}&  \\   
 \hline
Object &$\lambda_{\rm ref}$ &$t_{\rm exp}$  & airmass			\\   
 		&		[$\mu$m]  &[s] & 			\\   
HD~250550	&4.9948/4.6575/4.7363&120.	&1.33-1.42\\
&5.0087/4.6376/4.8219&120.	&1.52-1.93  \\
STD$_{\rm HD~250550}$	&4.9948/4.6575/4.7363/&40.	&1.36-1.39\\
&5.0087/4.6376/4.8219& \\
Hen~2-80		&4.9948/4.6575/4.7363&100.&1.29-1.32\\
&5.0087/4.6376/4.8219&100.&1.28-1.30\\
STD$_{\rm Hen~2-80}$&4.9948/4.6575/4.7363/&40.&1.28\\
&5.0087/4.6376/4.8219&\\
Hen~3-1227	&4.9948/4.6575/4.7363	&60.&1.27-1.38 \\
STD$_{\rm Hen~3-1227}$&4.9948/4.6575/4.7363&40.&1.28\\
HD~163296	&4.9948/4.6575/4.7363&60.&1.20-1.29\\
&5.0087/4.6376/4.8219&60.&1.14-1.19 \\
STD$_{\rm HD~163296}$	&4.9948/4.6575/4.7363/&40.&1.07-1.09\\
&5.0087/4.6376/4.8219& \\
\hline                           
\end{tabular}
\end{table}

\section{Data reduction and analysis} \label{sec:datared}
The science data were reduced using version 2.2.1 of the CRIRES data reduction pipeline\footnote{http://www.eso.org/pipelines/}. The observations were done in nodding mode: The spectra are nodded between two positions (A and B) 10" apart on the sky in the pattern ABBA so that the source is in different positions on the chip. The sky emission was removed by subtracting the nodded A and B spectra. One dimensional spectra were extracted at each grating setting for each individual source. The optimal extraction method\footnote{The spectral profile is fitted, for more details see the CRIRES Pipeline User Manual} was used. The spectra of the telluric standard stars were reduced in the same way. 
The telluric {standard star spectra} were divided by {a corresponding stellar atmosphere model of Kurucz (including line opacities)} to determine the instrument response function and atmospheric transmission curve. The optical depth of the telluric lines was then scaled to the depth of the science target and small adjustments in the wavelength were applied when needed.  {These adjustments in optical depth and wavelength are done by minimising the telluric residuals. This corrects for the small differences in wavelength calibration and atmospheric parameters between the science spectrum and the telluric standard.}
As a last step in the data reduction process, the science spectra were divided by the response function of the corresponding standard star (examples of the final telluric corrected science spectra can be seen in Figures \ref{fig:spec1} and \ref{fig:spec2} in the appendix).

\subsection{Flux calibration}
We have calibrated the continuum level of our sources to the corresponding spectroscopic standard (the collected telluric standards). For the standard stars we have collected the W2 magnitudes ($\lambda_{\rm c}$=4.60 $\rm \mu$m) from the WISE all-sky catalogue, HR6527 ($m_{\rm{M}}$=2.050$\pm$0.103 mag), HR4730 ($m_{\rm{M}}$=1.671$\pm$0.025 mag) and,  HR2421 ($m_{\rm{M}}$=1.521$\pm$0.037 mag). From this, the continuum flux at 4.6 $\mu$m of the three standard stars is respectively, 25.83 Jy, 36.62 Jy and 42.05 Jy.
There are small differences in the widths of the PSF for the science spectra and the telluric standard spectra (we expect these differences to be due to variations in the performance of the AO system). We do not correct for these differences but note here that the error expected from this in our flux calibration is <10\%. 
Furthermore, when viewing the full spectra for each source there are inconsistent variations in continuum level in different wavelength settings (or detectors) for overlapping or neighbouring regions. These variations are at the 10\% level. We apply flux shifts to the frames where this is needed to make the full spectrum fit one consistent continuum curve.

\subsection{Line detection} \label{line_selection}
All of our observed sources have clear detections of \ion{H}{I} recombination lines in emission: Pfund $\beta$ 7-5 and the Humphreys $\epsilon$ 11-6 line. For these lines individual line profiles are displayed in Fig. \ref{fig:HI} in the appendix. The full width half maxima (FWHM), the line centres and the derived doppler shifts, together with the 3 $\sigma$ error bars are shown for all sources in Table \ref{table:HI} in the appendix.
Spectra from four of our sources (HD~163296, HD~250550, Hen~2-80, Hen~3-1227) also show clear detections of CO ro-vibrational emission lines. {The CO emission from these four sources is not spatially resolved \citep[for more details on spatially resolving emission with CRIRES see][]{plas2014}.} CO ro-vibrational emission lines could also be present in MWC~137. The spectrum shows, what looks to be, many wide and often blended emission lines.

For the four sources with CO detections, individual CO lines were identified and extracted using wavelengths from \citet{chandra1996}.
For each source certain wavelength settings overlap, resulting in a few lines being detected twice.
Furthermore, in the cases of Hen~2-80 and Hen~3-1227 we have collected observations during two nights with the same wavelength settings. These doubly collected lines can be used for sanity checks and uncertainty estimates.
For the further analysis we consider our line selection carefully: Some lines are contaminated by telluric residuals and some lines are blended and cannot be separated for individual profile extraction. We thus manually select the best uncontaminated and unblended lines for further analysis (Table \ref{table:COlines}).

\subsection{CO line profiles} \label{sec:co_lines}
\begin{figure*}[!htbp]
\centering
\begin{minipage}[l]{0.4\textwidth}
   \includegraphics[width=\textwidth]{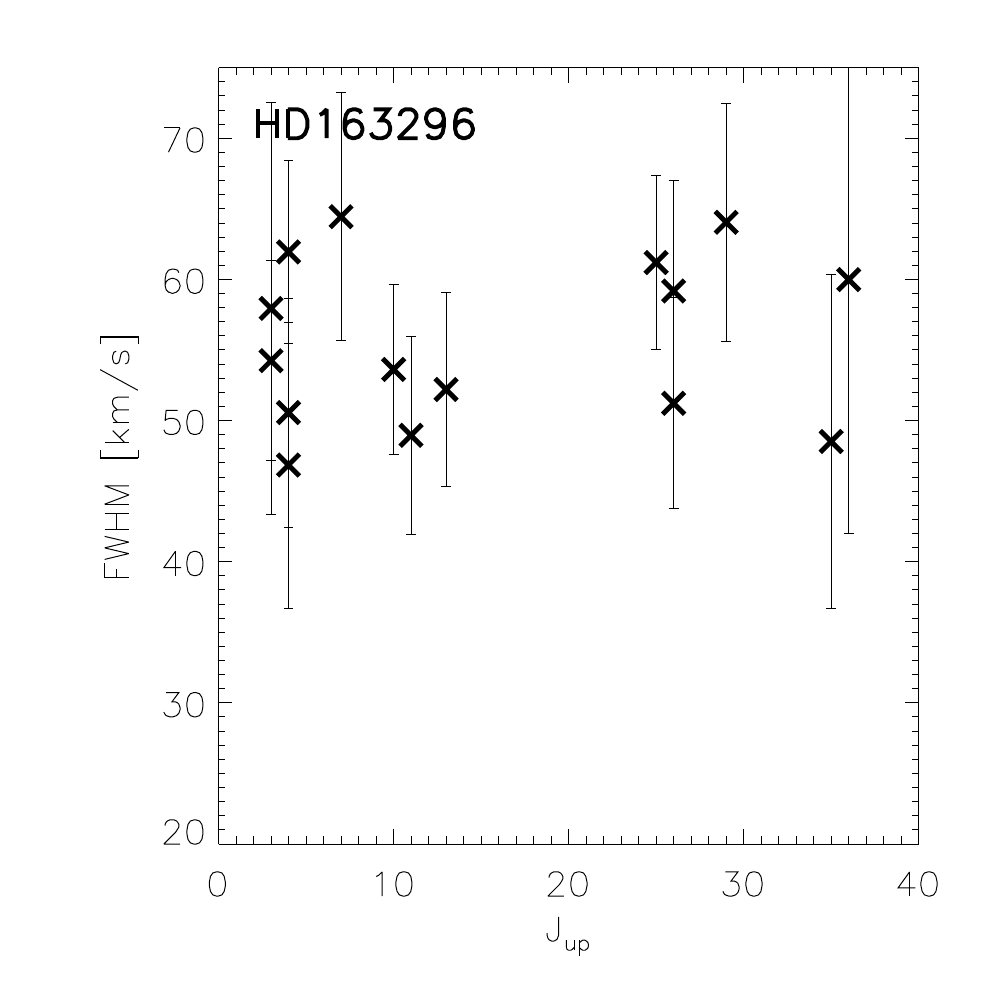}
\end{minipage}
\begin{minipage}[r]{0.4\textwidth}
   \includegraphics[width=\textwidth]{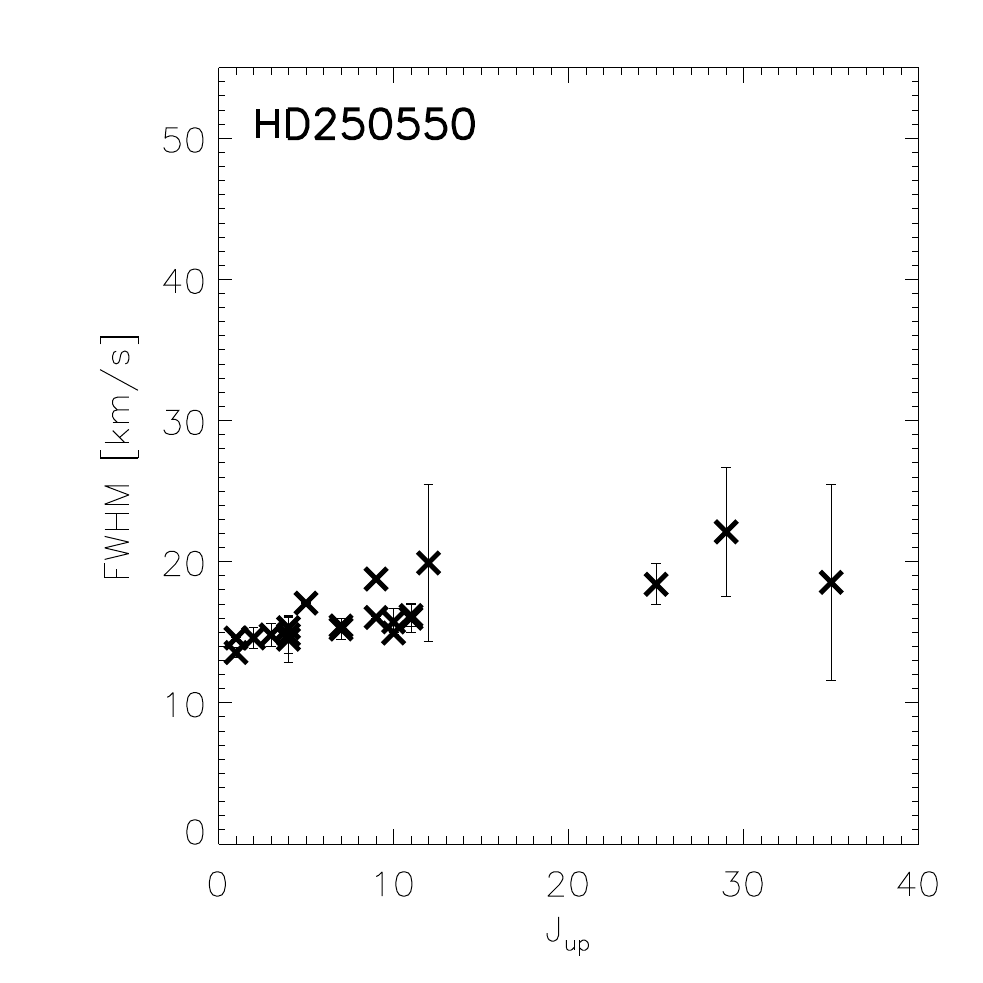}\\
\end{minipage}
\begin{minipage}[r]{0.4\textwidth}
   \includegraphics[width=\textwidth]{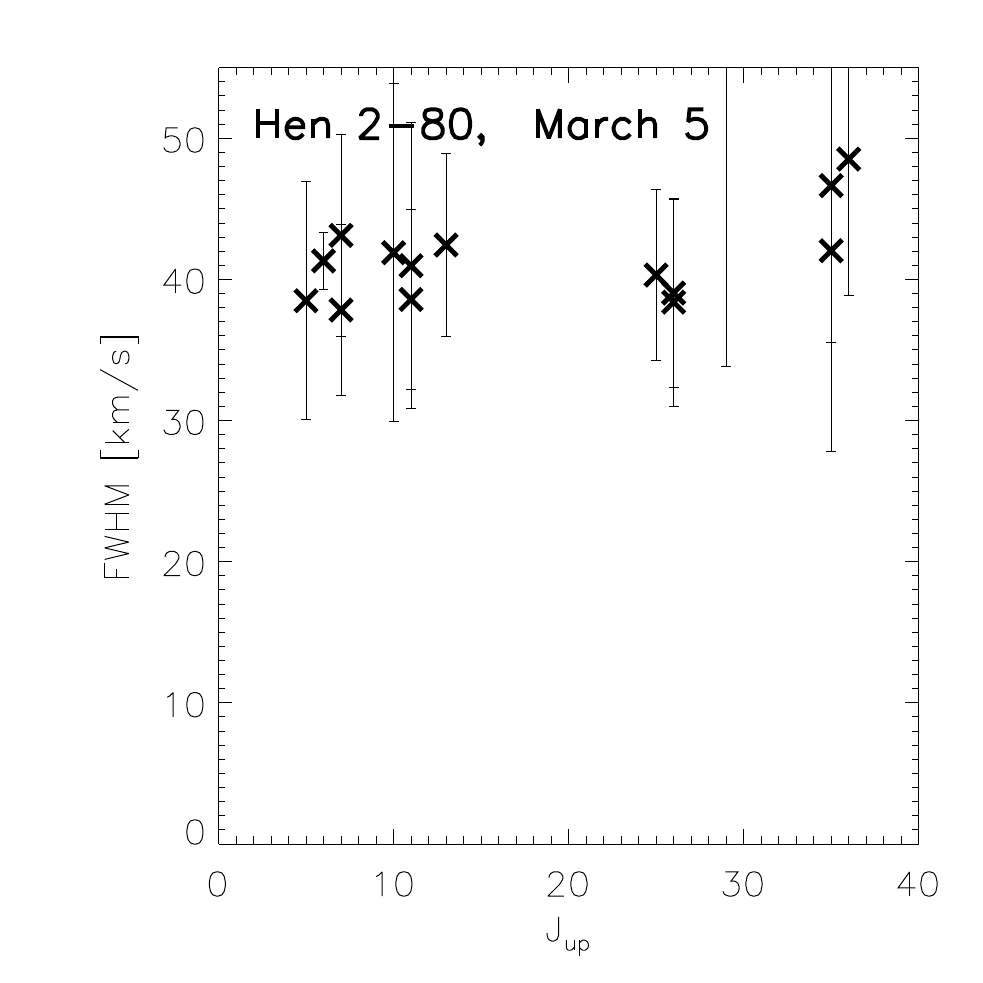}
\end{minipage}
\begin{minipage}[r]{0.4\textwidth}
   \includegraphics[width=\textwidth]{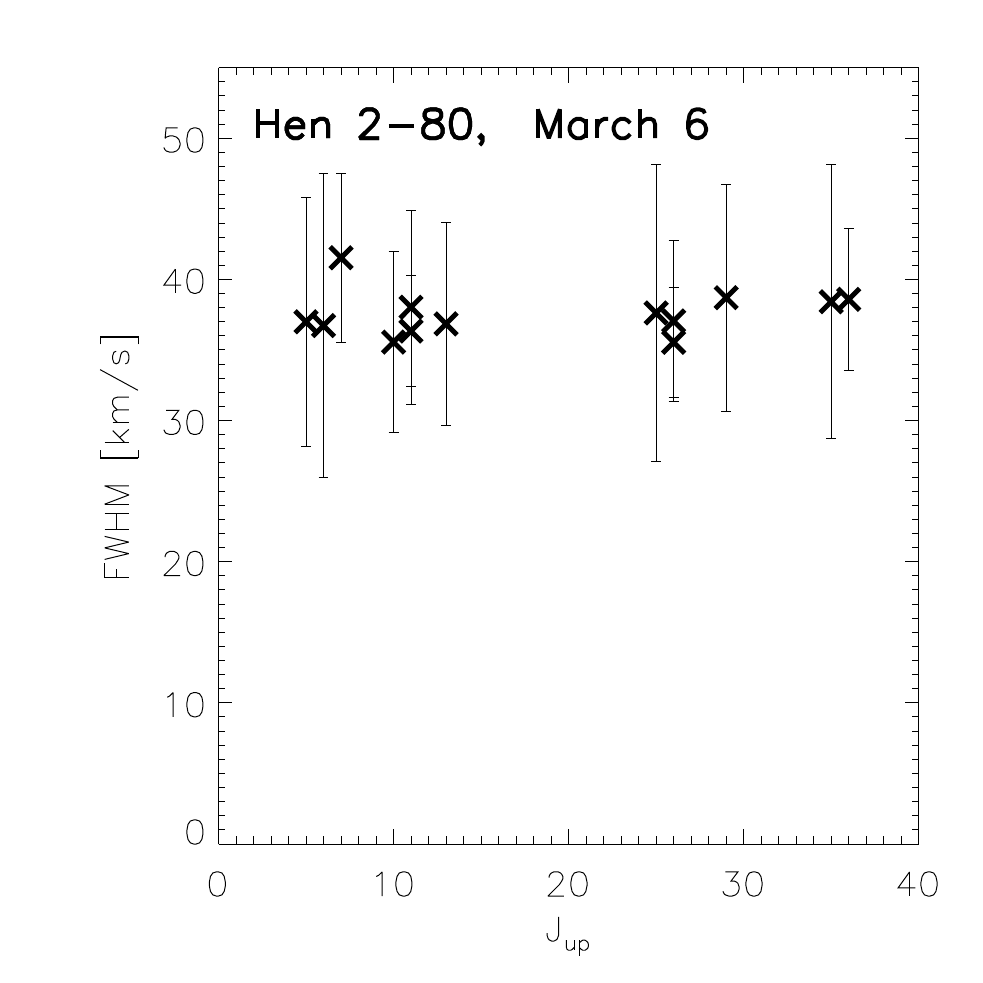}\\
\end{minipage}
\begin{minipage}[r]{0.4\textwidth}
   \includegraphics[width=\textwidth]{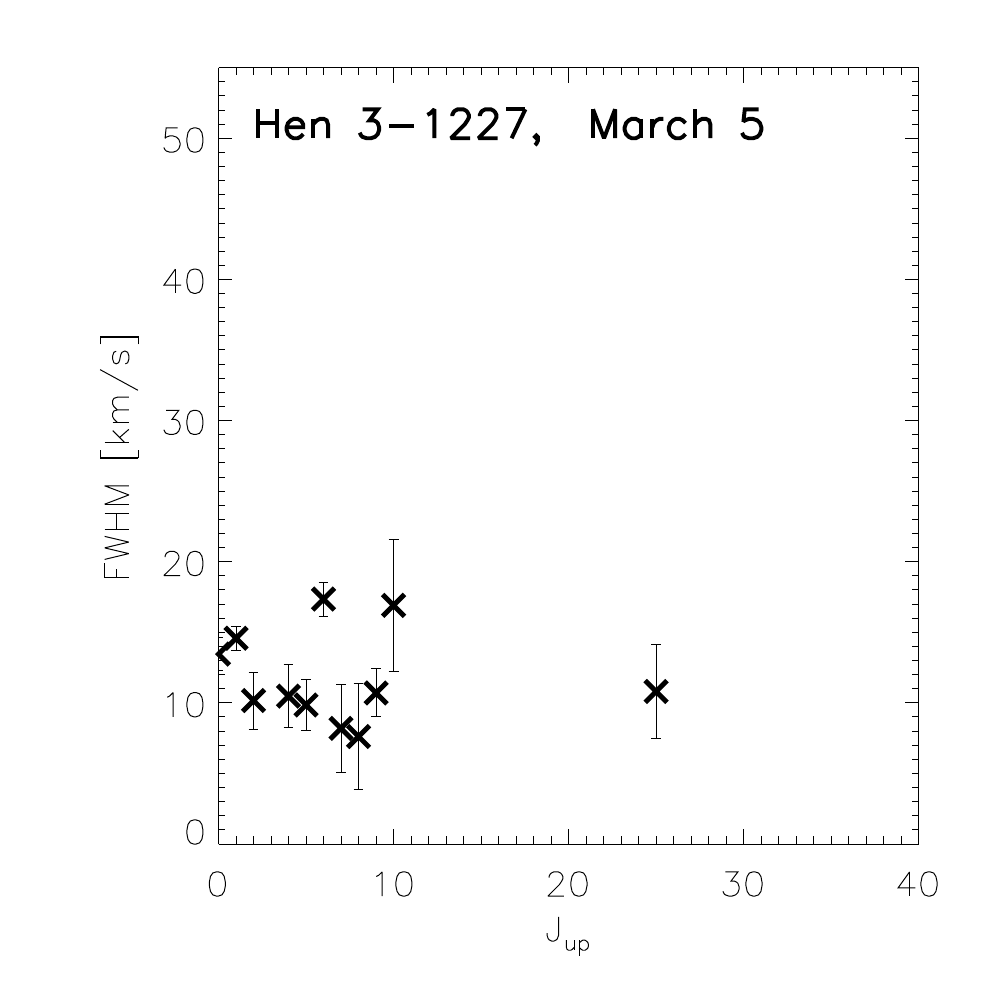}
\end{minipage}
\begin{minipage}[r]{0.4\textwidth}
   \includegraphics[width=\textwidth]{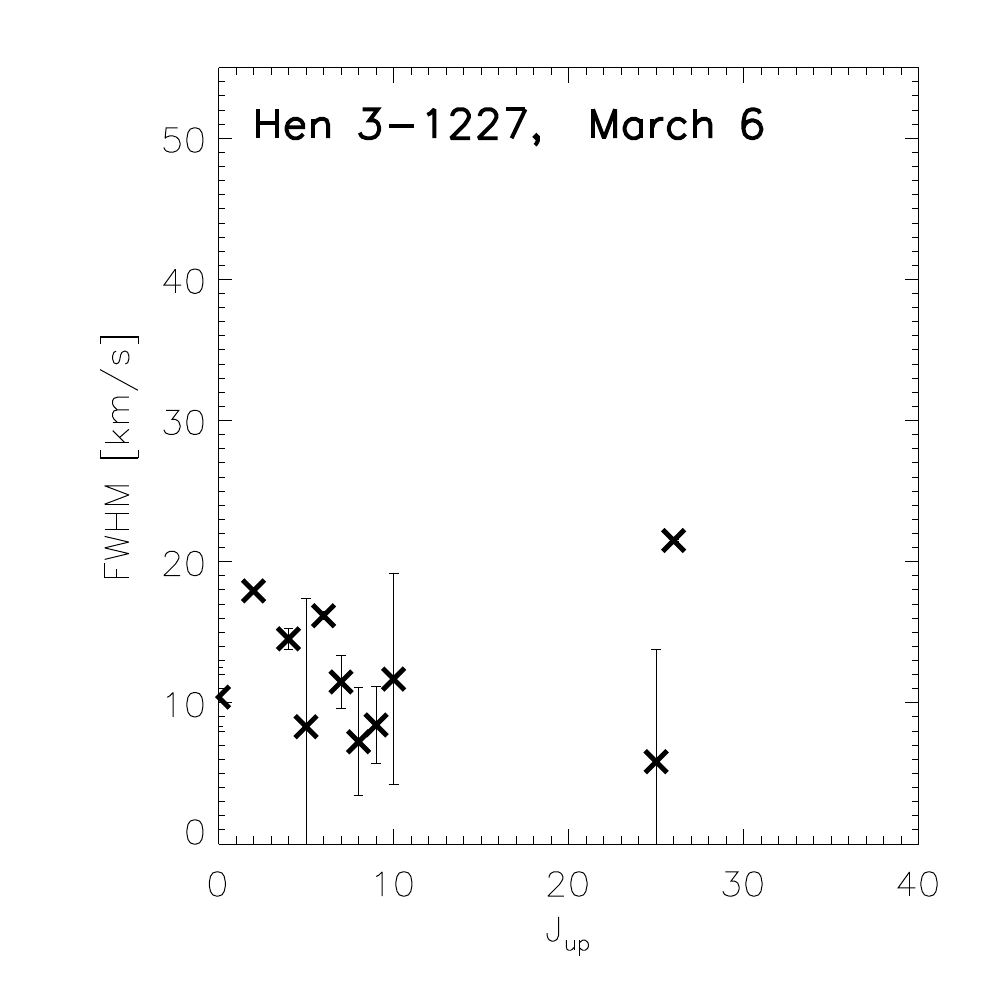}
\end{minipage}
\caption{ {FWHM as a function of $J_{\rm up}$ for the $v$=1-0 emission lines observed from our four CO detected sources. The source names are indicated on the plots. In the middle and bottom row FWHM collected for Hen~2-80 and Hen~3-1227 are shown for March 5, 2012 (left) and March 6, 2012 (right). The error bars are based on the variation that occurs in the FWHM when using the standard deviation of the nearby continuum as uncertainty for the continuum placement when fitting a Gaussian (in a few cases, where lines are very narrow, this approach leads to errors smaller than the plotting symbols).}
}
         \label{fwhm_all}
\end{figure*}

The CO ro-vibrational lines selected for analysis, are listed in Table \ref{table:COlines} (and shown in the appendix Figs. \ref{hd16_profile} to \ref{hen3v1_profile}). To quantify the line shape variations for each source, we show plots of  FWHM versus $J_{\rm up}$ in Fig. \ref{fwhm_all}.
The  FWHM plotted here is the FWHM of the fitted Gaussian. {A Gaussian does not always match the detailed line shapes well (however, a good line shape match is obtained in all cases except Hen~2-80), but is a good and consistent measurement of the width because the entire line profile is evaluated to derive a FWHM. A manual measurement of the FWHM, would be dependent on personal choice of the maximum location and hence difficult to reproduce.} For HD~250550, we fit a two component Gaussian (a central main component and a smaller blue shifted component) due to the presence of a shoulder on the blue side of the profile. The  FWHM values in Fig. \ref{fwhm_all} are those from the main Gaussian. {We do not show the FWHM of the secondary component since we are interested in emission from the disc and the origin of the secondary component is unknown.}

 The error bars in Fig. \ref{fwhm_all}, for all sources, are based on the variation in FWHM when shifting the continuum by using the standard deviation of the nearby continuum (for each line). The Gaussian overestimates the error for the FWHM of Hen~2-80. For Hen~3-1227, the error is underestimated due to the very narrow and peaky shape of the Gaussian.

From the  FWHM versus $J$ plots, we see that only the lines from HD~163296 and HD~250550, show differences in line profile shape through the line samples. 
To asses whether a correlation between FWHM and $J$-value is present for any of the sources, we have used the ASURV (Astronomy SURvival analysis) code \citep{isobe1986}, to perform Kendall's Tau test for each CO detected source. For HD~250550 we found a probability below $10^{-4}$ that a correlation is not present (even if we disregard the three high $J$ lines a correlation is still highly probable) while the other three sources (HD~163296, Hen~2-80, and Hen~3-1227) show probabilities of 70\%-84\% that a correlation is not present. Thus, there is a very strong correlation between FWHM and $J$-value in the data from HD~250550.

\begin{figure*}[!htbp]
\centering
\begin{minipage}[l]{0.4\textwidth}
   \includegraphics[width=\textwidth]{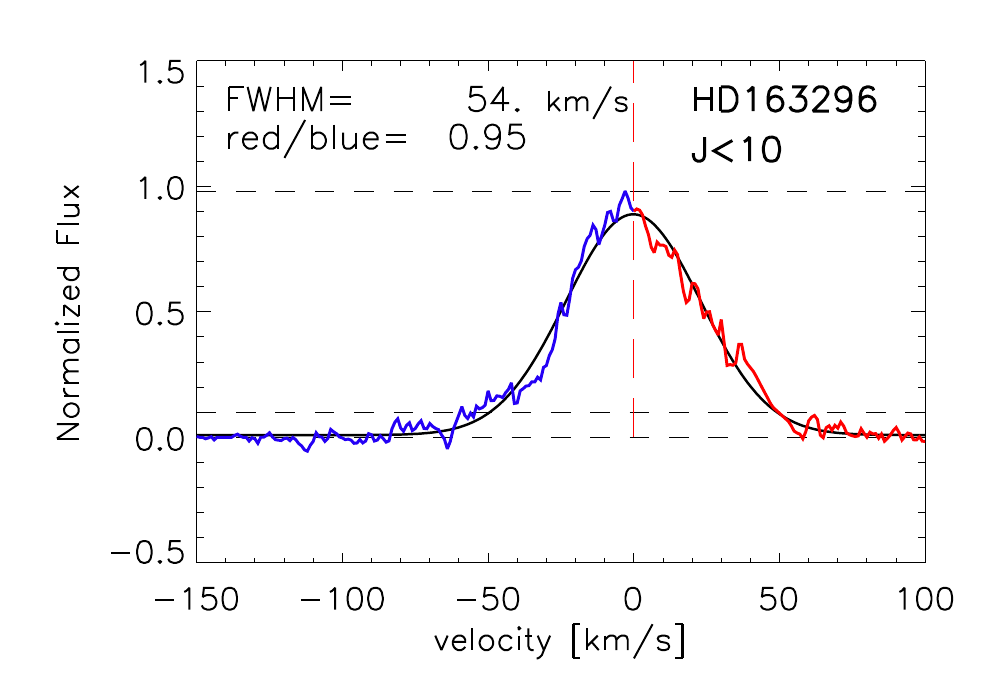}
\end{minipage}
\begin{minipage}[r]{0.4\textwidth}
   \includegraphics[width=\textwidth]{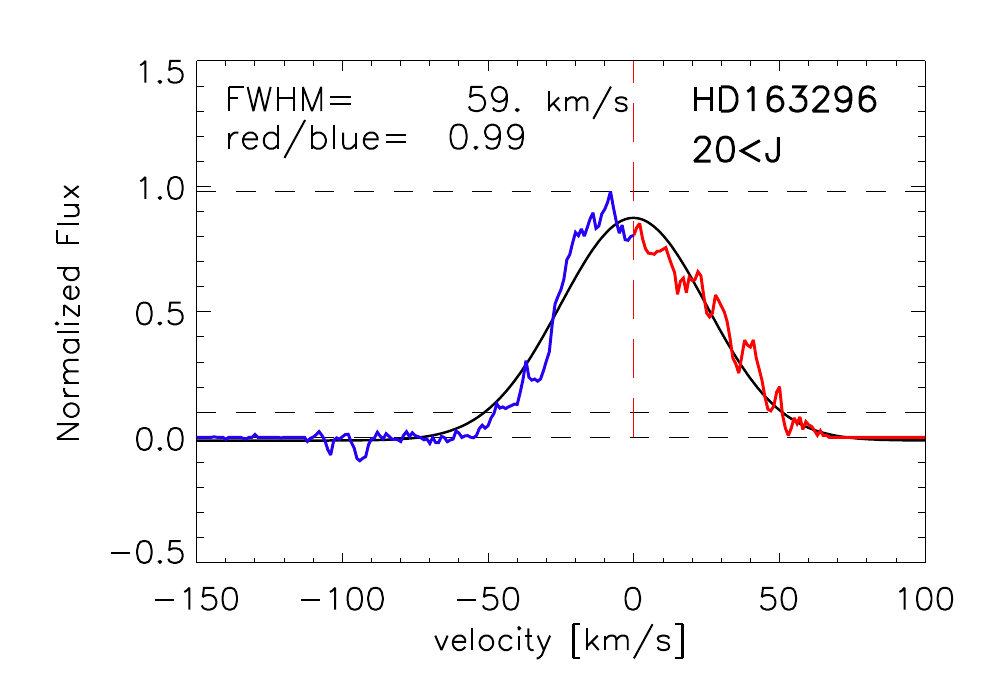}\\
\end{minipage}
\begin{minipage}[r]{0.4\textwidth}
  \includegraphics[width=\textwidth]{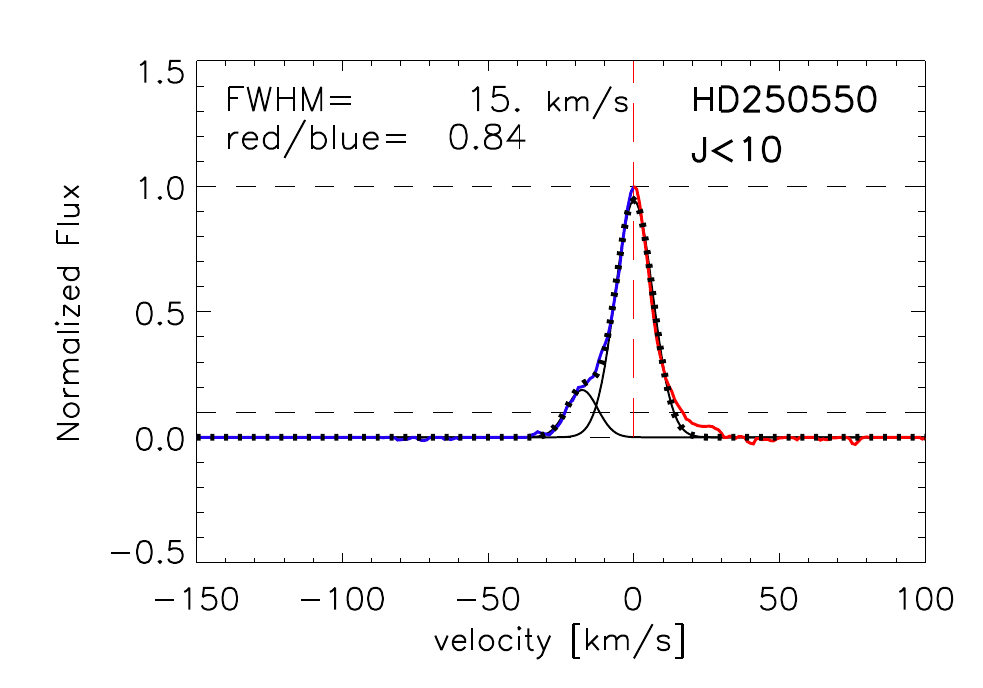}
\end{minipage}
\begin{minipage}[r]{0.4\textwidth}
   \includegraphics[width=\textwidth]{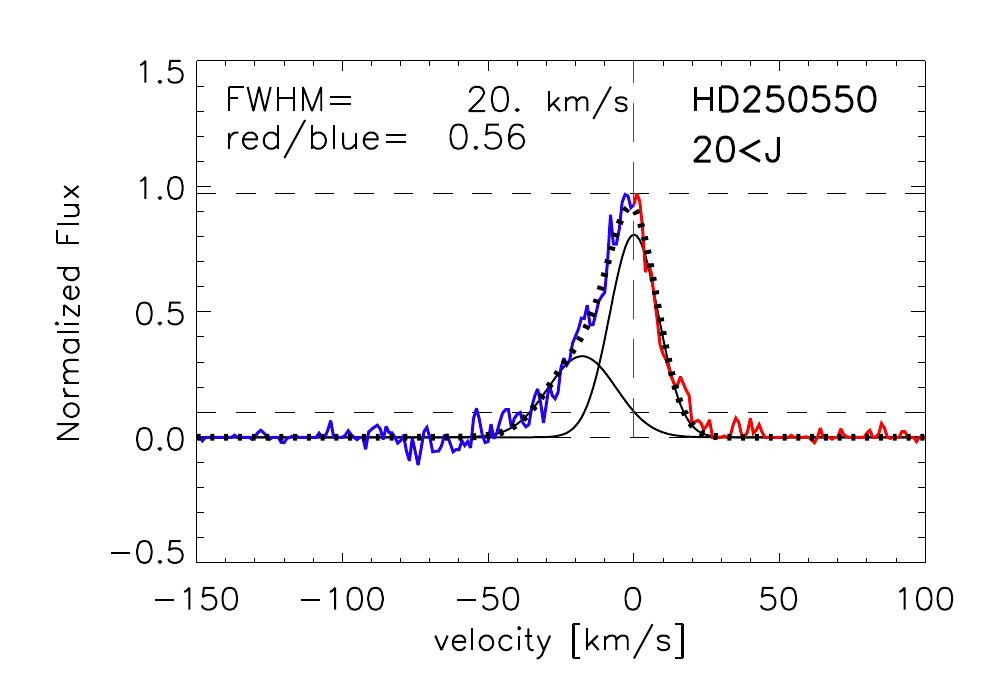}\\
\end{minipage}
\begin{minipage}[r]{0.4\textwidth}
   \includegraphics[width=\textwidth]{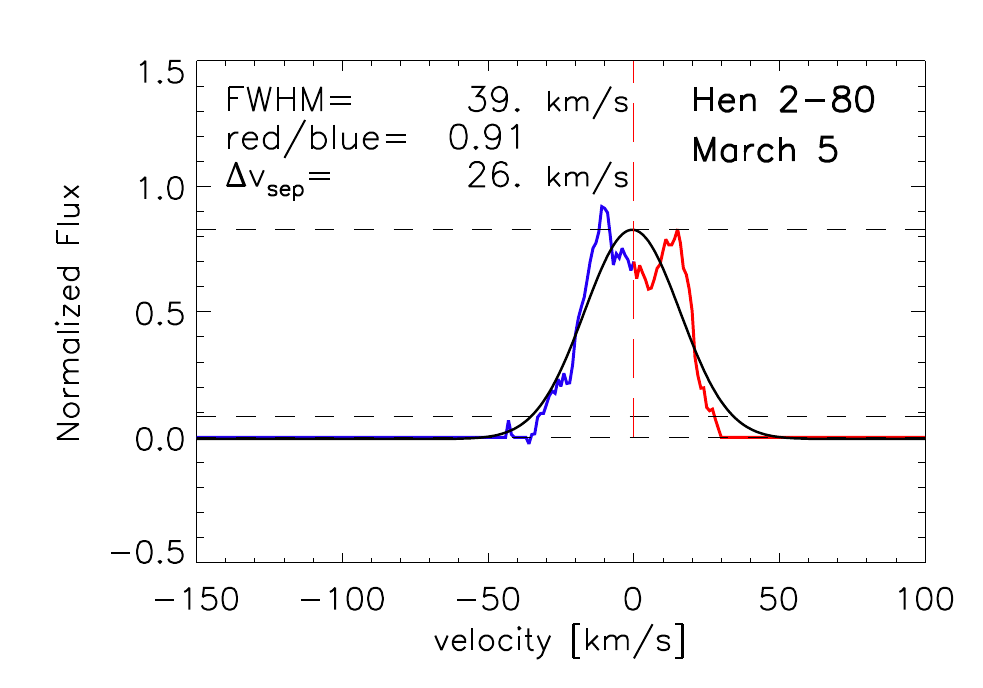}
\end{minipage}
\begin{minipage}[r]{0.4\textwidth}
   \includegraphics[width=\textwidth]{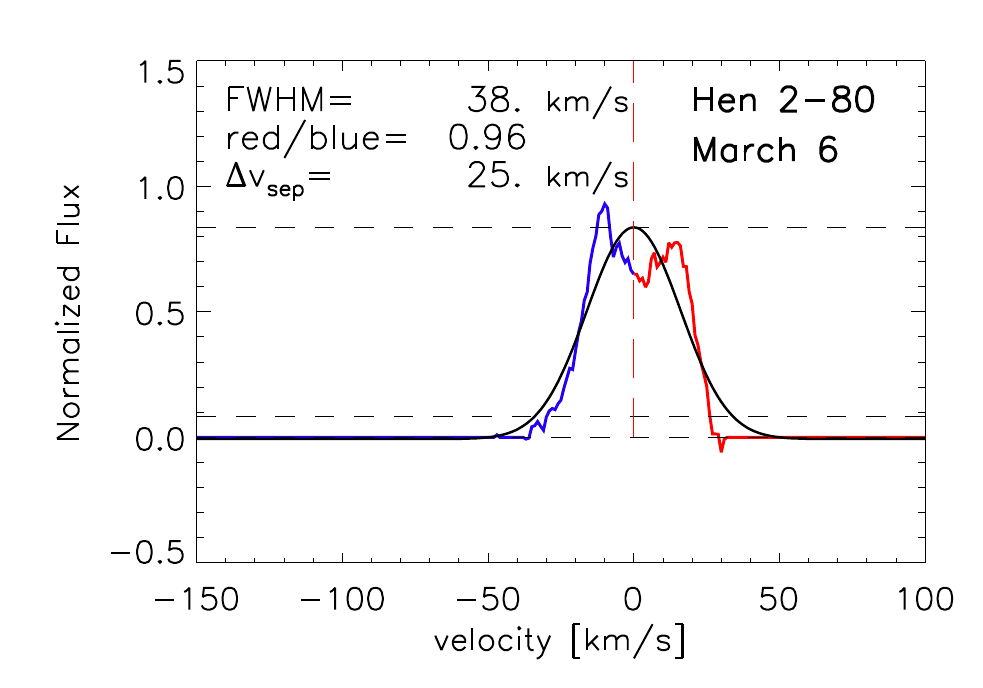}\\
\end{minipage}
\begin{minipage}[r]{0.4\textwidth}
   \includegraphics[width=\textwidth]{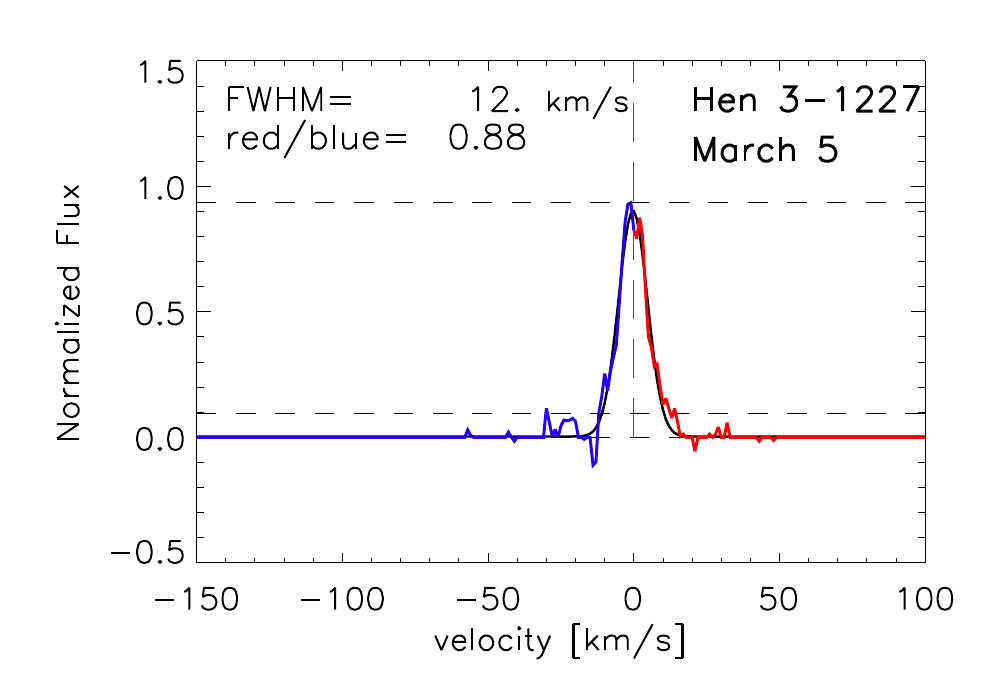}
\end{minipage}
\begin{minipage}[r]{0.4\textwidth}
   \includegraphics[width=\textwidth]{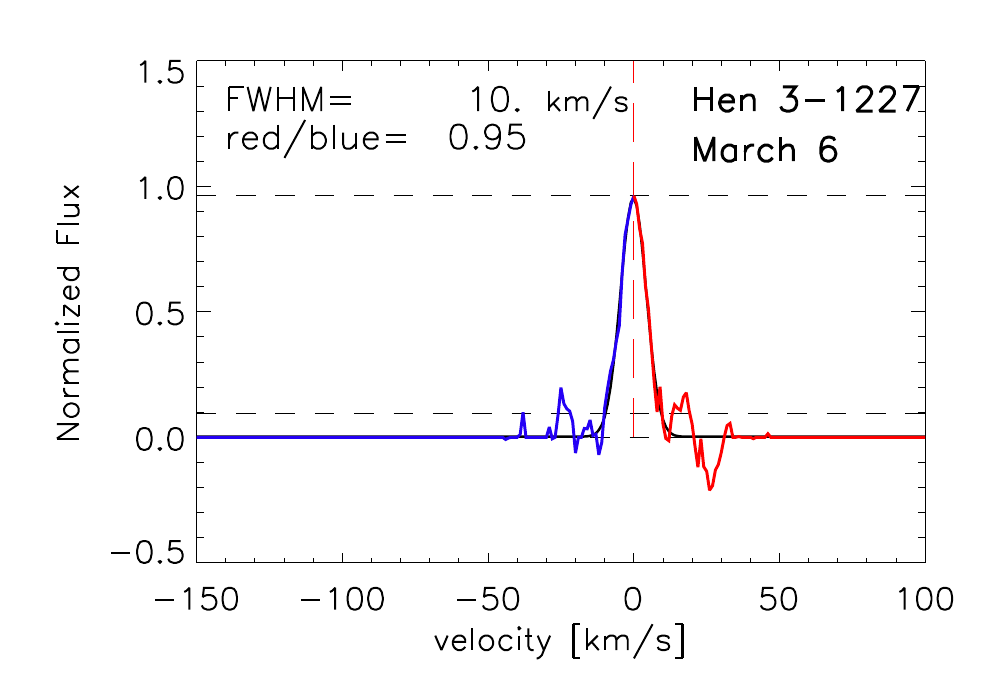}
\end{minipage}
\caption{Normalised median of all \element[][12]{CO} $v$=1-0 lines observed from the four sources where CO is detected. The source names are indicated on the plots. In the first two rows (for HD~163296 and HD~250550), we show low $J$ ($J$<10) medians on the left and high $J$ ($J$>20) medians on the right. In the lower two rows (for Hen~2-80 and Hen~3-1227), we show medians for all $J$ collected on two separate nights: March 5, 2012 on the left, March 6, 2012 on the right. In all frames, the solid black line shows the Gaussian fit, the red dashed line is the centre of the Gaussian or zero velocity, and the red- and blue-shifted side of the line profile are indicated with a red and blue lines respectively. The  FWHM of the Gaussian is printed at the top left of each frame together with the ratio of the integrated red and blue side of the line. For Hen~2-80 the peak separation of the median is also shown. For HD~250550 the sum of the two Gaussians is shown as a dotted black line.}
         \label{fig_median}
\end{figure*}

In Fig. \ref{fig_median}, we show the medians of the $v$=1-0 lines from each source. 
For HD~250550, splitting high and low $J$ into two separate medians is useful since the FWHM increases with $J$. For HD~163296, the FWHM does not have a clear correlation with $J$, however, the shape of the single peak is asymmetric for high $J$ lines, while symmetric for low $J$ lines. {Hence, even though there is no significant change of the FWHM over $J$ (within error bars or as shown by the Kendall's Tau test), we also split the high and low $J$ into two separate medians for HD~163296}. For the last two sources (Hen~2-80 and Hen~3-1227) high and low $J$ medians do not show any significant differences, instead separate medians (including all $J$) from the two observation nights are shown. In Table \ref{table:fwhm} we show the FWHM results from the line profile analysis and in Tables \ref{tab:hd16} to \ref{tab:hen3} in the appendix we show line luxes for all the collected lines.


\begin{table*}[!htbp]
\caption{Results from the line profile analysis. }             
\label{table:fwhm}      
\centering                          
\begin{tabular}{cc|cccccccccc}        
\hline    
\rule{0pt}{2.2ex} {\element[][12]{CO}}\\
\hline    
 Object&Group&FWHM&$3\sigma_{\rm}$&FWHM&$3\sigma_{\rm}$ &FWHM&$3\sigma_{\rm}$&  FWHM	&$3\sigma_{\rm}$	\\
&& $v$=1-0, ($J$<10)&	& $v$=1-0, ($J$>20)&	&	  $v$=2-1&	&\element[][13]{CO} $v=1-0$\\
&&	[km/s]&	&	[km/s]&&	[km/s]&\\
\hline            
HD~163296& II		&54.5&3.9		&59.2&4.4	&1275$_{-950}$	&--\\
HD~250550& I	&15.2&0.2		&19.7&2.	&21.5&2.6&14.1&1.9	 \\
Hen~2-80  05&	I	&38.9&3.2		&42.4&4.6	&43.7&4.8&42.9&4.2	\\
Hen~2-80  06&	I	&39.3&4.4		&38.9&3.1		&43.0&3.8&39.3&4.2	 \\
Hen~3-1227 05&I	&11.5&0.7 		&10.8&3.3		&--&&12.4&15.8			\\
Hen~3-1227 06&I	&10.6&1.3		&10.4&4.0		&--&&19.2&6.8\\
\hline                           

\end{tabular}
\tablefoot{The averaged FWHM with errors, are derived from the selected lines (listed in Table \ref{table:COlines}) for \element[][12]{CO} $v$=1-0 (FWHM from high and low $J$ separately), \element[][12]{CO} $v$=2-1, and \element[][13]{CO}  $v$=1-0.}
\end{table*}

\section{Discussion of the observational sample} \label{sec:discus}

\subsection{{CO emitting region and disc classification}} \label{trend}
Table \ref{tab:allHAE} presents an overview of observed properties from the discs in our sample, the discs in the sample by {\citet{plas2014}} and discs from \citet{salyk2011}.
\begin{sidewaystable*}
\caption{Source overview from this work and the literature. }             
\label{tab:allHAE}      
\centering                          
{\small 
\begin{tabular}{llllllllllllllllllll}        
\hline            
ID			&  Group  	& Ref. 	&  FWHM$_{\rm  J<10}$  &  FWHM$_{\rm 10<J<20}$ & FWHM$_{\rm J>20}$ 	& FWHM  	&  $R_{\rm CO 10\%}$  	& $i$           & [30/13.5]  & dust/gas        & $M_{*}$       & $\log(L_{*})$   &$\log(T_{\rm eff})$  &Sp. type  \\
                            &                &                & [km/s]                               & [km/s]                                     & [km/s]                              & vs $J$   & [au]                                    & [$^\circ$] & [K]             &	consistent? & [M$_\odot$] &                          &                                   &                \\
\hline            
HD~100546	&	Ia	&vdP	&16.1$\pm$1.3	&16.2$\pm$1.2	&16.0$\pm$1.9	&constant		& $6.8\pm1.6$	 &	$42\pm5$		&3.5$\pm$0.2&yes& 2.4 & 1.62&4.02&B9Vne\\
HD~97048  \hspace{-2mm} 		&	Ib	&vdP	&18.2$\pm$2.4	&16.0$\pm$1.4	&16.1$\pm$1.8	&constant	  \hspace{-2mm} 	& $10.1\pm1.8$	& $42.8^{+0.8}_{-2.5}$		&5.9$\pm$0.4&yes& 2.5 & 1.42&4.00&A0pshe\\
HD~179218  \hspace{-2mm} 	&	Ia	&vdP	&16.0$\pm$5.1	&13.7$\pm$10.3	&17.7$\pm$5.4	&constant \hspace{-2mm} 		& $9.2\pm1.5$		& $57\pm2$		&2.4$\pm$0.2&yes& 2.7 & 1.88&4.02&B9e\\
HD~135344B  \hspace{-2mm} 	&	Ib	&vdP	&16.6$\pm$2.6	&12.1$\pm$1.5	&27.7$\pm$3.4	&increasing \hspace{-2mm} 	& $0.4\pm0.2$	& $14\pm3$		&10.9$\pm$0.3&no& 1.7 & 1.01&3.82&F8V\\
HD~101412  \hspace{-2mm} 	&	II	&vdP	&		---		&	---			&		---		&	---		& $0.6\pm0.1$	& $80\pm7$				&0.92$\pm$0.03& yes & 2.3 & 1.40&4.02&B9.5V	\\
HD~190073  \hspace{-2mm} 	&	II	&vdP	&15.6$\pm$2.6	&  ---	&		16.4$\pm$2.4		&	constant		& $1.9\pm2.8$		& $23^{+15}_{-23}$	&0.75$\pm$0.02&yes& 2.85 & 1.92&3.95&A2IVpe\\
HD~98922 \hspace{-2mm} 		&	II	&vdP	&18.1$\pm$0.9	&18.6$\pm$8.8	&26.1$\pm$4.1	&increasing \hspace{-2mm} 	& $4.2\pm2.2$	& $45$		&0.75$\pm$0.03&no& 2.2 & 2.95$^\ddagger$&4.02&B9Ve\\
HD~95881 \hspace{-2mm} 		&	II	&vdP	&34.4$\pm$4.8	&29.2$\pm$6.9	&47.2$\pm$4.4	&	increasing		& $1.9\pm0.6$		& $55$	&0.77$\pm$0.05& yes	& 2.0 & 1.34&3.95&A2III/IV  	\\
HD~150193 \hspace{-2mm} 	&	II	&vdP	&54.1$\pm$1.8	&		---		&	$60.0\pm2.2$		& increasing	& $0.5\pm0.2$ & $38\pm9$			&1.42$\pm$0.05&yes& 2.3 & 1.19&3.95&A1Ve\\
HD~104237 \hspace{-2mm} 	&	II	&vdP	&		---		&	---			&	---			&	---		& $0.3\pm0.5$ & $18^{+14}_{-11}$						&1.28$\pm$0.03& --- & 1.96 & 1.53&3.92&A4Ve+sh  		\\
\hline
Hen~2-80	 \hspace{-2mm} 	&	I	&B15&39.1$\pm$2.7	&37.6$\pm$2.9	&40.7$\pm$2.8	&constant	 \hspace{-2mm} 	&1.9$\pm1.1$ & $45\pm15^{\ast}$					& --- &no& $5.1^\ast$ & >2.7&4.15&B6Ve\\
HD~250550 \hspace{-2mm} 	&	Ia	&B15&15.2$\pm$0.2	&16.8$\pm$1.9	&19.7$\pm$2.8	&increasing \hspace{-2mm} 	&0.5$\pm2.4$	& $10$				&2.5$\pm$0.1&no& 3.6 & 1.3&4.03&B8-A0IVe\\
HD~163296 \hspace{-2mm} 	&	II	&B15&54.5$\pm$3.9	&50.9$\pm$4.9	&59.2$\pm$4.4	&	constant		&0.4$\pm0.1$	& $46$				&2.0$\pm$0.1&yes& 2.3 & 1.40&3.94&A3Ve\\
Hen~3-1227 \hspace{-2mm} 	&	I	&B15&11.0$\pm$0.7	& ---		&10.6$\pm$2.6	&constant	 \hspace{-2mm} 	&4$\pm6$	 & $15\pm10^{\ast}$		& --- &			yes& $7^\ast$ & >2.0&4.30&Be\\
\hline            
ID			& Group &Ref.   	& FWHM$_{\rm  J<10}$	& FWHM$_{\rm 10<J<20}$	& FWHM$_{\rm J>20}$	 & FWHM  & $R_{\rm CO gauss}$  &   $i$         & [30/13.5]  & dust/gas      & $M_{*}$       & $\log(L_{*})$   & $\log(T_{\rm eff})$  & Sp. type   \\
                            &             &          &   [km/s]                               &  [km/s]                                         & [km/s]                        &   vs $J$   &  [au]                               & [$^\circ$] & [K]            & consistent? & [M$_\odot$] &                          &                                    &       \\
\hline            
AB Aur	 \hspace{-2mm} 	&	Ia	&B4,S11&	$21.5\pm1.7$		&	$21.6\pm1.7$		& $27.5\pm4.1$			&	constant	&0.59	& $21$			&4.50$\pm$0.10&no& 2.4 & 1.67&3.98 &B9neqIV/V   \\
LkHa 330	 \hspace{-2mm} 	&	?	& S11&			&			&28			&	---	&2.10	& $42\pm10$			& --- &	---	& 2.5 	&1.20&3.77 & G2\\
MWC480 \hspace{-2mm} &II	&B4,S11&	 $56.0\pm11.3$		&	$58.4$		& $90.9\pm21.8$			&	increasing	&0.08	& $26\pm 7$			&1.19$\pm$0.03&		yes& 1.65 &	1.06	&3.94 &A3Ve   \\
MWC758 \hspace{-2mm} &Ia	&B4,S11&	 $23.4\pm2.4$		&	$26.6$		& $31.0\pm2.1$			&	increasing	&0.19	& $16\pm 4$		&4.1$\pm$0.2&		no& 1.80 &	1.04	&3.89 &A5IVe  \\
VV Ser \hspace{-2mm} 		&	II	&B4,S11&	$43.0\pm2.9$		&	$43.5\pm3.0$		& $55.9\pm4.7$			&	increasing	&0.72	& $70\pm5$				&0.79$\pm$0.02&	yes&	 2.6  & 1.69	&3.95 &B6\\
\hline                           
\end{tabular}
}
\tablefoot{Collected observed properties of known HAeBe discs, including the discs from our sample=B15, the discs from {\citet{plas2014}}=vdP and discs from \citet{salyk2011}=S11, where the FWHM have been taken from \citet{blake2004}=B4 if indicated. The $R_{\rm CO10\%}$ is the radius derived from the Keplerian velocity at the half width at 10\% of the maximum in the averaged line profiles. The column labeled \textit{dust/gas consistent?}, indicates whether there are inconsistencies between the dust inner radii based on the SED classification and gas radii based on the CO ro-vibrational emission lines. For HD~101412 and HD~104237, CO ro-vibrational lines were detected, but due to blending and telluric residuals, the FWHM could not be measured reliably. [30/13.5] values are taken from Table 1 of \citet{maaskant2014b}. An asterisk behind the inclination and mass indicates that in absence of a measurement, an typical inclination consistent with the line profile shape and a stellar mass estimated from the spectral type was assumed. $^\ddagger$ \citet{blondel2006} note the distance ambiguity and that this star is likely a spectroscopic double-star.}
\end{sidewaystable*}

\begin{figure}[!htbp]
\begin{center}$
\begin{array}{cc}
  \includegraphics[width=0.46\textwidth]{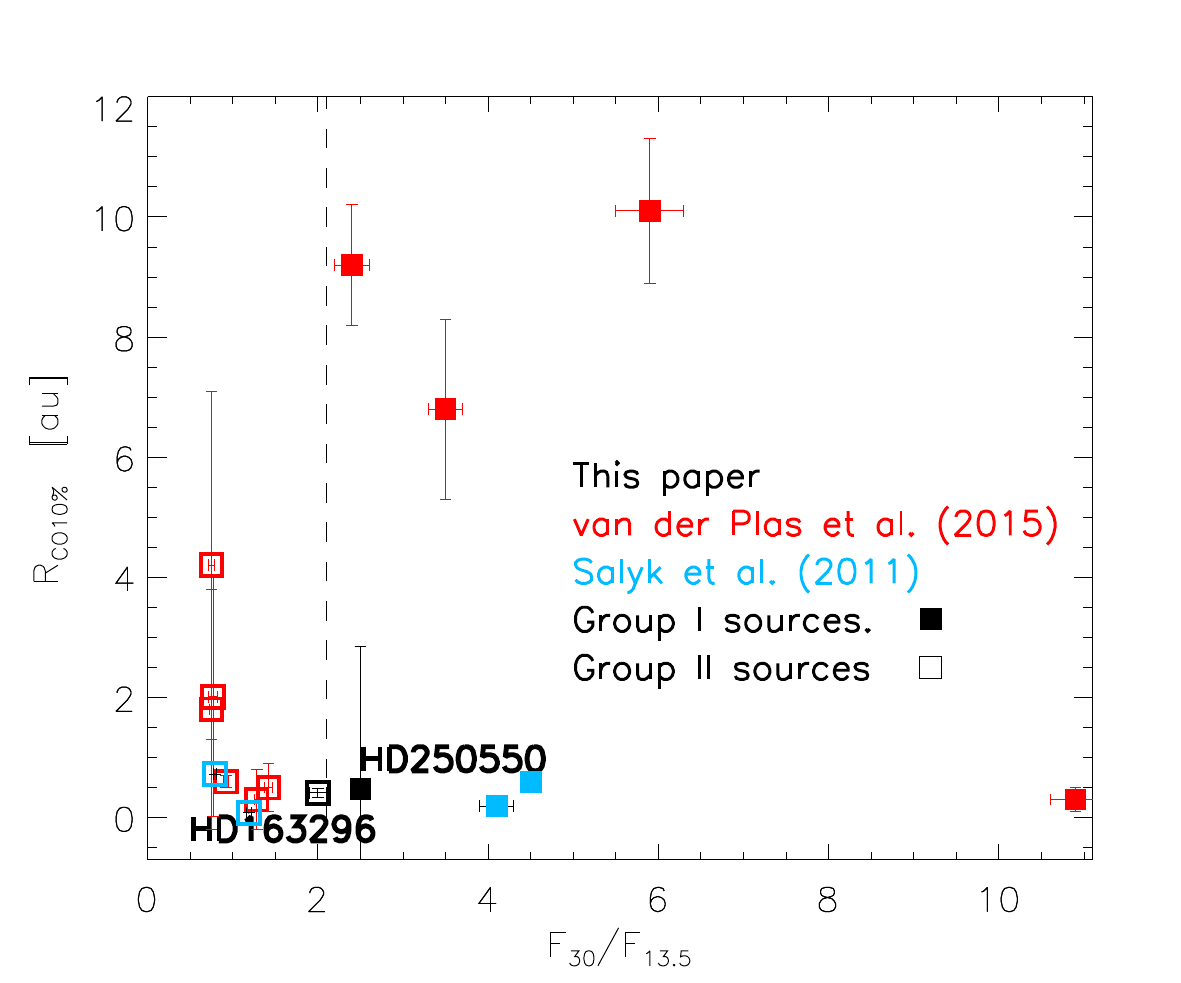}
\end{array}$
\end{center}
\caption{$R_{\rm CO10\%}$ versus the  30~$\mu$m over 13.5~$\mu$m continuum flux, from this study (black), together with those from studies by {\citet{plas2014}} (red) and \citet{salyk2011} (blue). Group I discs are plotted as filled squares while group II discs are plotted as open squares.}
 \label{R10_F30}
\end{figure}

A trend discussed by {\citet{plas2014}}, is the tendency for group II discs to be emitting CO ro-vibrational emission from the dust sublimation radius (wide profiles) and group I discs to be emitting from much further out (narrow profiles).
To explore this trend, we plot in Fig. \ref{R10_F30} the $R_{\rm CO10\%}$ versus the 30/13.5 $\mu$m continuum flux for the sources listed in Table \ref{tab:allHAE}. {The $F_{30/13.5}$ ratio measures the MIR excess slope seen in the SEDs and separates discs that are classified as group I or group II into two regions. Group I discs are seen to fall above the $F_{30/13.5}$=2.1 limit while discs classified as group II are seen to fall below this limit \citep{maaskant2014b,acke2010}. This scheme is in agreement with the classification from the $L_{NIR}/L_{IR}$ versus [12]-[60] colour diagram but we only have the $F30/F13.5$ ratio for two of our observed sources (HD~163296 and HD~250550). The reason we use the $F30/F13.5$ ratio here (instead of the $L_{NIR}/L_{IR}$ versus [12]-[60] colour diagram) is that it offers us a simple way to display in one diagram the comparison between group I/group II classification and inner radius of the CO ro-vibrational emitting region.
$R_{\rm CO10\%}$ is an estimate for where the onset of the CO emission is located; it is derived from the half width at 10\% of the maximum (HW10M) in the median line profiles
\begin{equation}
{R_{\rm CO10\%}=\frac{G\cdot M_*}{\rm HW10M^2} \,{\rm sin}^2i} \quad .
\end{equation}
Here $G$ is the gravitational constant, $M_*$ is the stellar mass, and $i$ is the inclination of the disc. The HW10M are calculated from the Gaussian fits.} If we exclusively consider the van der Plas sample, there was indeed a sharp jump in radii for the onset of CO emission (with only one {exception}, HD~135344B) when comparing group I and group II discs (left and right of the dashed line in Fig. \ref{R10_F30}). Considering now also the new sources from our own sample and those from \citet{salyk2011}, the {previously found sharp jump disappears}. Three more sources (four in total) with CO emitting regions close to the star, fall in the regime of discs classified as group I from the SED. 
Thus, comparing a larger sample of sources \citep[e.g.][and this paper]{salyk2011, plas2014} leads to a much more diverse picture from these discs than seen from a smaller sample, the \citet{plas2014} sample alone. {The distribution of group I and II sources over $R_{\rm CO10\%}$ is very similar except for the three sources with $R_{\rm CO10\%}>6$~au. These are interestingly the three sources for which \citet{plas2014} find fluorescence from the vibrational CO temperatures being much larger than the rotational ones ($T_{\rm vib} \gg T_{\rm rot}$).}

Discs whose gas diagnostics do not correlate with their SED group classification, could be discs {where the radial dust and gas distribution differ}. {Such cases} have been noticed before. \citet{fedele2008} found evidence of discs flared in the gas, but self-shadowed in the dust when comparing [\ion{O}{I}] emission, and 10 $\mu$m dust emission {(vertically different distribution of dust and gas)}. \citet{brown2009} resolved dust gaps in three discs (HD~135344B, LkH$\alpha$ 330, SR 21N) with continuum observations (880~$\mu$m), while gas observations (CO, 4.7~$\mu$m) show gas still present in the inner regions of the same discs. Furthermore, PAHs are observed from within dust gaps \citep{maaskant2014b} and ALMA observations also reveal gas in dust depleted regions \citep[e.g.][]{casassus2013, bruderer2014}.
The presence or absence of a dust gap cannot necessarily be deduced from the SED alone \citep[see e.g. RY Tau,][]{Isella2010}, hence, imaging data could be useful to help clarify the geometry of the dust in the inner disc. In addition, comparisons between SED and gas lines from other sources that seem to show dust/gas {radial} de-coupling could also add to our understanding.

\subsection{Silicate features and disc geometry}
\label{subsect:silicate}

{In the following, we compare the inner dust radii based on SED classification with those found for the CO ro-vibrational emission. In Table  \ref{tab:allHAE}, we identify} six sources {where these two do not agree.} Five group I sources show CO ro-vibrational transitions emitted from close to the star {($< 2$~au, smaller than typical group~I dust gap/hole sizes)}, and one group II source shows CO emission from {$4.2$~au, which is well beyond the dust condensation radius}. 

{Out of the five group I sources which show $R_{\rm CO10\%}<2$~au,} three are labeled group Ia and one is labeled group Ib. Group Ia sources display emission features due to amorphous and crystalline silicates in their mid-infrared spectra whereas group Ib sources do not have these features. Group II sources are in this context all labeled group IIa since they all have silicate features.
{Thus, three of the sources that are inconsistent in SED classification and onset radii for CO} show a prominent silicate feature. 

\citet{maaskant2013} and \citet{khalafinejad2015} showed that the presence of large gaps or holes in the inner discs will cause these silicate features to disappear. Consequently the presence of  a prominent silicate component observed from group Ia sources could indicate a gap/hole small enough to keep the dust at temperatures required for the silicate emission feature. {HD~100546 and HD~179218 are two group Ia sources with a confirmed gap and a detection of the silicate feature. In HD~100546 that feature arises from the wall at 10-13 au; in HD~179218, the inner radius of the outer disk component is $~\sim 15$~au \citep{fedele2008}.}  On the other hand \citet{khalafinejad2015} found a very weak tentative detection of silicate emission features from HD~100453, thus a source on the border of a group Ia/Ib identification. This disc has a larger gap of $\sim$17 au and thus lower temperatures ($T\sim$160 K) in the inner regions of the outer disc. The distance of the inner wall of the outer disc can thus be critical {for the presence of a silicate feature, or turned around, the presence or absence of this feature may distinguish between small and large dust gaps/holes in group~I sources. We use the above discussion to devise from the SEDs and $R_{\rm CO10\%}$  in the following paragraphs a finite set of possible inner disk geometries.}

\begin{figure}[!htbp]
\begin{center}$
\begin{array}{cc}
  \includegraphics[width=.47\textwidth]{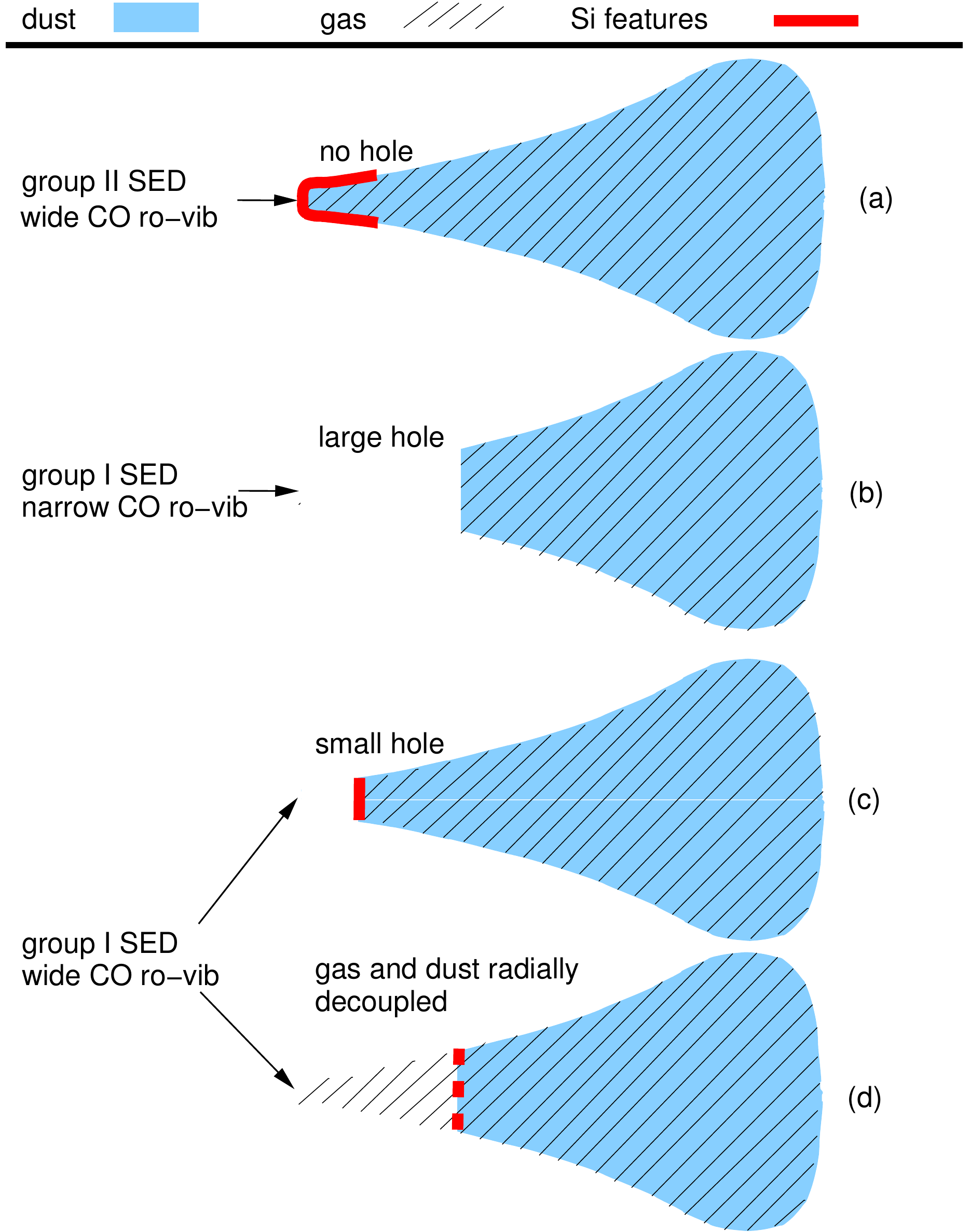}
\end{array}$
\end{center}
\caption{A sketch of various possible disc geometries: a) A continuous disc without a hole/gap, resulting in a group II SED with {silicate features} and broad CO ro-vibrational lines emitted from close to the star. b) A disc with a large hole, resulting in a group I SED without {silicate feature} (group Ib) and with narrow CO ro-vibrational lines emitted from the outer disc wall. c)  A disc with a small hole, resulting in a group I SED with {silicate feature} (group Ia) and with broad CO ro-vibrational lines emitted from the outer disc wall (in this case close to the star). d)  A disc with a (small or large) dust hole, resulting in a group I SED (with or without silicate features) and broad CO lines. The {radial gas distribution differs from that of} the dust and the CO lines are thus emitted from within the hole}
 \label{fig:Si_gI/gII}
\end{figure}

In Fig. \ref{fig:Si_gI/gII}, we sketch four possible inner disc geometries and link them to the SEDs: a) A no hole disc with silicate features present (group IIa SED). b) A disc with a large hole and no silicate features (group Ib SED). c) A disc with a small hole and silicate features present (group Ia SED). d) A disc {where the radial dust and gas distribution differs} (group I SED). Depending on the size of the dust hole, silicate features may or may not be present.

In the general view, group I sources with broad CO lines detected could be interpreted in one of two ways: 1) These discs have small dust+gas gaps/holes meaning that silicate features should be detectable and that CO lines are emitted from the disc wall beyond the small gap/hole  (sketch c in Fig. \ref{fig:Si_gI/gII}). 2) These discs have gaps/holes only in the dust and therefore display group I behaviour in the SED, while CO lines are emitted from inside the dust gap. The dust and gas are in this case {distributed differently} (sketch d in Fig~\ref{fig:Si_gI/gII}). 
HD~250550,  AB~Aur and MWC~758 (all group Ia sources) are consistent with interpretation number one, since they have silicate features present. For Hen~2-80, the presence of the silicate feature is unknown. For this source both interpretations are possible. Observations of the silicate feature would thus be very helpful to place this object into the above suggested scheme.
HD~135344B, however, is labeled as a group Ib source and thus has no silicate features, implying that the dust gap is large. Thus, the onset of CO emission close to the star seems to require gas to be present within the large gap \citep[30 au,][]{maaskant2013}, and interpretation number two is more likely here. Meanwhile, \citet{carmona2014} recently constrained the structure of this disc with modelling, using multi-wavelength gas and dust observations, and indeed concluded that the CO emission originates inside the gap.

HD~98922 is the only group II source with CO emitted further out than expected, $\sim$4.2$^{+4}_{-2}$ au. However, considering the high luminosity of the central star ($\sim$890 $L_{\astrosun}$) the inner dust rim could be further out than typically seen for group II sources, consistent with the onset of the CO emission. {Unless the gas surface density is strongly depleted (CO column densities $N_{\rm CO} < 10^{15}$~cm$^{-2}$), CO can survive the harsh UV radiation field of the star, because the dissociating bands become optically thick. For example, CO bandhead emission at $2.3~\mu$m is frequently seen in the inner discs (inside 10 au) of young massive stars with $L_\ast>10^3$~L$\odot$ \citep[e.g.][]{ilee2013}. However, the interpretation based on the high luminosity hinges on the distance which is not well determined; in addition, HD~98922 could also be a spectroscopic double-star \citep{blondel2006}. Gaia should clarify this issue soon.}

\section{Modelling} \label{sec:mod}
{To support the discussion of the link between disc geometry and FWHM of the CO ro-vibrational lines, we explore modelled CO ro-vibrational lines from two distinctly different inner disc geometries (a disc with a gap and a disc without a gap) produced from two pre-existing ProDiMo models (these two models are presented as examples of geometries and are not meant to specifically match any of the observed discs in our sample).} With these models we produce samples of CO ro-vibrational lines to tentatively test the connection between inner disc geometry and FWHM versus $J$ behaviour.

ProDiMo is a thermo-chemical disc code that solves the gas heating and cooling and the gas chemistry self-consistently \citep{woitke2009}. The two models are parametrized disc structures originating from the radiative transfer codes \mbox{MCFOST} \citep{Pinte2006,Pinte2009} and \mbox{MCMax} \citep{min2009}.
\mbox{MCFOST} and \mbox{MCMax} are three dimensional Monte Carlo continuum radiative transfer codes that calculate the dust temperature and continuum radiation field in the disc.
The coupling between ProDiMo and \mbox{MCFOST} or \mbox{MCMax} is explained in \citet{Woitke2010}.
To model the CO ro-vibrational lines with ProDiMo we use the complete CO ro-vibrational molecular model described in \citet{thi2012}.

The first model is a flaring disc without a gap (Model $\#$1): We run ProDiMo on top of a previously published \mbox{MCMax} model that was fitted to the Q- and N-band images of the disc around HD~97048 \citep{maaskant2013}. In that study \citet{maaskant2013} modelled a series of discs without and with gaps of varying size to asses the need for gaps to match the observed SED and images. Model $\#$1 is a ProDiMo model run on top of the continuous disc model (no gap) from \citet{maaskant2013}, and with the PAH (Polycyclic Aromatic Hydrocarbons) abundance set  to $f_{\rm PAH}$=0.7.

\begin{figure*}[!htbp]
\begin{center}$
\begin{array}{cc}
   \includegraphics[width=.47\textwidth]{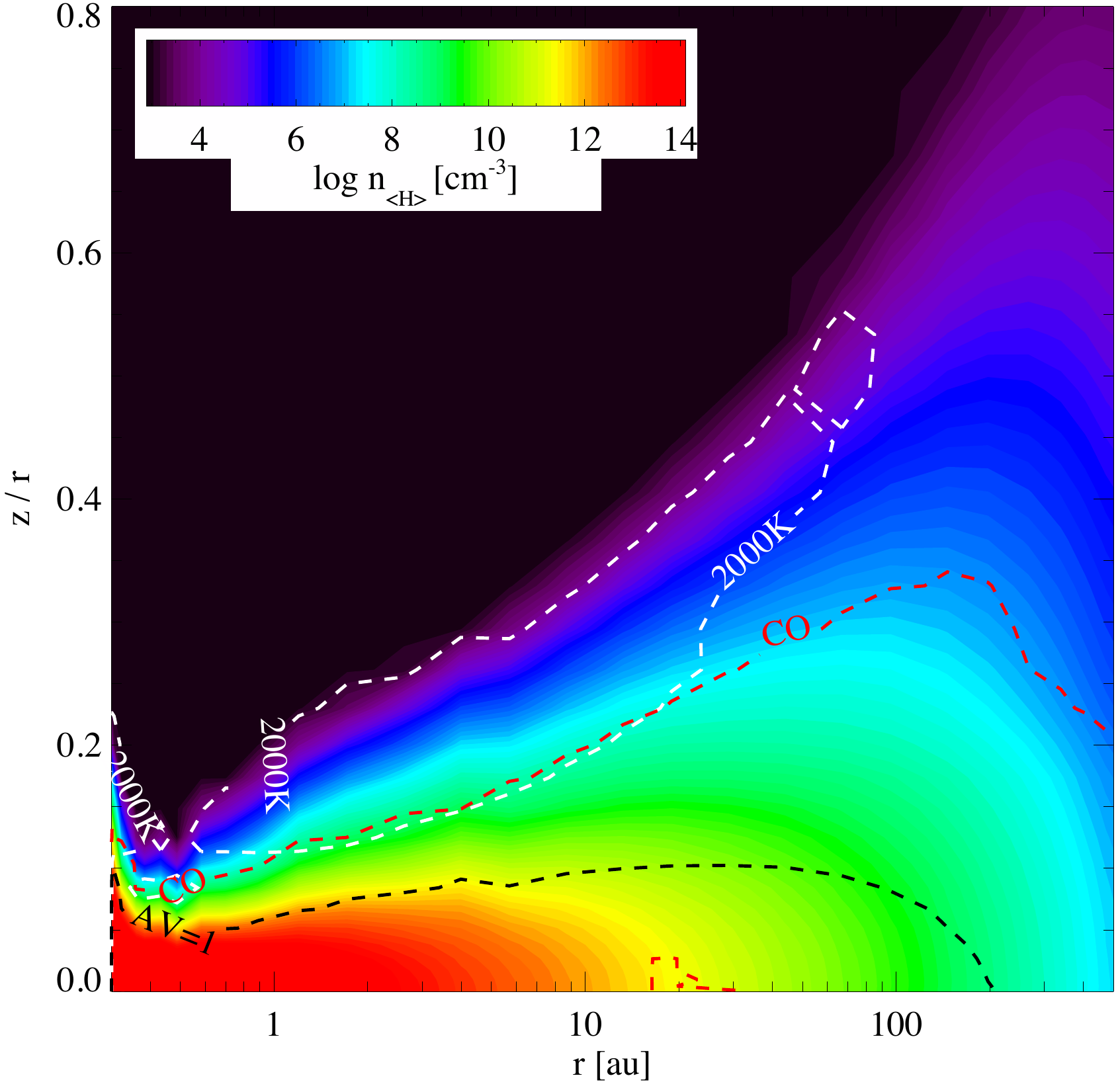}
   \includegraphics[width=.47\textwidth]{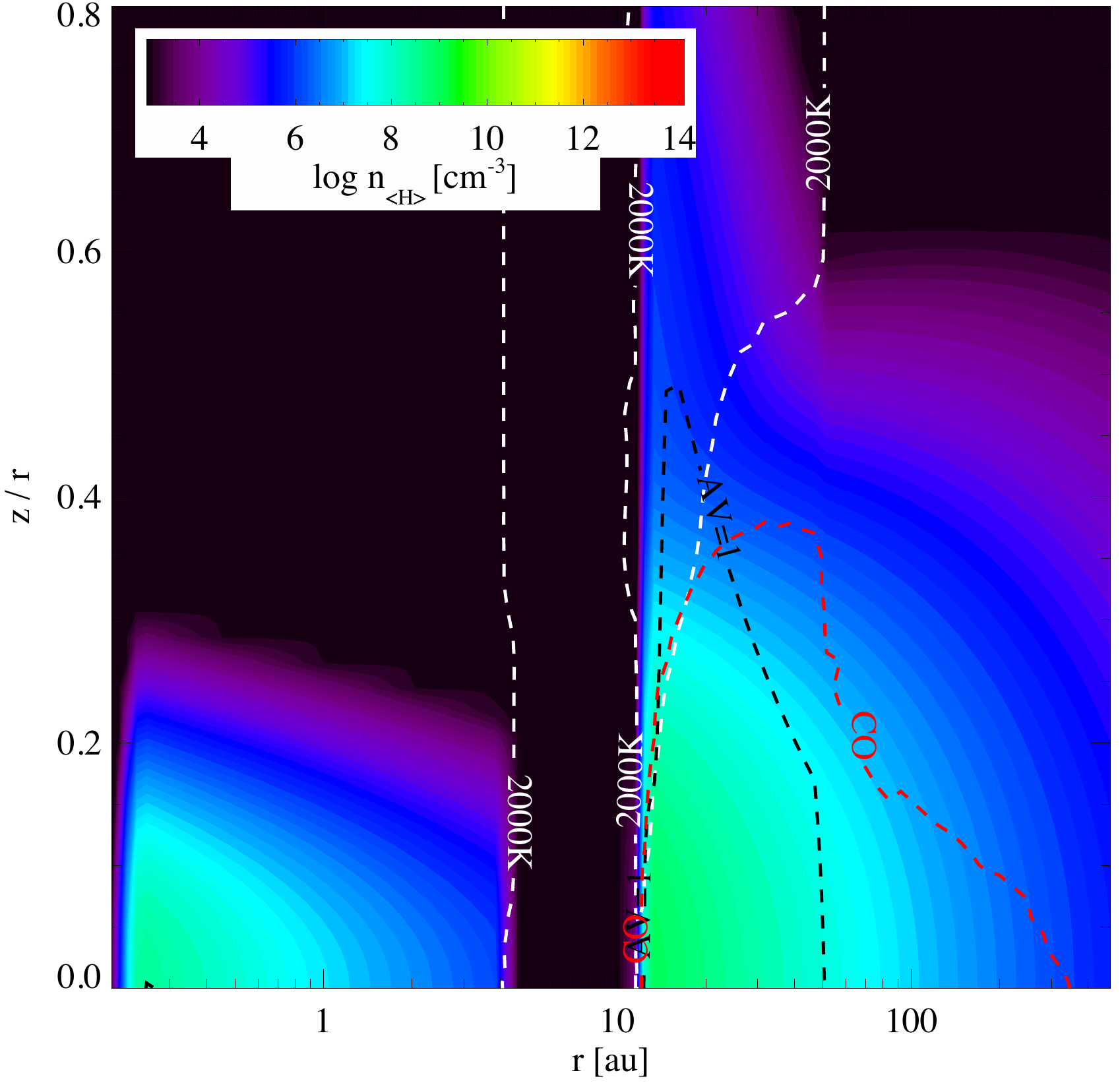}
 \end{array}$
\end{center}
\caption{Gas density distribution of the two disc models. Model $\#$1 is on the left and model $\#$2 is on the right. Contour lines showing $T_{\rm gas}$=2000 K (white),  min(A$_{v,\rm ver}$,A$_{v,\rm rad}$) = 1.0 (minimum of the radial and vertical dust extinction) (black), and the CO abundance of 10$^{-5}$ (red) are overplotted.}
         \label{nH}
\end{figure*}

The second model is a flaring disc with a gap from 4-13 au (Model $\#$2): We use a previously published model computed for the disc around HD~100546 \citep{thi2011,hein2014}. This is a ProDiMo model on top of an \mbox{MCFOST} dust model  \citep{tatulli2011,benisty2010} constrained by the observed SED and near- and mid- infrared interferometric  data. For this disc, UV fluorescence has been identified as the dominant excitation mechanism, and the {CO ro-vibrational emission starts} at the outer disc wall. 
The gas density distributions for both models are shown in Fig. \ref{nH}, and their key parameters are listed in Table \ref{tab:mod}. For more details on the two models, see \citet{maaskant2013} and \citet{hein2014}.

\begin{table}
\caption{{Key parameters of the two modelled disc geometries}}             
\label{tab:mod}      
\centering                          
\begin{tabular}{lll}        
\hline            

&Model $\#$1	& Model $\#$2		\\   
\hline                           
 $R_{\rm in}$&0.3 au	&0.19 au 		\\   
  $R_{\rm out}$&500 au	& 500 au		\\   
 $L_{*}$&40.70 $L_{\astrosun}$	& 	26 $L_{\astrosun}$	\\   
$M_{*}$ &1.84 $M_{\astrosun}$ 	& 2.4 $M_{\astrosun}$		\\   
$M_{\rm{disc}}$ & 6$\cdot10^{-4}$ $M_{\astrosun}$ 	&5.6 $\cdot10^{-4}$ $M_{\astrosun}$		\\   
Gap& none& 4-13 au\\
\hline 

\hline                           
\end{tabular}
\tablefoot{See Fig. \ref{nH} for the gas density distribution of the two models.}
\end{table}

\section{A comparison of models and observations} \label{sec:modeldiscus}
\subsection{{FWHM versus $J$}} \label{indiv}
To investigate the FWHM versus $J$ behaviour for the CO ro-vibrational lines in different disc types, {we use the two modelled examples of disc geometry. These two models show profoundly different FWHM versus $J$ behaviour (Fig. \ref{fwhm_hd97}). For model  $\#$1 (disc without a gap), the FWHM grows steadily with $J$ similar to the behaviour observed for HD~250550. For model $\#$2 (gap at 4-13 au), no $J$ dependence is seen for the FWHM of the lines. The width is constant for all transitions, indicating that the emission is always coming from the same narrow region, similar to the behaviour observed for Hen~2-80. 
For both models we performed tests, switching off the UV fluorescence. There are no significant changes in line fluxes for Model \#1; thus, UV fluorescence has no impact on the line fluxes in this modelled flaring disc without a gap. For Model $\#$2, UV fluorescence does have a large impact on the line fluxes of the higher $v$-bands ($v_{\rm{u}}$>1). This can be explained by the lack of CO gas at temperatures of thousands of Kelvin (due to the lack of CO in the inner 10 au), which are needed to excite CO to the higher $v$-bands via thermal excitation. UV fluorescent excitation, however, can occur at any gas temperature. Thus, for the modelled disc with a gap, UV fluorescence is an important excitation mechanism.
From these two models, we see that UV fluorescence does not always affect line fluxes in flaring discs. However, links between constant FWHM versus $J$ behaviour, the presence of a dust gap and the excitation mechanism could be possible. This would be consistent with the conclusions of the study by \citet{brittain2007} who found that UV fluorescence can be an important excitation mechanism in dust-depleted discs and the study of \citet{maaskant2013} who found evidence that discs classified from the SED as group I might all be discs with gaps.

{We can split our sample of sources in Table~\ref{tab:allHAE} in half, if we choose $R^{\rm crit}_{\rm CO10\%}\!=\!1$~au. Six out of eight sources with $R_{\rm CO10\%}\!>\!1$~au show constant FWHM versus $J$ behaviour, while this is true only for two out of eight sources with $R_{\rm CO10\%}\!<\!1$~au. One of these two sources is HD~163296 for which \citet{blake2004} and \citet{salyk2011} find strong profile differences between low and high $J$ lines (single versus double peaked). \citet{hein2016} found from a comparison between NIRSPEC and CRIRES spectra taken ten years apart possible variability related to a disc wind component in the CO ro-vibrational line profiles; this will affect the FWHM measurements. The other source is AB~Aur, where the high $J$ lines are systematically wider than the low $J$ lines, but within the error bars, the source is consistent with a constant FWHM versus $J$ behavior. Spectra with higher S/N are needed to clarify this case. The two sources showing increasing FWHM versus $J$ while having $R_{\rm CO10\%}\!>\!1$~au are HD~98922 and HD~95881. The stellar luminosity of HD~98922 ($\log L=2.95$) can push the dust condensation radius much further out than in normal Herbig discs (see Sect.~\ref{subsect:silicate}). We conclude that this trend in FWHM versus $J$ bears additional diagnostic value and we will investigate it further in an upcoming modeling paper.}

\subsection{{Disc inclination}} \label{inclination}
The single peaked shape and the narrow FWHM of lines from e.g. HD~250550 or Hen~3-1227 could be connected to a low inclination of the discs. With Model \#1, we tested the line behaviour for different inclinations. The $v_{\rm sep}$ versus inclination behaviour for a high and a low $J$ modelled line is shown in Fig.~\ref{incl_test}. At low inclination the peak separation drops below the instrument detection limit and we would observe a single peak. In Fig. \ref{fwhm_hd97} we furthermore see, that low inclination yields smaller FWHM and an overall flatter FWHM versus $J$ behaviour. 
An inclination of <35$\degree$ could already produce single peaked line profiles for low $J$ lines, but the model indicates that an inclination of <20$\degree$ is necessary in order to also see the higher $J$ lines as single peaks.

An inclination of this size is consistent with the inclination assumed for HD~250550 by \citet{fedele2011}. However, at such low inclinations, the FWHM of the lines in the model become much smaller than observed for HD~250550 (Fig. \ref{fwhm_hd97}). If the single peaks from HD~250550 are due to low inclination, the CO emission needs to onset very close to the star (with the main contribution from this region), in order to be consistent with the observed FWHM of 15-20 km/s.
Model \#1 has an extended emitting region for the CO, with $\sim$30\% of the line flux coming from 20-70 au for low $J$ lines, and $\sim$20\% of the line flux coming from 20-50 au for high $J$ lines. HD~250550 could also have an extended emitting region since the  FWHM versus $J_{\rm up}$ behaviour shows a steady increase in FWHM with $J$ level just like the model (Fig. \ref{fwhm_hd97}). In the model, this extended emitting region means that the peak separation of low $J$ lines becomes quite narrow already at higher inclinations. 
 Another explanation for the single peaked line profiles could be the presence of a slow molecular wind component in the line profiles \citep[see e.g.][]{bast2011,pontoppidan2011}.

\begin{figure}[!htbp]
\begin{center}$
\begin{array}{cc}
   \includegraphics[width=0.43\textwidth]{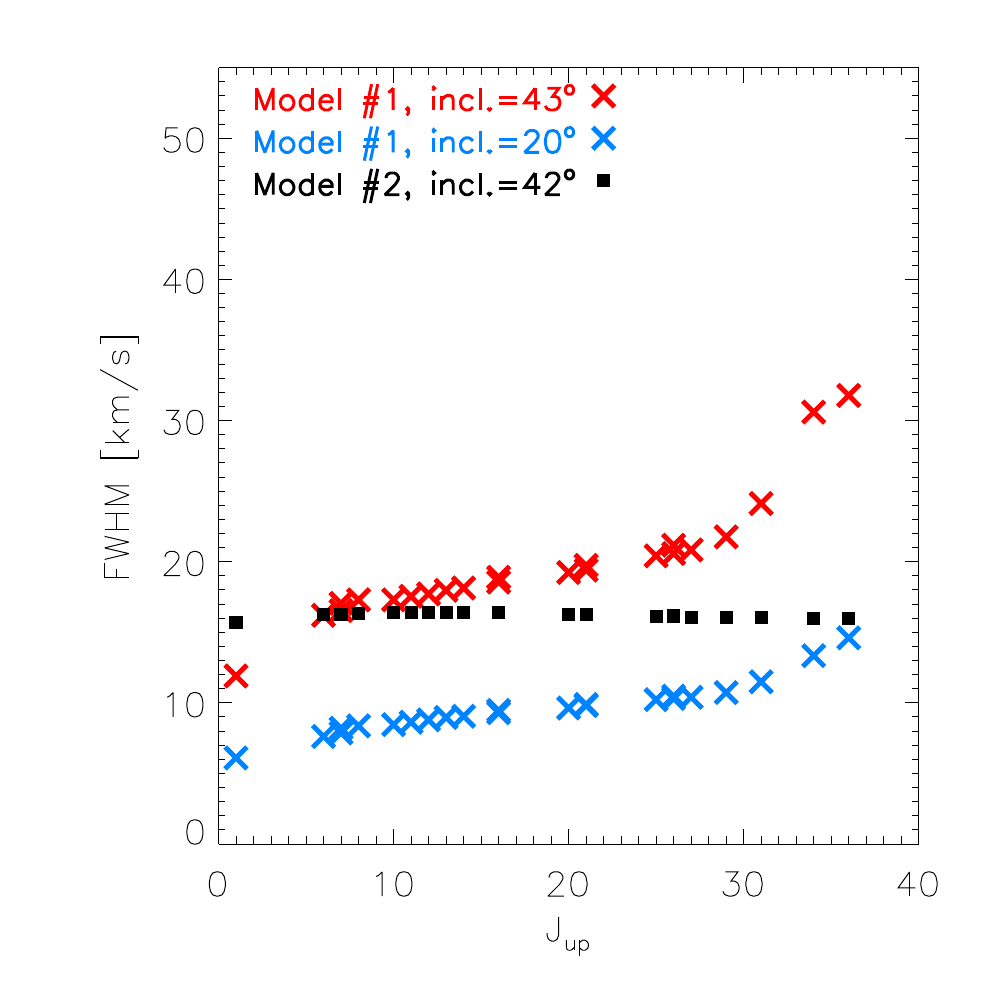}
\end{array}$
\end{center}
\caption{ FWHM as a function of $J_{\rm up}$ for $v$=1-0 emission lines collected from: Model $\#$1, a Herbig Ae/Be ProDiMo model with the inner disc starting at 0.3 au (two different inclinations shown); Model $\#$2, HD~100546, a disc with a gap.}
 \label{fwhm_hd97}
\end{figure}

\begin{figure}[!htbp]
\begin{center}$
\begin{array}{cc}
   \includegraphics[width=0.43\textwidth]{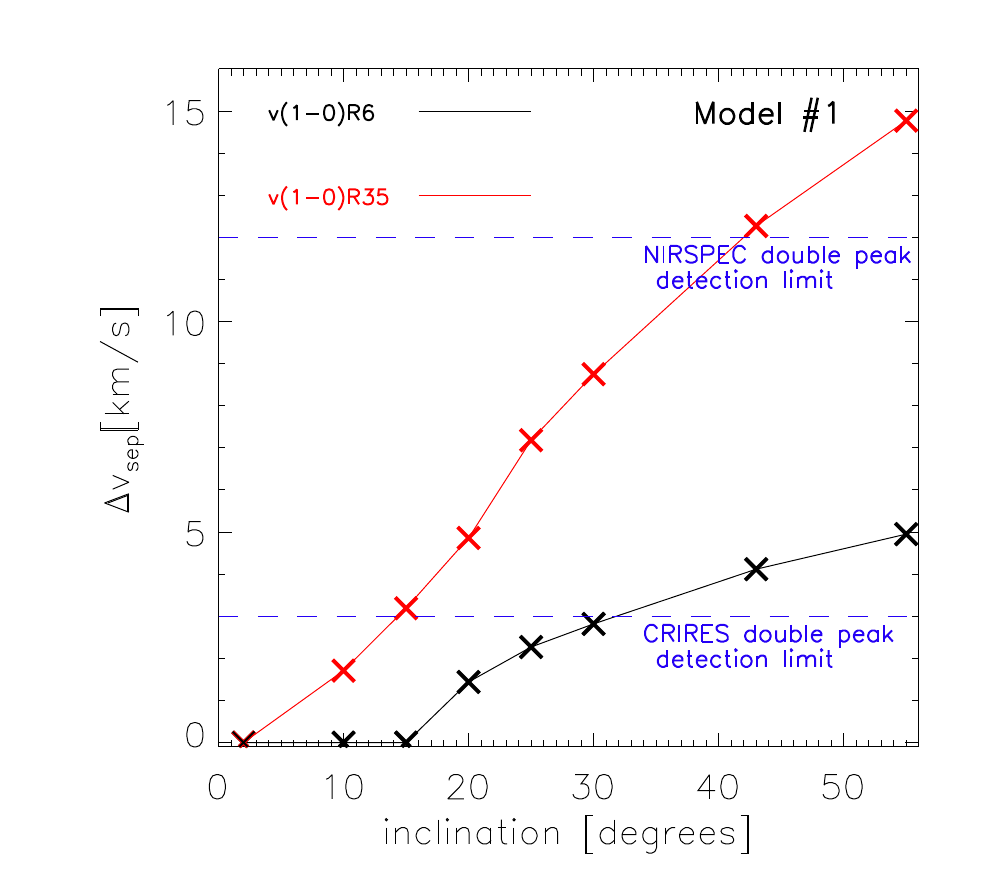}
 \end{array}$
\end{center}
\caption{Inclination versus peak separation for a modelled v(1-0)R6 line and v(1-0)R35 line from a flaring disc model with no gap (Model \#1).}
         \label{incl_test}
\end{figure}

\section{Conclusions} \label{conclusion}
{We have collected high-resolution IR spectra from six additional Herbig Ae/Be stars} and have confirmed detections of CO ro-vibrational emission lines from four of these sources (HD~163296, HD~250550, Hen~2-80, and Hen~3-1227).} 

For HD~250550, the FWHM of the CO lines is increasing with $J$ value, suggesting an extended emitting region that moves closer to the star for higher $J$ levels. Hen~2-80 shows a constant FWHM versus $J$ behaviour, suggesting that all transitions are emitted from the same narrow region. 
Results from {two modelled disc geometries} suggest a connection between dust gaps in the inner disc and a constant FWHM versus $J$ behaviour (possibly connected to a dominant fluorescent excitation mechanism), while an increasing  FWHM versus $J$ behaviour could be connected with a continuous disc without a gap (possibly connected to a dominant thermal excitation mechanism).
This could eventually turn into an additional observational diagnostic.
If the FWHM increases steadily with $J$, the gas would be continuously distributed in a disc reaching from the dust sublimation radius to the maximum allowed by the excitation mechanism. Meanwhile, if the FWHM stays constant and the line is fairly narrow, it could suggest a gap/hole since large parts of the region, where the lines normally originate are missing. Consequently the emitting area becomes very small and the same for all lines. This would be consistent with the interpretation of group I discs as discs with gaps \citep{maaskant2013}. 

Our observed sample together with previously observed sources does not support the simple dichotomy where discs labeled group I from the SED have narrower CO ro-vibrational emission lines, i.e. are emitting further out in the disc. Many group I discs do have narrow profiles, but several {exceptions} exist. We suggest that these {exceptions} might represent cases, where either gas and dust are {radially not co-spatial}, or the gap in the disc, responsible for the group I classification, is small (silicate features detected). 

Further investigations, {such as additional examples of observed discs that show increasing FWHM versus $J$ for the CO ro-vibrational lines, complementary data from HD~250550 that can reveal or exclude the presence of a dust gap, data to clarify the inclination of HD~250550, and detailed modelling}, are required to better understand both the trend of FWHM versus $J$ behaviour, its link with disc geometry (flat, flaring, gap, no gap) and the suggested {spatial} decoupling of dust and gas (dust gap versus dust+gas gap). The significance of these mechanisms for the appearance of the CO ro-vibrational lines will be investigated in a modelling grid in a forthcoming paper.

\begin{acknowledgements}
The authors thank the ESO staff on Paranal for their help in the observations on which this paper is based. {The authors thank the anonymous referee for a careful report that helped to improve the paper.}
The authors thank Andres Carmona for many interesting discussions and for sharing spectra of the OI line observed from Hen~2-80. 
The authors thank Koen Maaskant for supplying the \mbox{MCMax} model for HD~97048. 
IK, WFT, and PW acknowledge funding from the EU FP7-2011 under Grant Agreement no. 284405. 
Gvdp acknowledges support from the Millennium Science Initiative (Chilean Ministry of Economy) through grant Nucleus P10-022-F and also acknowledges financial support provided by FONDECYT following grant 3140393.
\end{acknowledgements}

\bibliographystyle{aa}
\bibliography{reference}

\begin{appendix}
\section{Observed sample} \label{sample_app}

\begin{figure}[!htbp]
\begin{center}$
\begin{array}{cc}
   \includegraphics[width=.45\textwidth]{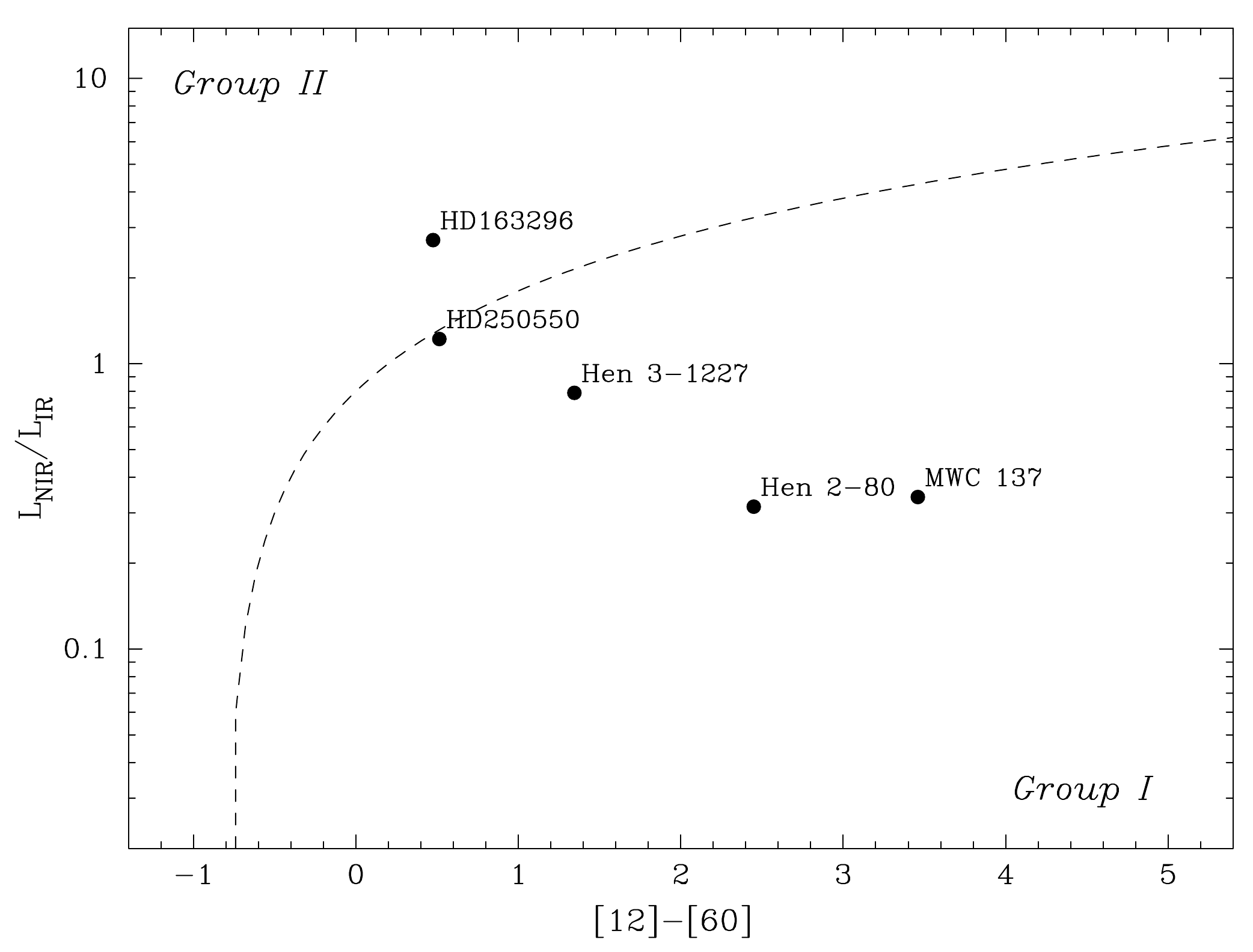}
 \end{array}$
\end{center}
\caption{$L_{\rm{NIR}}/L_{\rm{IR}}$ versus [12]-[60] colour plot {\citep{boekel2005}} displaying all sources in our sample except T~Ori (no 12 or 60 micron data available). Sources on the left of the empirical line are defined as group II and sources on the right as group I. }
         \label{fig:nir_miras}
\end{figure}

\subsection*{HD~163296}
HD~163296 is a well studied Herbig Ae star of spectral type A3Ve \citep{gray1998}, with a protoplanetary disc. In the $L_{NIR}/L_{IR}$ versus [12]-[60] colour diagram it falls in the range of group II discs (see Fig. \ref{fig:nir_miras}). This classification is supported by the lack of PAHs in its infrared spectrum \citep{acke2004}. However, HD~163296 has properties of both a flat and a flaring discs, in particular a scattering disc is observed \citep{grady2000} which is not expected for a self-shadowed disc.
\citet{montesinos2009} estimated the stellar mass in the system to $M_*\sim$2.3 $M_{\astrosun}$ and the age to 5 Myr. 
From spatially resolved sub-mm data, \citet{Isella2007} estimated the disc around HD~163296 to have an inclination of 46$\degree\pm$4$\degree$ and a position angle of 128$\degree\pm$4$\degree$. The disc is located at a distance of 119 pc (Hipparcos).
The disc has been observed in scattered light out to a radius of $\sim$500 au \citep{grady2000}.

The presence of a giant planet or brown dwarf orbiting in the outer disc has been suggested by \citet{grady2000}, to explain an annulus of reduced scattering around 325 au in the optical choronagraphic images, from the Space Telescope Imaging Spectrograph (STIS) on the Hubble Space Telescope (HST). The reduced scattering is consistent with a cleared zone like that caused by a substellar companion.

\citet{Isella2007} studied the disc with interferometric observations, from 0.87 to 7 mm, both in the continuum and of CO emission lines. From these observations the \element[][12]{CO} and \element[][13]{CO} was found to be optically thick. From the continuum dust emission the authors found an outer radius of 200$\pm$15 au while CO lines imply an outer radius of 550$\pm$50 au. The authors suggest that it is necessary to introduce a sharp drop in continuum emission of a factor $>$30 at a radius of 200 au, to explain the lack of continuum emission from radii larger than 200 au. HD~163296 thereby differs from the standard transitional disc where it is the inner disc that is dust depleted. 

Recent ALMA band 7 data, at $\sim$850 $\mu$m, sets the outer dust disc radius at 240 au and the CO outer radius at 575 au \citep{gregorio2013}. These observations show a well resolved dusty disc and no suggestion of gaps or holes at R>25 au. The authors find that one cannot fit a standard tapered-edge model with one unique density profile to both data sets simultaneously. The CO channel maps require a thicker gas disc, while the mid- and far-infrared SED require a flatter dust disc than previously presented in \citet{Tilling2012}.

HD~163296 shows variations in NIR brightness on timescales of years \citep{sitko2008}.
The source has a bipolar jet (HH 409), discovered in choronagraphic images from the STIS \citep{grady2000}. The high velocity gas in the jet has radial velocities of 200-300 km/s.
\citet{ellerbroek2014} found that the jet displays periods of intensified outflow activity at regular intervals of 16.0$\pm$0.7 years.
They also identify transient optical fading, together with enhanced NIR excess, and conclude that this is consistent with a scenario where dust clouds are launched above the disc plane.

\subsection*{HD~250550}
HD~250550 is a confirmed Herbig Ae/Be star of spectral type B8--A0IVe \citep{gray1998}, with a protoplanetary disc. The SED of HD~250550 is similar to that of HD~163296 and it has a lack of PAHs in its infrared spectrum \citep{acke2006_b}.  In our $L_{NIR}/L_{IR}$ versus [12]-[60] colour diagram this disc falls (on the group I side) close to line dividing group I and group II discs (see Fig. \ref{fig:nir_miras}). In previous literature using standard classification methods it has been labeled as a group I source \citep[e.g.][]{acke2006_b, fedele2011, maaskant2014b}, and in this paper we will thus consider this a group I source.
The central star has an estimated mass of $M_*\sim$3.6 $M_{\astrosun}$ and age of 1 Myr \citep{martin2004}. 
The object is located at a distance of 280 pc \citep{canto1984,kawamura1998}.
In FUSE observations, emission lines from \ion{C}{iii} and \ion{O}{vi} were detected \citep{bouret2003}.

\subsection*{Hen~2-80 (SS73 34)}
Hen~2-80 was recently confirmed to be a Herbig Be star of spectral type B6Ve \citep{carmona2010}. In our $L_{NIR}/L_{IR}$ versus [12]-[60] colour diagram it falls clearly on the group I side (see Fig. \ref{fig:nir_miras}). However, this is the earliest spectral type in our sample. In a study of discs around Herbig Be stars \citet{verhoeff2012} concluded that generally the SEDs of Herbig Be stars do not show flaring geometries but rather indicate self-shadowed geometries. In a study of photo evaporation of discs as a function of stellar luminosity, \citet{gorti2009} concluded that hot luminous young stars should evaporate their outer discs on relatively short timescales. In a simplistic view this would mean that more luminous stars lose their outer disc fast and the chance of seeing a flaring disc around a massive pre main sequence star is not very large.

Hen~2-80 is located at a distance larger than 750 pc (minimum distance based on luminosity).
\citet{carmona2010} observed the source using high-resolution spectroscopy in the optical, covering the wavelength region 3500-9200 \AA\ ($R\sim$45000). The spectrum is flat and exhibits only a few absorption lines. The observed emission lines in the spectrum include H$\alpha$, He\,{\sc i} lines, Mg\,{\sc ii}, H\,{\sc i} P(17), O\,{\sc i}, S\,{\sc ii}, N\,{\sc ii}, and Ca\,{\sc ii}.
\citet{carmona2010} published spectra of the H$\alpha$ line showing narrow absorption centred on the line.

\subsection*{MWC~137}
MWC137 was classified as a Herbig Be star of spectral type B0ep \citep{sabaddin1981, finkenzeller1984}. The SED of MWC137 resembles that of a flaring disc around a Herbig Ae/Be star (see Fig \ref{fig:sed}), and it falls clearly on the group I side in our $L_{NIR}/L_{IR}$ versus [12]-[60] colour diagram (see Fig. \ref{fig:nir_miras}). However, the listed luminosity ($L_{*}$>4.1) and effective temperature ($T_{\rm eff}$=30000) are unexpectedly high for a Herbig star and \citet{esteban1998} favoured it to be a B[e] supergiant. Meanwhile, the N-band mid-IR spectrum published by \citet{verhoeff2012} looks very similar to that of MWC297 which is believed to be a massive Herbig Be star.

MWC~137 is located at a distance larger than 1000 pc (minimum distance based on luminosity).
The object appears to be spectroscopically variable \citep{zickgraf2003}. An upper limit of 0.007 $M_{\astrosun}$ for the disc mass has been found, using continuum images at 1.4 and 2.7 mm from IRAM and at 1.3 and 0.7 cm from the NRAO VLA \citep{fuente2003}.

\citet{verhoeff2012} performed N-band imaging and long slit spectroscopy with VISIR/VLT. They conclude that the observed PAH emission originates both from a circumstellar disc and from a surrounding nebula.

In a survey of massive evolved stars, including MWC~137, \citet{oksala2013} obtained K-band spectroscopic data with the SINFONI at the VLT. Their observations of the first overtone CO band head emission is the first detection of CO emission from this source. \element[][13]{CO} emission was also detected. Their model suggests that the CO emitting region is located in a detached disc or ring structure and not in a continuous disc.
\citet{kraus2009} note that pre-main sequence levels of $^{13}$C should not produce visible \element[][13]{CO} band head emission. This lead \citet{oksala2013} to conclude from their \element[][12]{CO}/\element[][13]{CO} band head ratio that MWC137 could be an evolved, post-main sequence object. 
\citet{esteban1998} investigate the nebula (S 266) around MWC137, using narrow-band H$\alpha$ image and high resolution spectroscopy and suggest it to be a ring nebula; in the continuum subtracted H$\alpha$ image, they see a clear filamentary shell structure and a fainter and more diffuse emission filling its inner regions.
{The classification of this object remains uncertain, thus we consider it a non-confirmed Herbig-candidate.}

\subsection*{T~Ori}
T~Ori is a confirmed Herbig Ae star of spectral type A3IVe \citep{mora2001}. T~Ori has not been plotted in our $L_{NIR}/L_{IR}$ versus [12]-[60] colour diagram, since the 12 and 60 micron data for this source is contaminated by emission from nearby bright sources. However, its SED appears typical of a self-shadowed or flat disc (see Fig. \ref{fig:sed}). This classification is supported by the absence of PAH emission in the infrared spectrum of T~Ori \citep{acke2006_b}. Furthermore, T~Ori shows strong photometric variations in the optical of the type that is seen around UX Orionis stars and in a modelling study by \citet{dullemond2003} it was shown that UXOR-type phenomenon should only occur in self-shadowed discs. 
The star is located at a distance of 510 pc \citep{zeeuw1999} and is reported as a spectroscopic binary \citep{shevchenko1994}.

\subsection*{Hen~3-1227 (Hen 2-174, SS73 62, WRAY 15-1520)}
Hen~3-1227 is a star of spectral type Be \citep{pereira2003}. It is located at a distance larger than 400 pc (minimum distance based on luminosity). In our $L_{NIR}/L_{IR}$ versus [12]-[60] colour diagram (see Fig. \ref{fig:nir_miras}), Hen~3-1227 is located on the group I side. However, the SED does not resemble a typical disc around a Herbig star (neither group I nor group II). 

This source has been studied in several surveys and was classified as a possible planetary nebula \citep{webster1966, henize1967}. However, this classification was later rejected by \citet{stenholm1987} and \citet{acker1987}. 
Its spectrum shows only H$\alpha$, H$\beta$ and [\ion{O}{i}] (6300 \AA) emission \citep{swings1973}.
\citet{pereira2003} collected spectra from the ESO 1.52 m telescope in La Silla (3500--7500 and 3100--5100\AA) and reported a continuum of a reddened Be star with H$\alpha$, H$\beta$, H$\delta$ and [\ion{O}{i}] (6300 and 6363 \AA) in emission and H$\gamma$ in absorption.
\citet{miroshnichenko2007} labels Hen~3-1227 as an FS CMa-type candidate, but the somewhat broad Balmer lines seen in the optical spectra suggests a high surface gravity, which could indicate an object close to the main sequence. The classification remains uncertain, but from the optical spectra we cannot exclude it as a pre-main sequence object. {As for \mbox{MWC 137} we therefore also consider \mbox{Hen 3-1227} a non-confirmed Herbig-candidate.}

\section{CRIRES spectra}
\begin{figure*}[!h]
\centering
\begin{minipage}[l]{0.4\textwidth}
   \includegraphics[width=\textwidth]{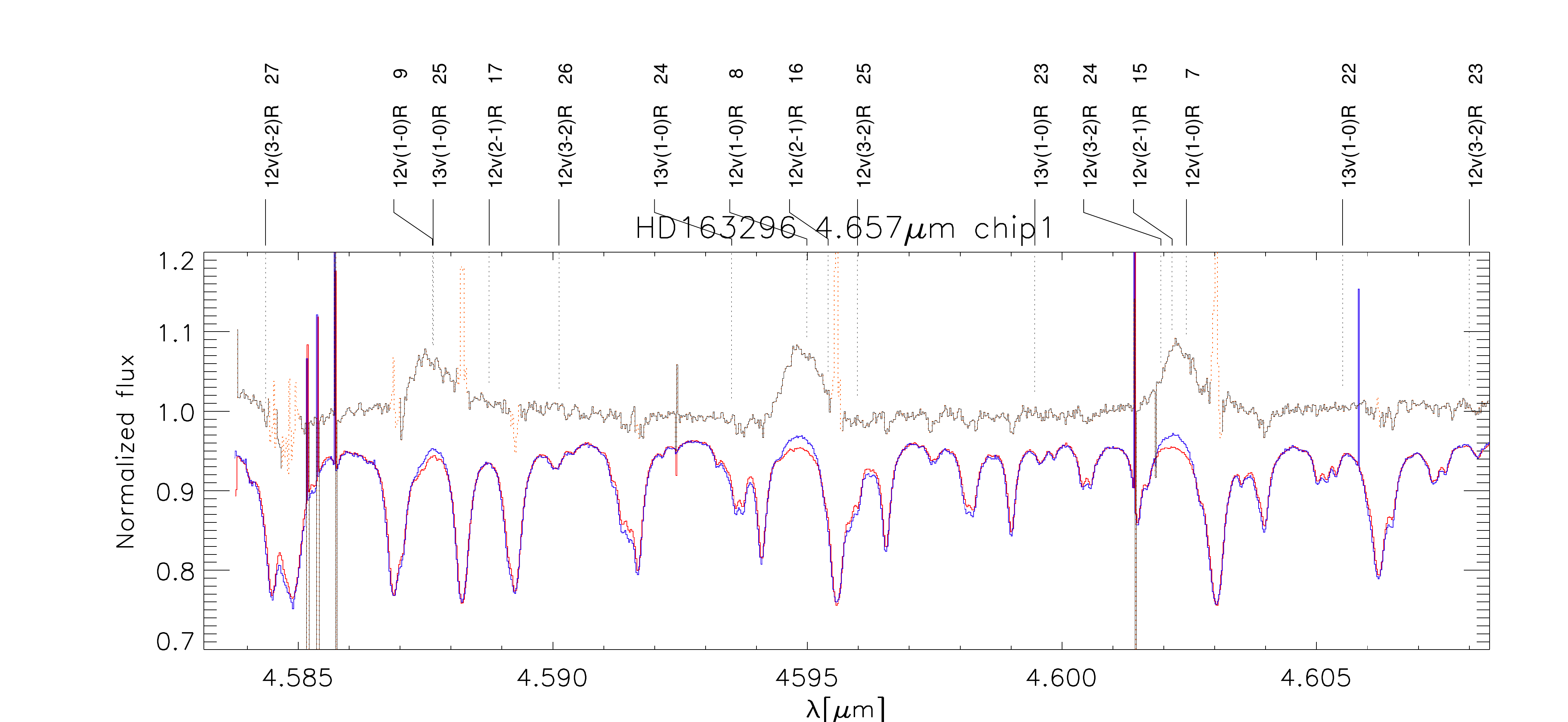}
\end{minipage}
 \begin{minipage}[l]{0.4\textwidth}
  \includegraphics[width=\textwidth]{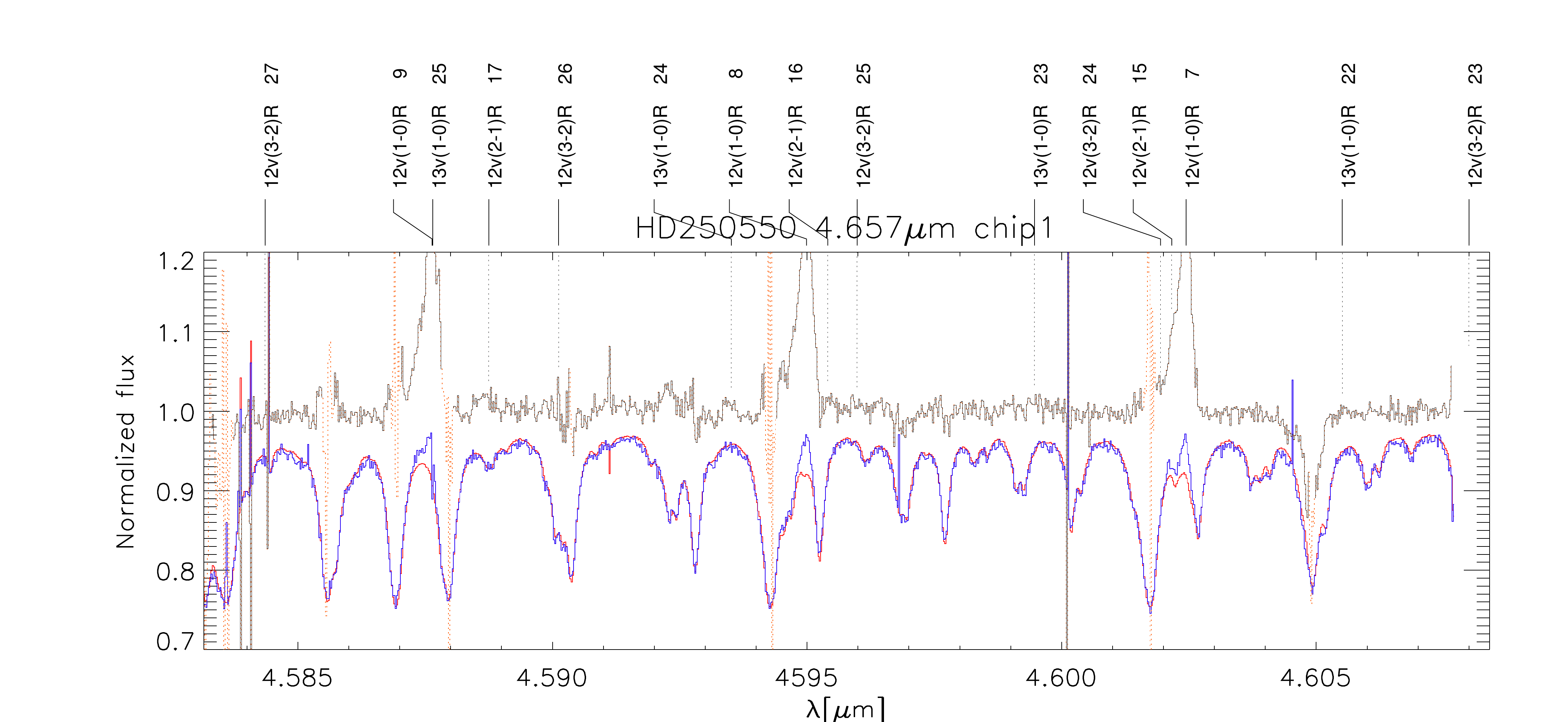}\\
\end{minipage}
 \begin{minipage}[l]{0.4\textwidth}
  \includegraphics[width=\textwidth]{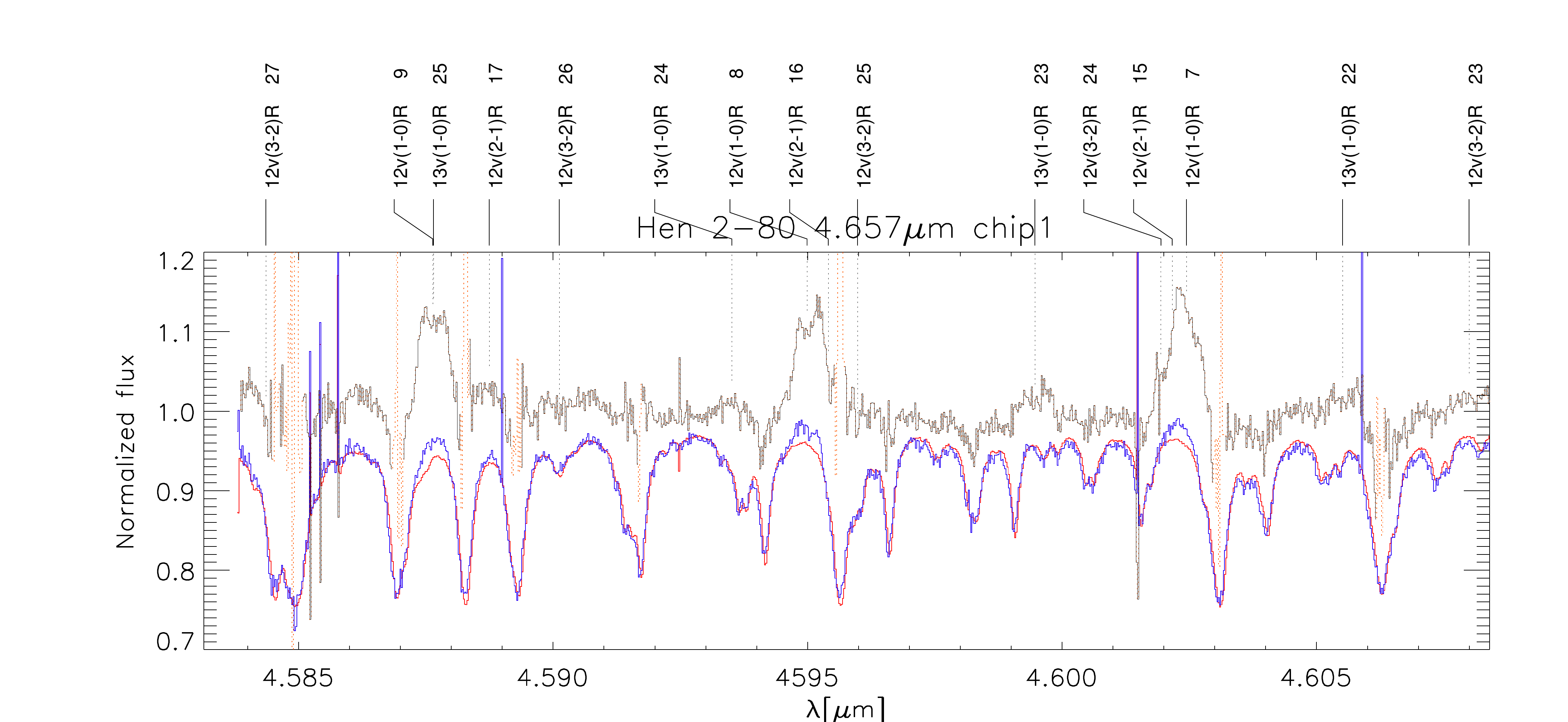}
\end{minipage}
 \begin{minipage}[l]{0.4\textwidth}
  \includegraphics[width=\textwidth]{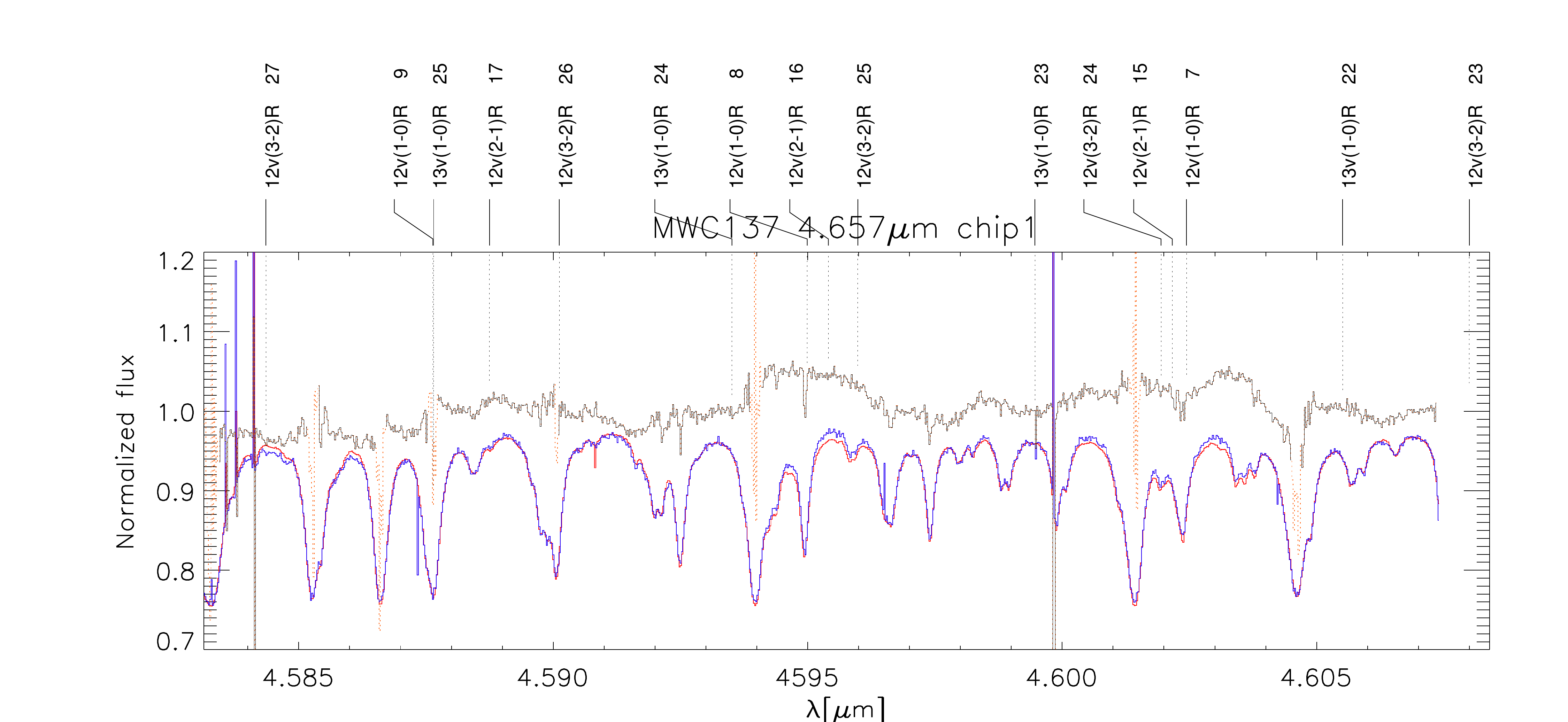}\\
\end{minipage}
 \begin{minipage}[l]{0.4\textwidth}
  \includegraphics[width=\textwidth]{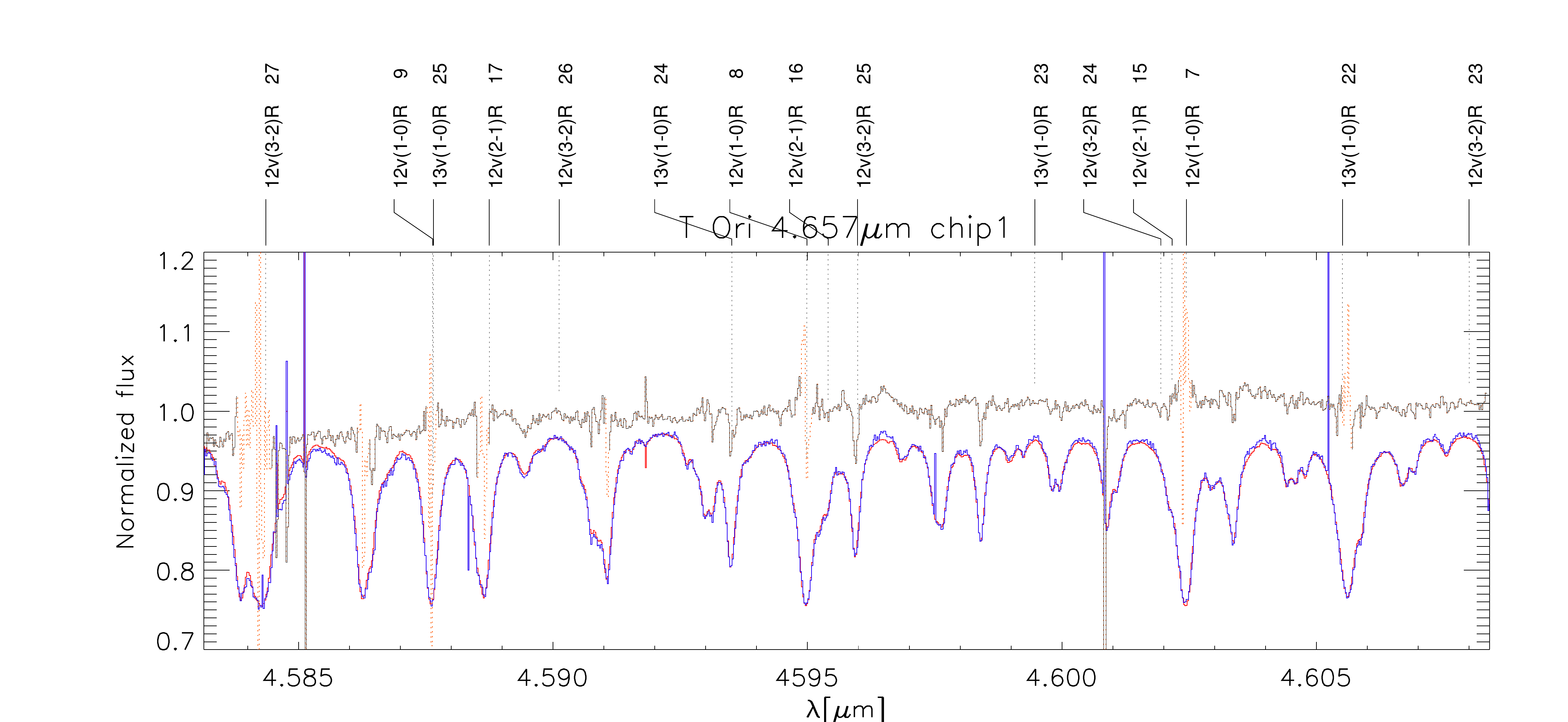}
\end{minipage}
 \begin{minipage}[l]{0.4\textwidth}
  \includegraphics[width=\textwidth]{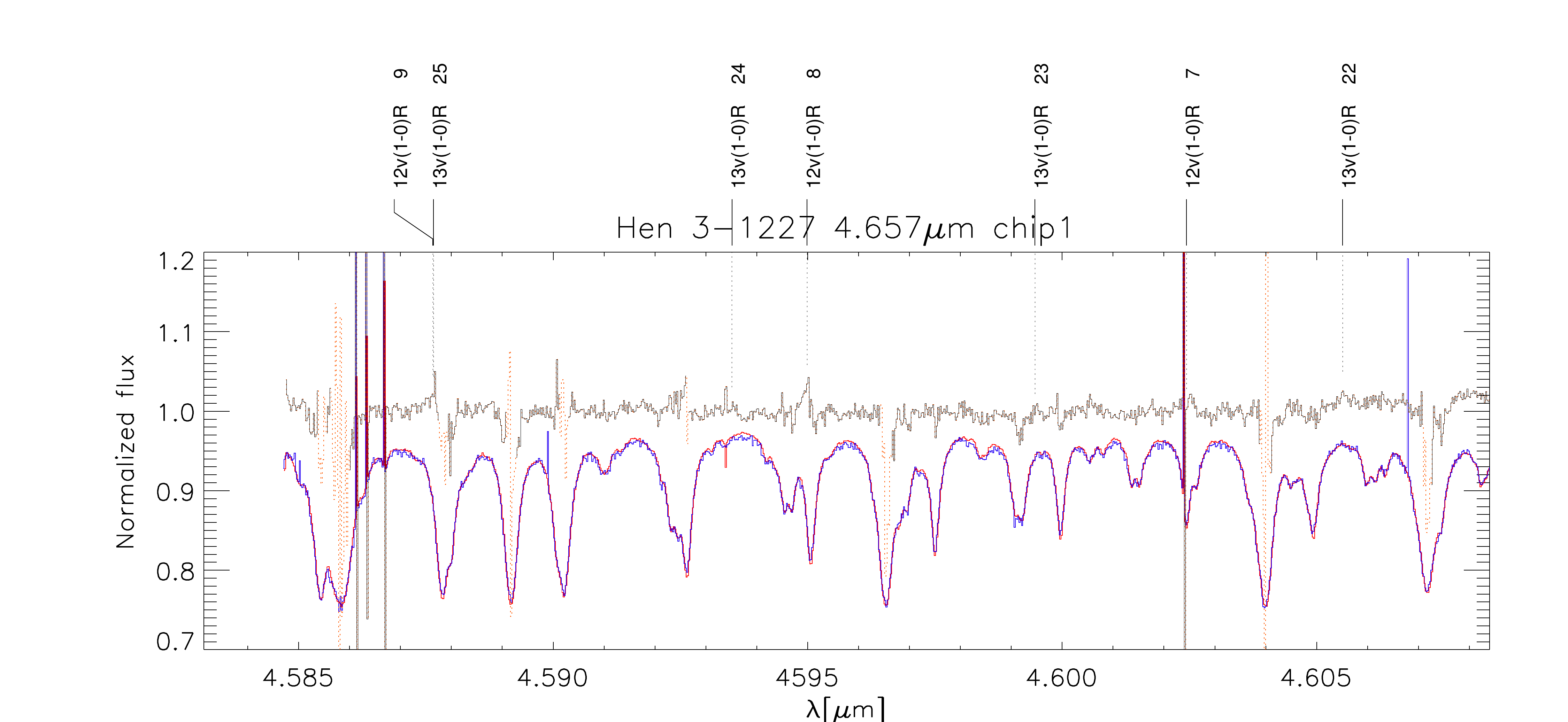}\\
\end{minipage}
\begin{minipage}[l]{0.4\textwidth}
   \includegraphics[width=\textwidth]{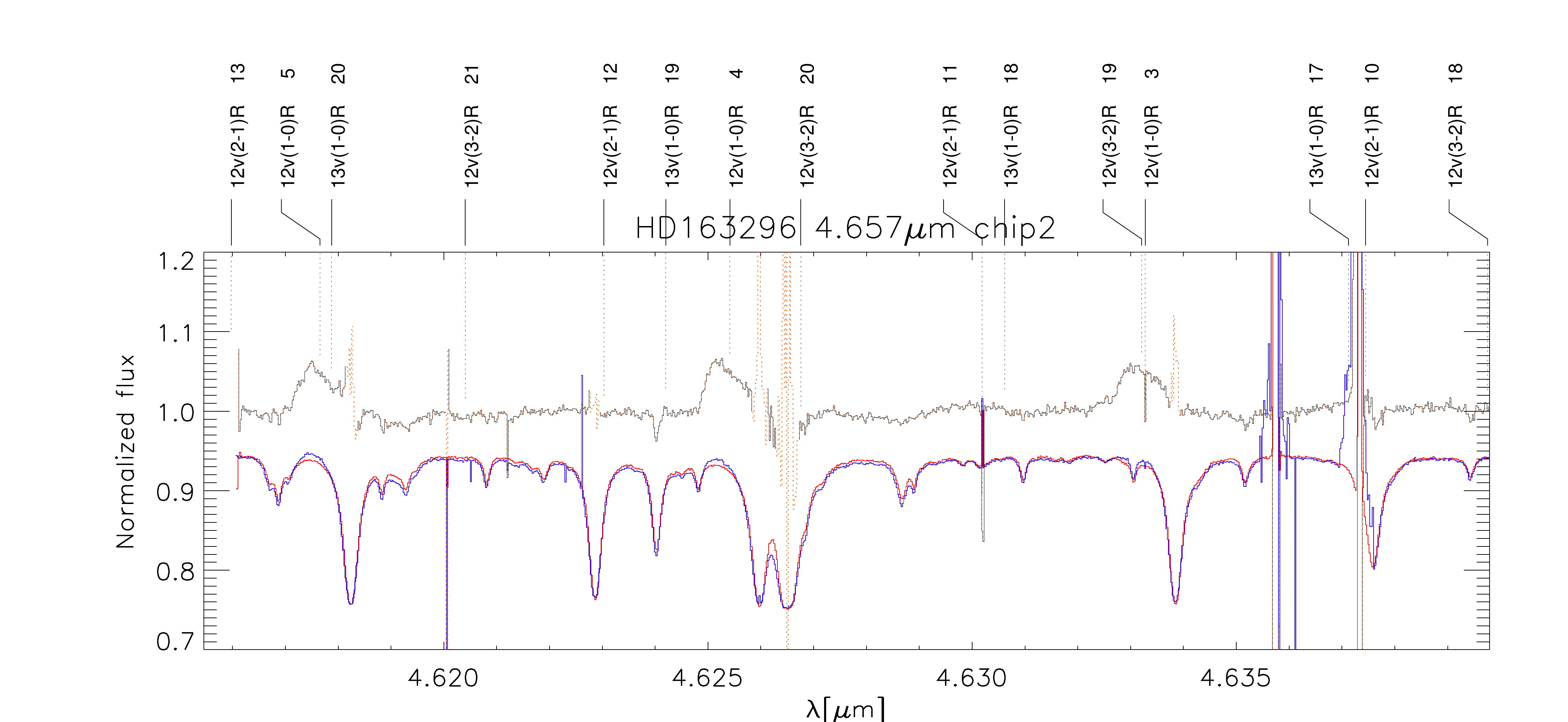}
\end{minipage}
 \begin{minipage}[l]{0.4\textwidth}
  \includegraphics[width=\textwidth]{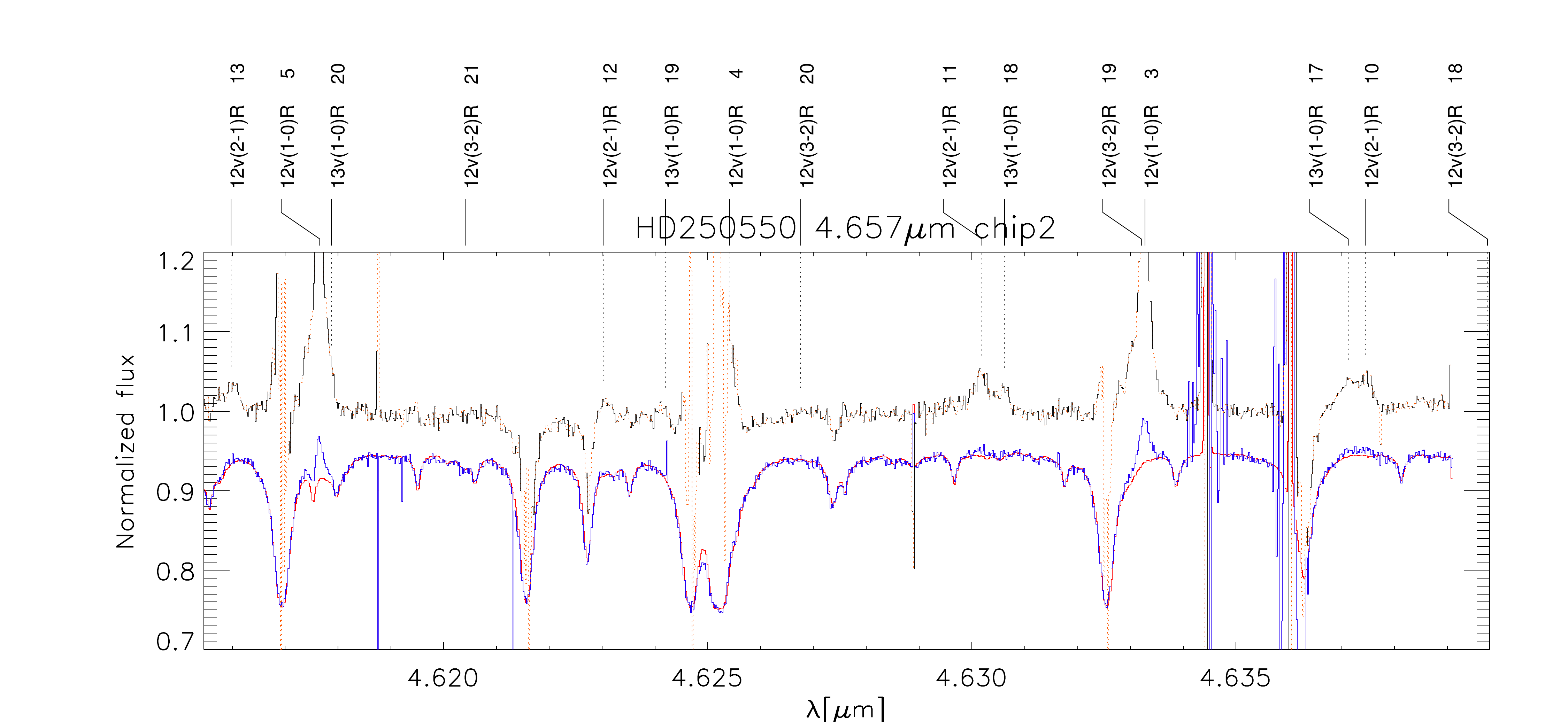}\\
\end{minipage}
 \begin{minipage}[l]{0.4\textwidth}
  \includegraphics[width=\textwidth]{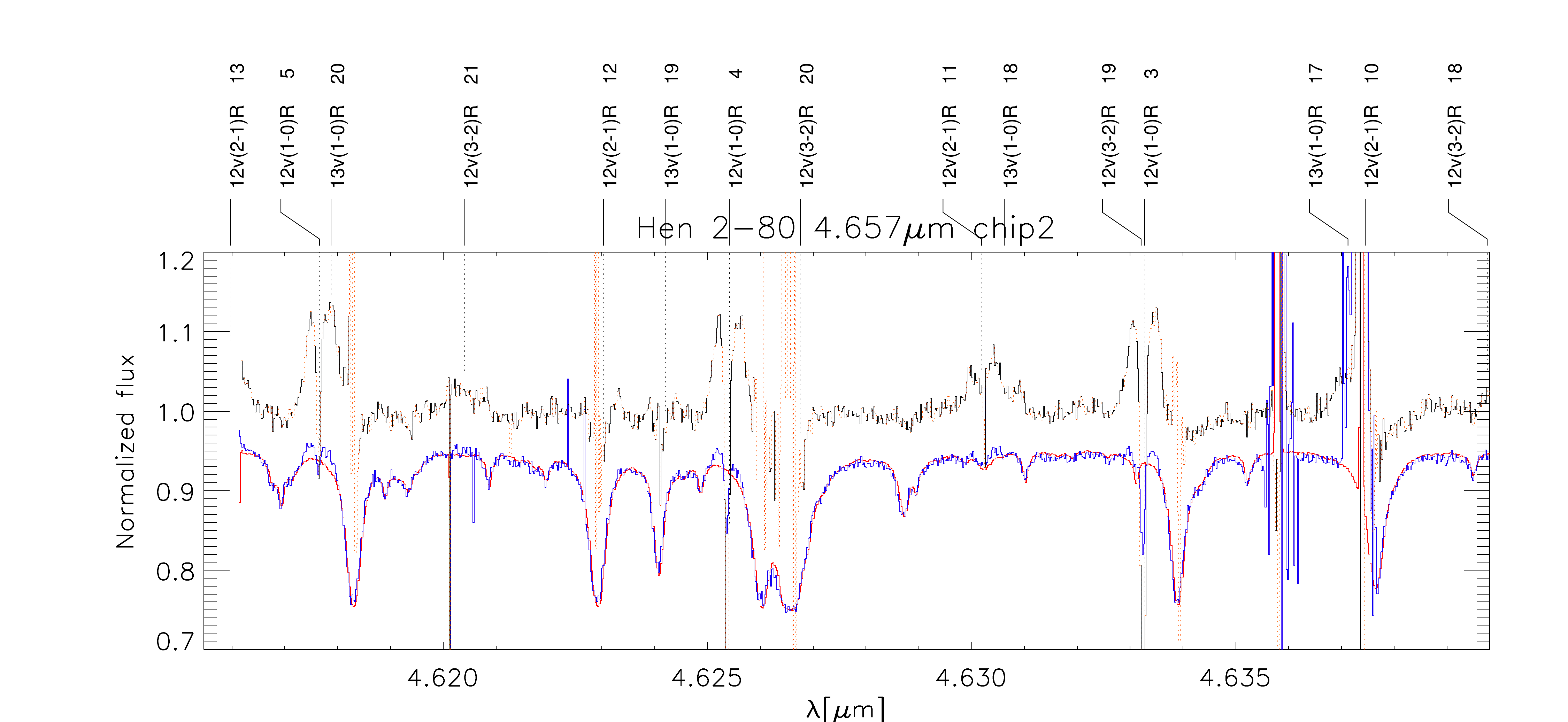}
\end{minipage}
 \begin{minipage}[l]{0.4\textwidth}
  \includegraphics[width=\textwidth]{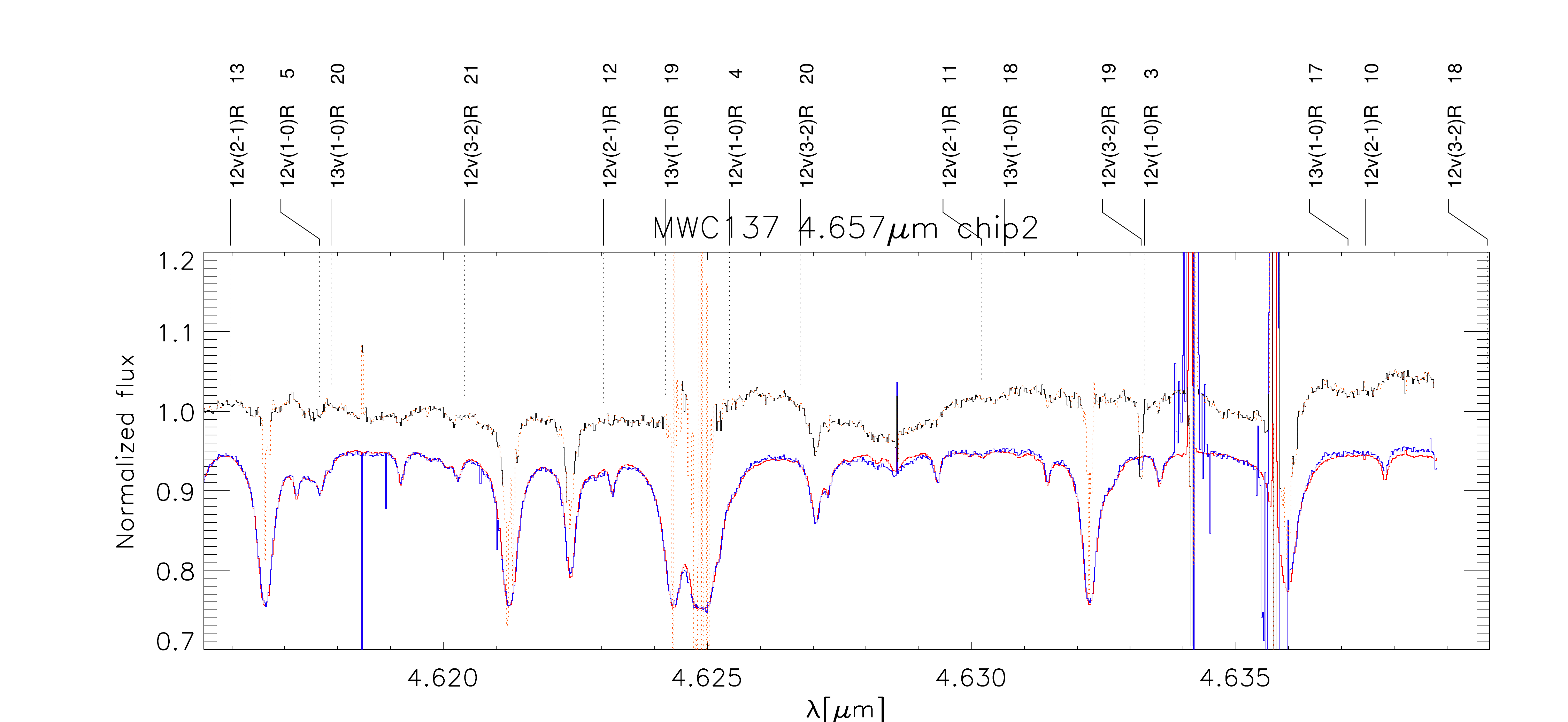}\\
\end{minipage}
 \begin{minipage}[l]{0.4\textwidth}
  \includegraphics[width=\textwidth]{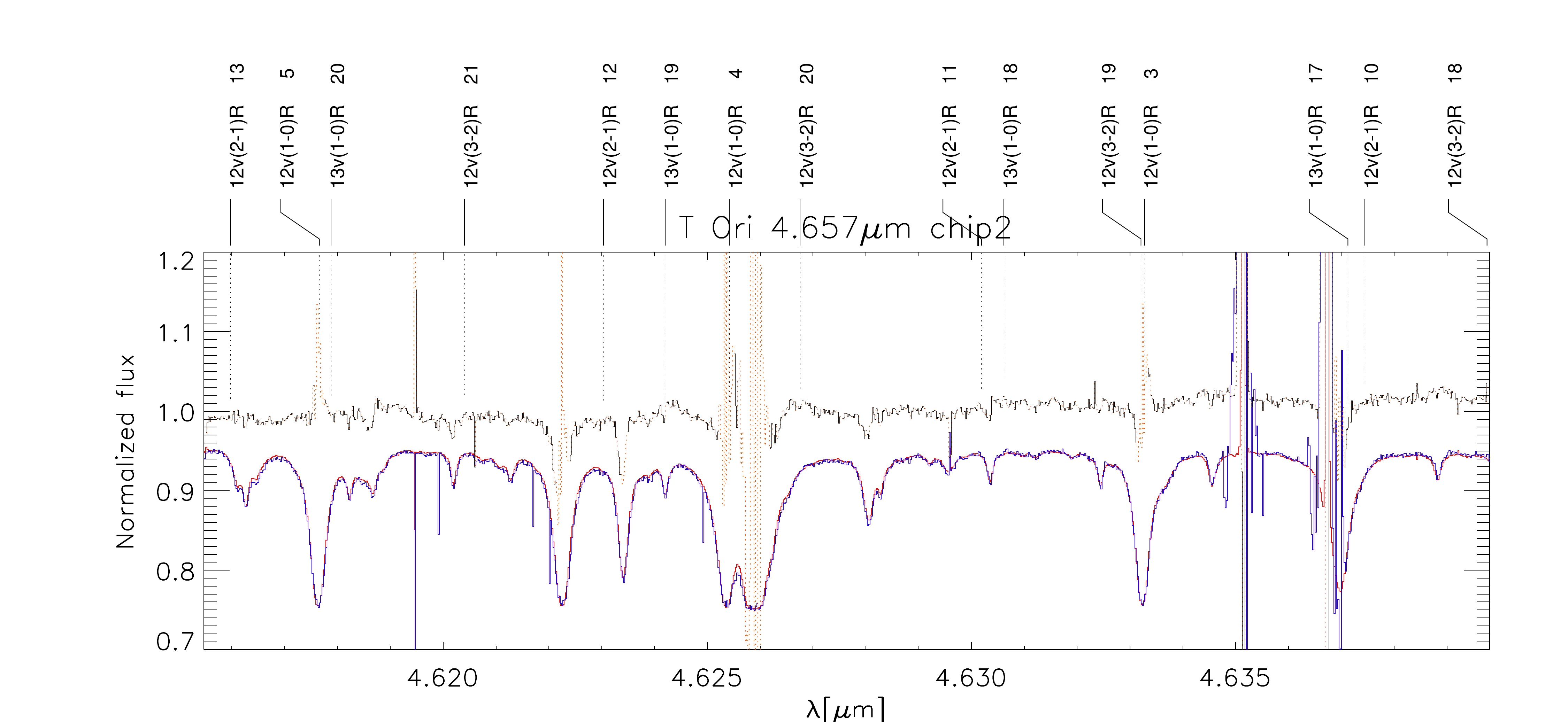}
\end{minipage}
 \begin{minipage}[l]{0.4\textwidth}
  \includegraphics[width=\textwidth]{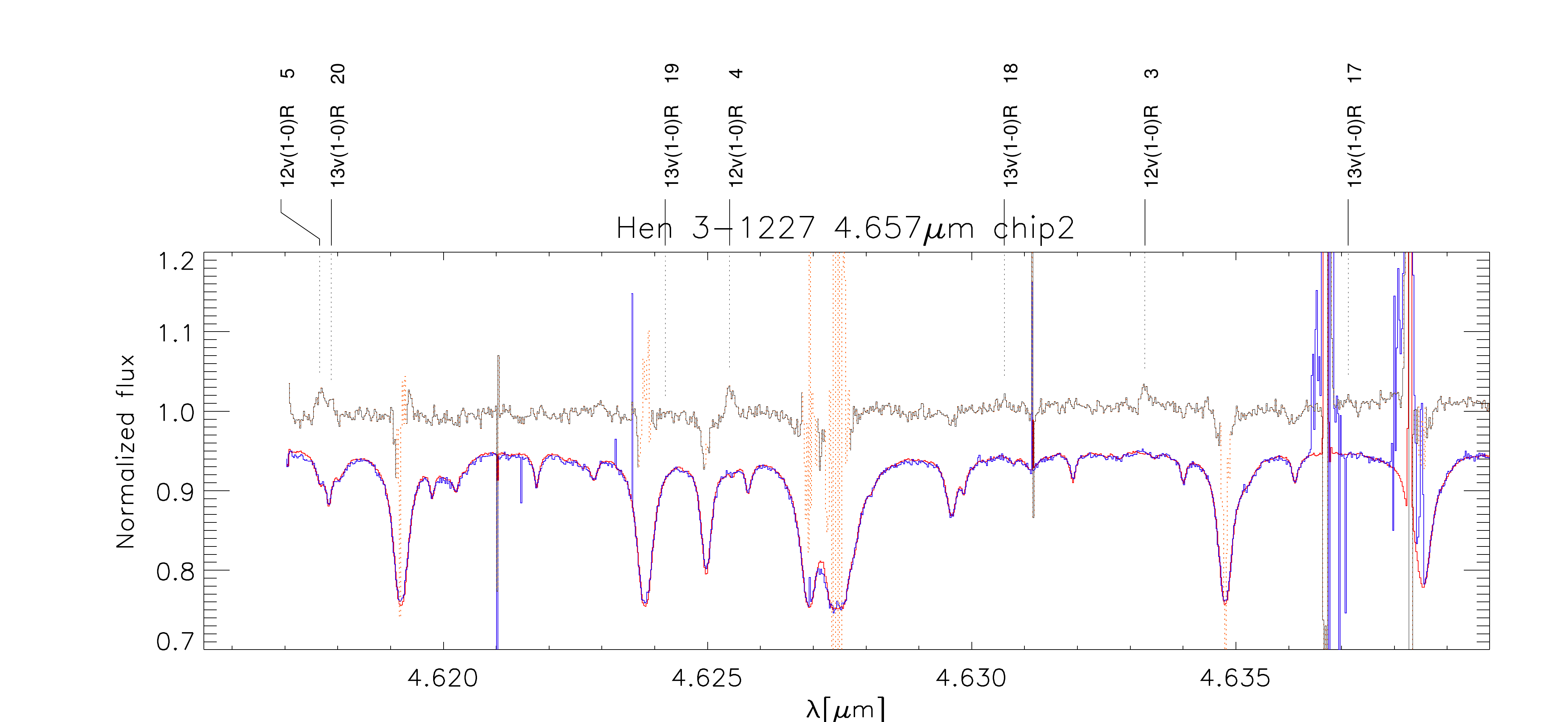}\\
\end{minipage}
\caption{Spectra from all observed sources in our sample (sources names are indicated in the plot titles). These are from chip 1 and chip 2 (indicated in the plot titles) of the spectra taken at wavelength setting 4.657 $\mu$m. The spectra are wavelength calibrated using the H\,{\sc i} lines.}
         \label{fig:spec1}
\end{figure*}

\begin{figure*}[!h]
\centering
\begin{minipage}[l]{0.4\textwidth}
   \includegraphics[width=\textwidth]{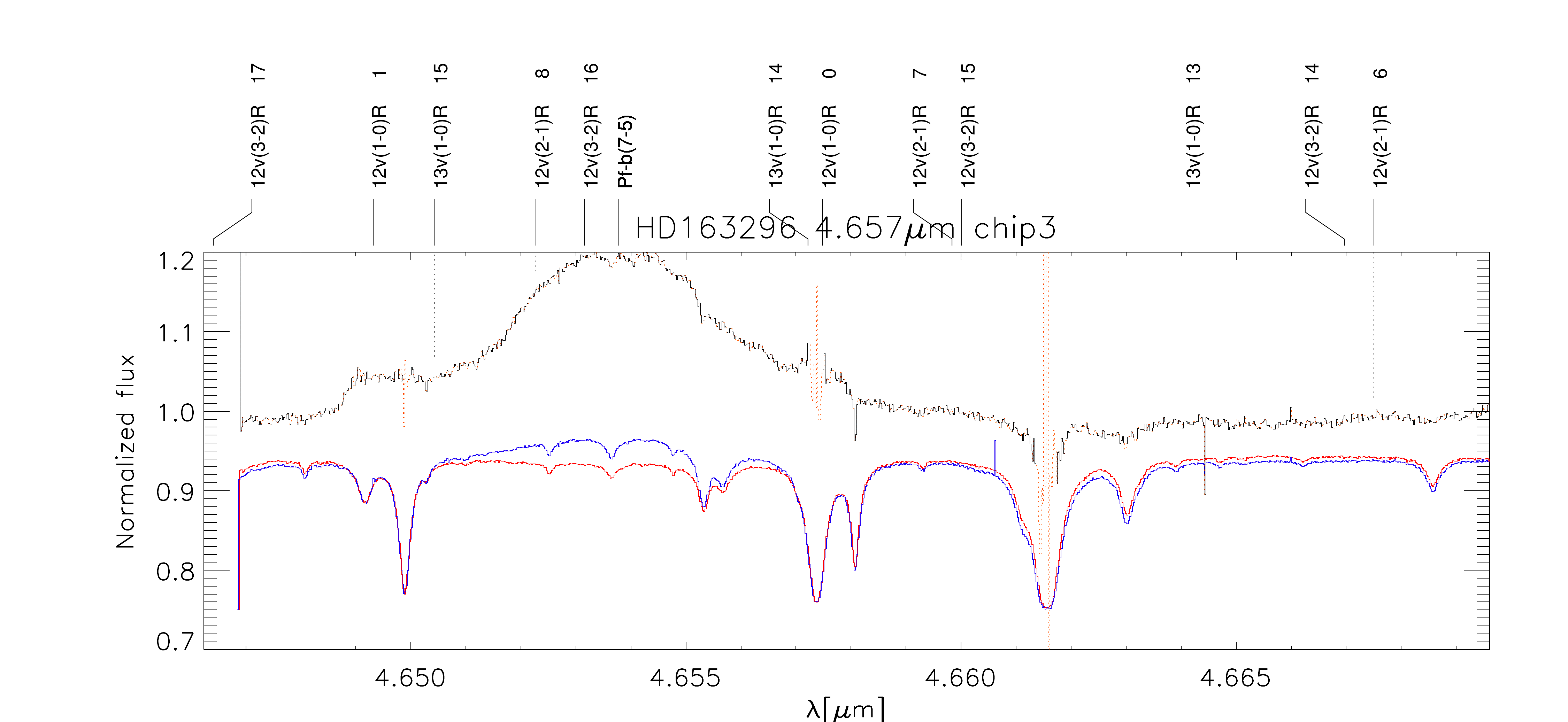}
\end{minipage}
\begin{minipage}[l]{0.4\textwidth}
   \includegraphics[width=\textwidth]{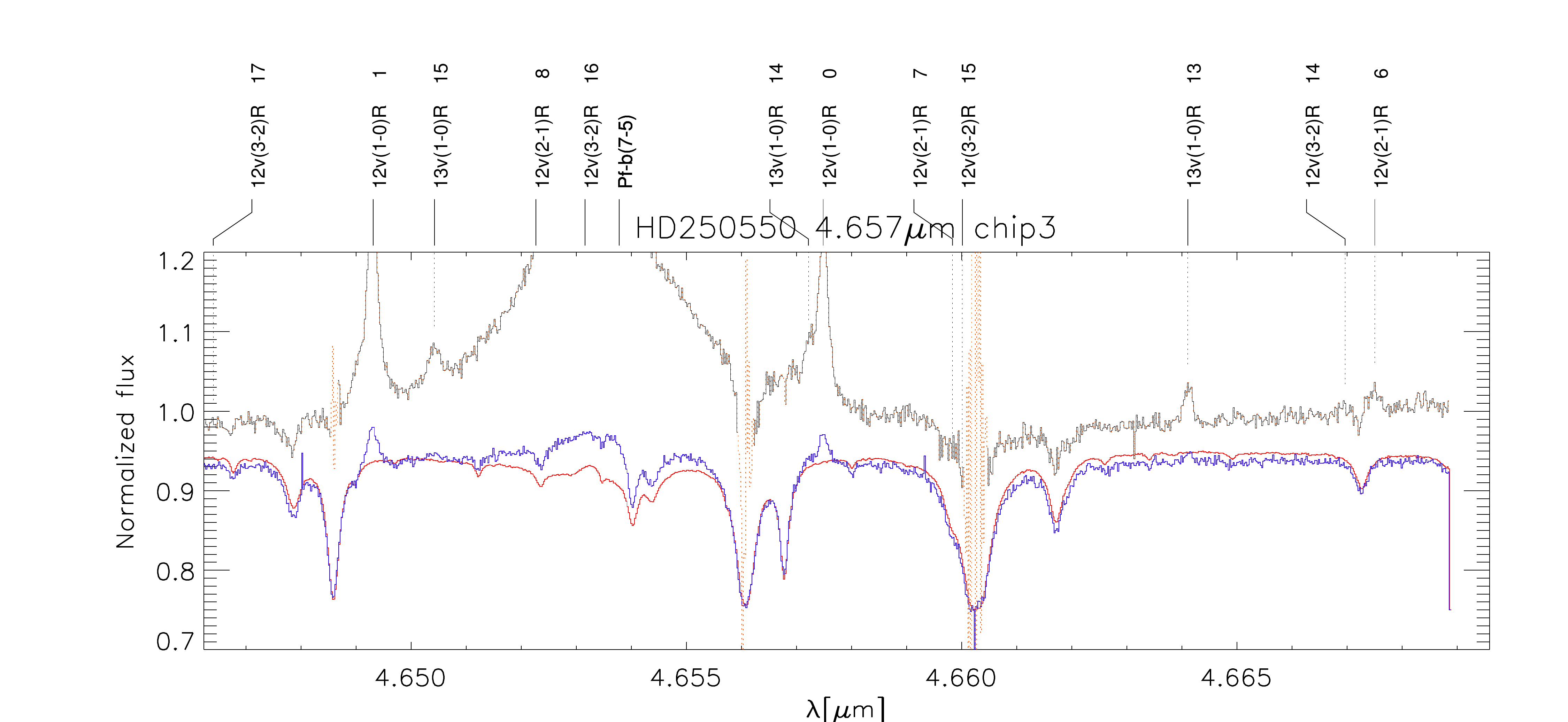}\\
\end{minipage}
\begin{minipage}[l]{0.4\textwidth}
   \includegraphics[width=\textwidth]{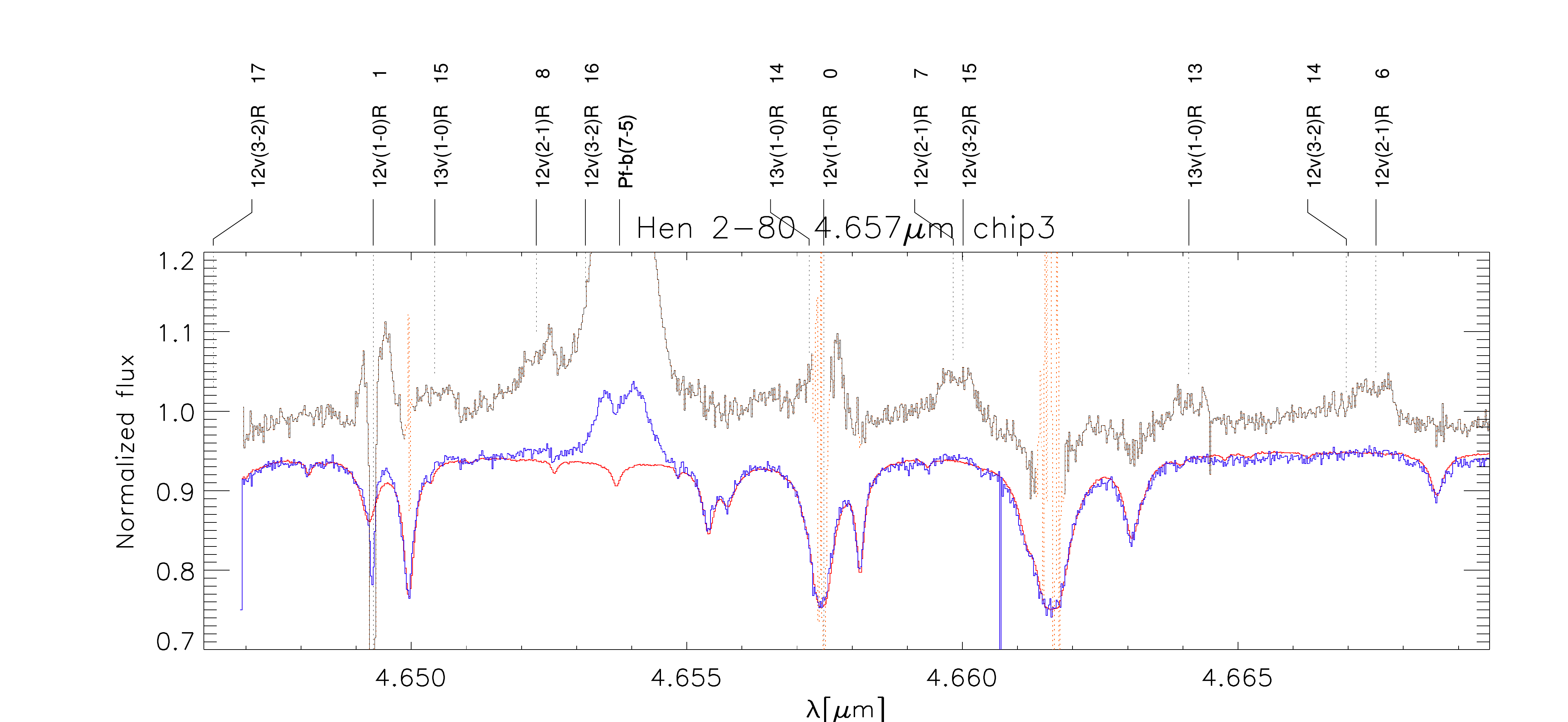}
\end{minipage}
\begin{minipage}[l]{0.4\textwidth}
   \includegraphics[width=\textwidth]{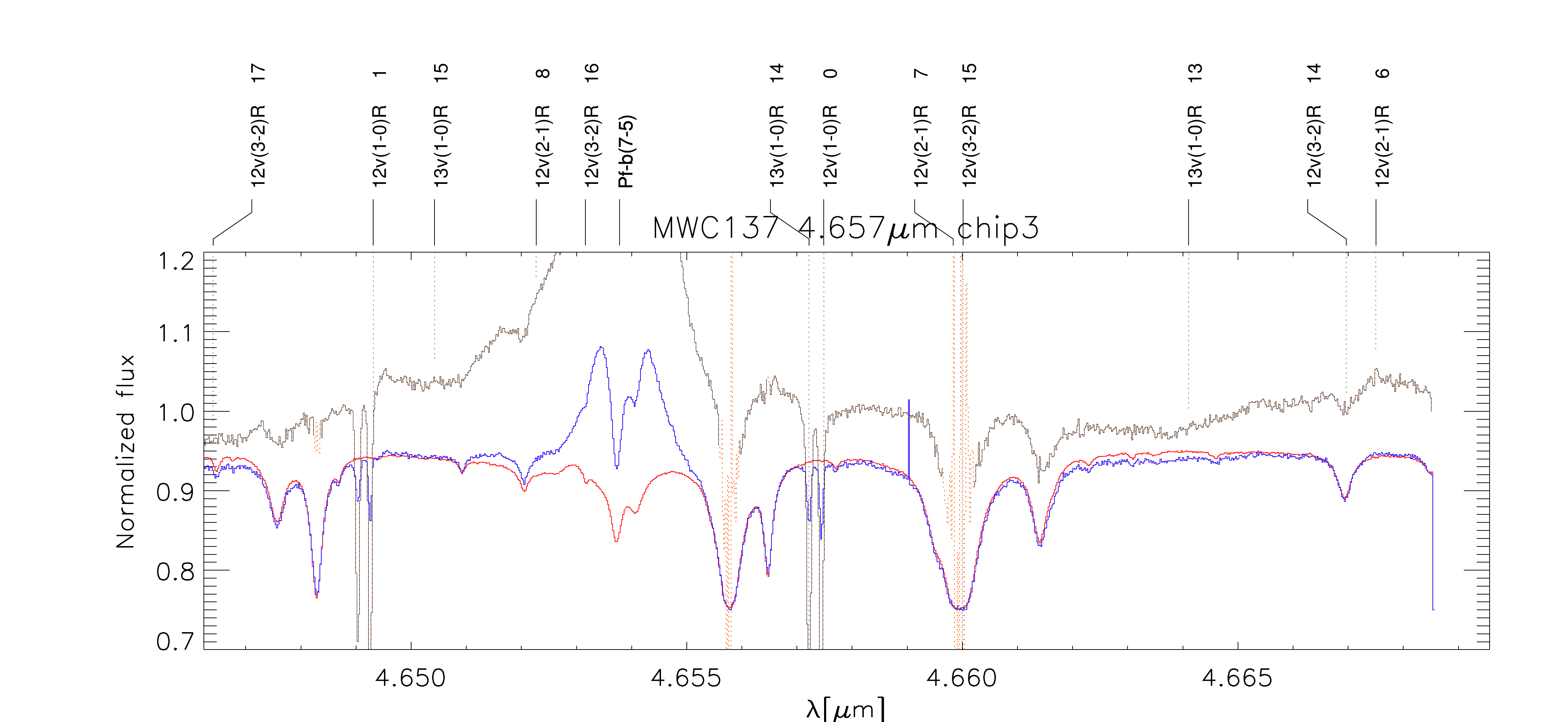}\\
\end{minipage}
\begin{minipage}[l]{0.4\textwidth}
   \includegraphics[width=\textwidth]{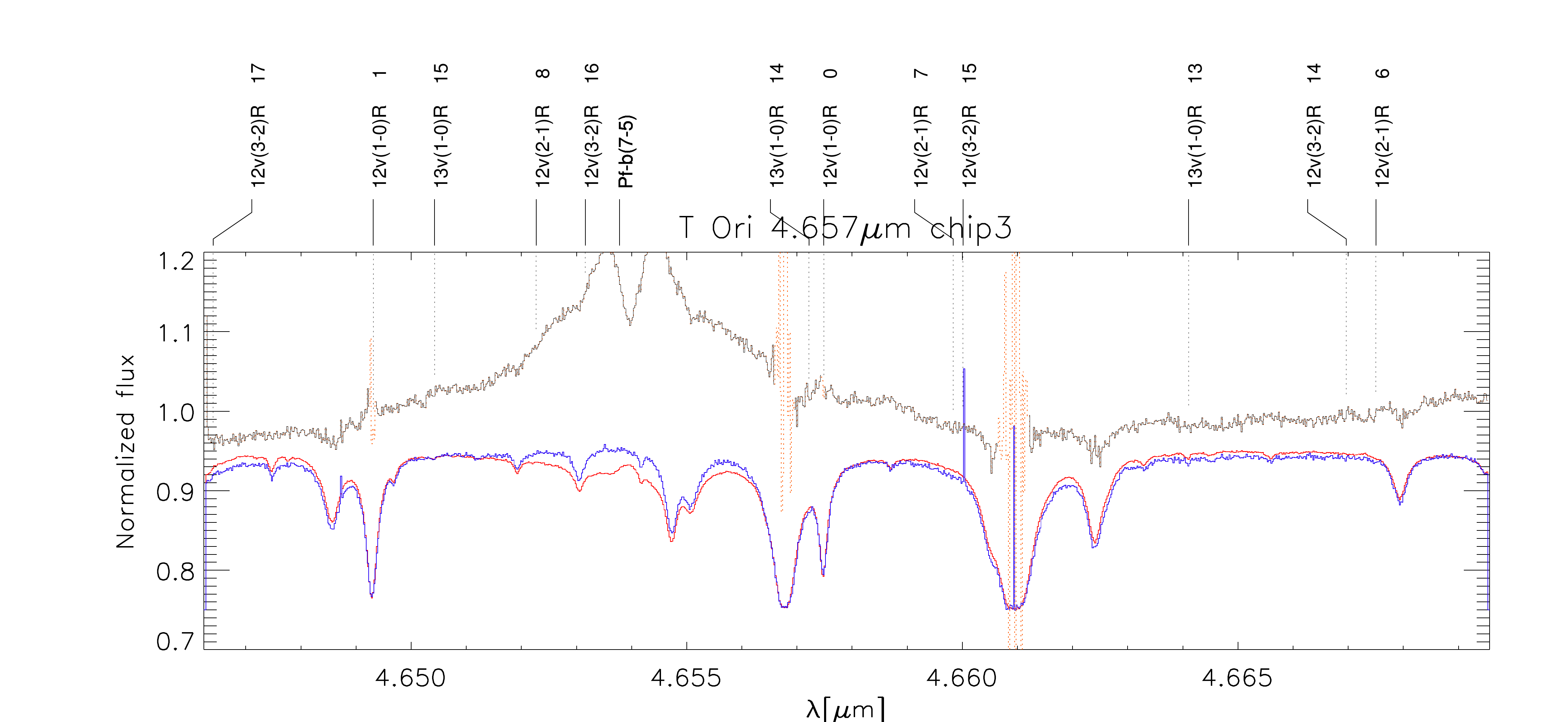}
\end{minipage}
\begin{minipage}[l]{0.4\textwidth}
   \includegraphics[width=\textwidth]{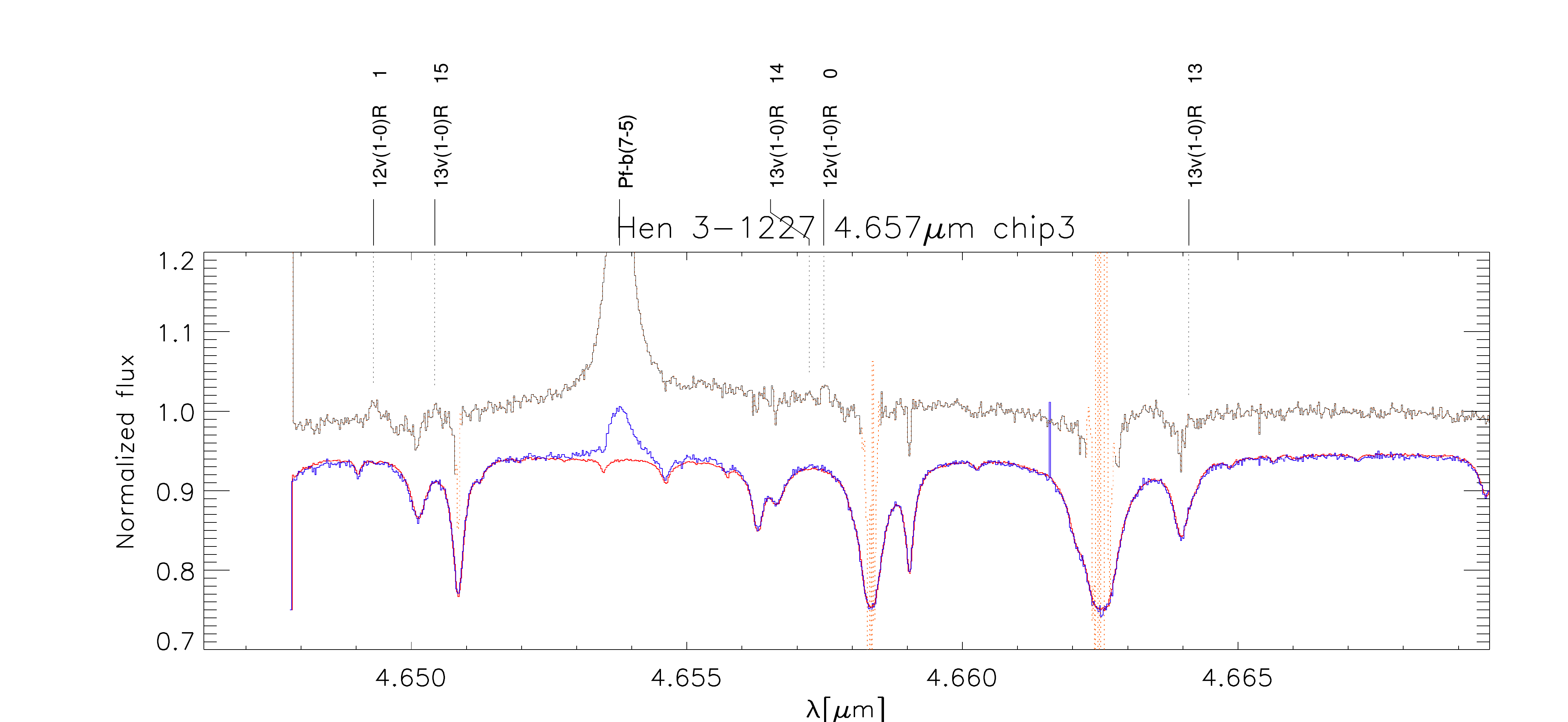}\\
\end{minipage}
\begin{minipage}[l]{0.4\textwidth}
   \includegraphics[width=\textwidth]{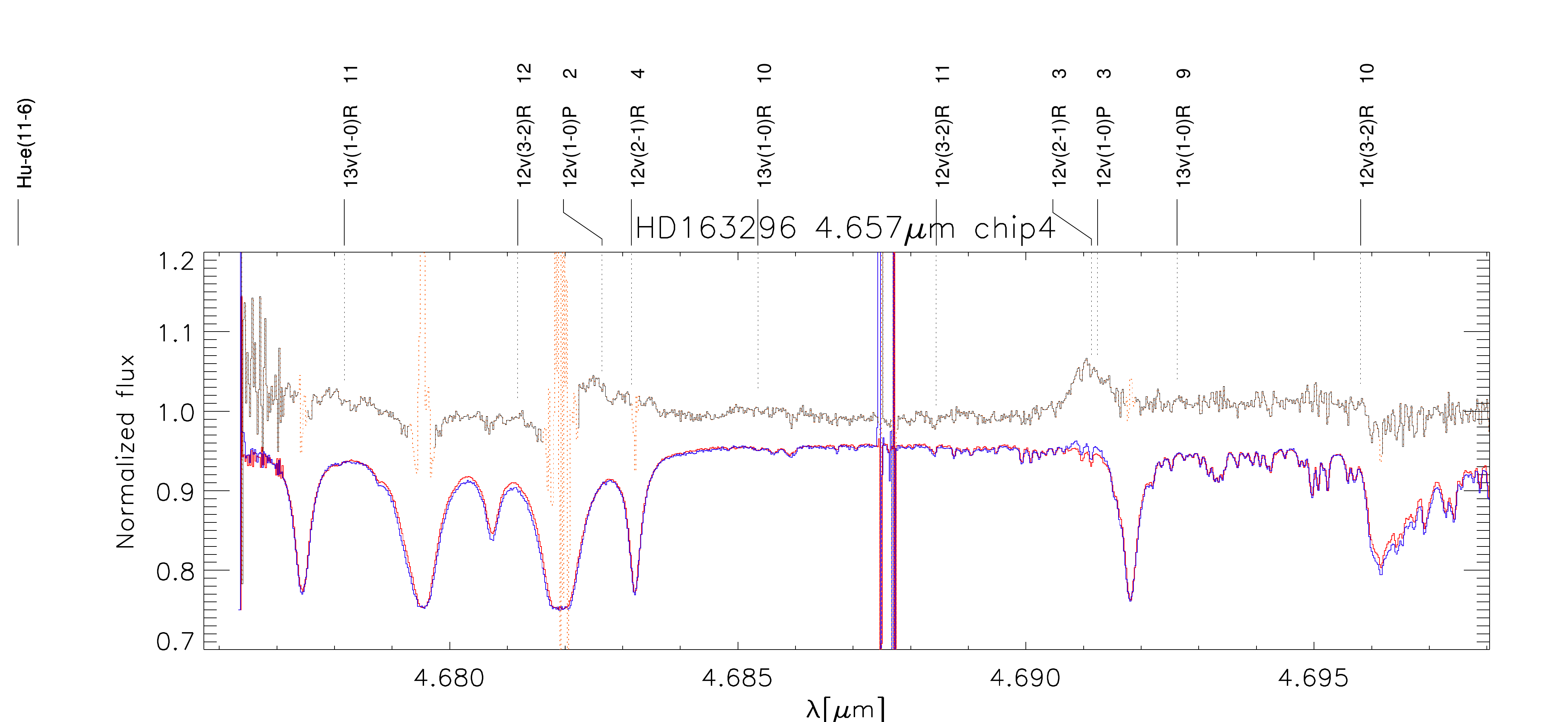}
 \end{minipage}
\begin{minipage}[l]{0.4\textwidth}
  \includegraphics[width=\textwidth]{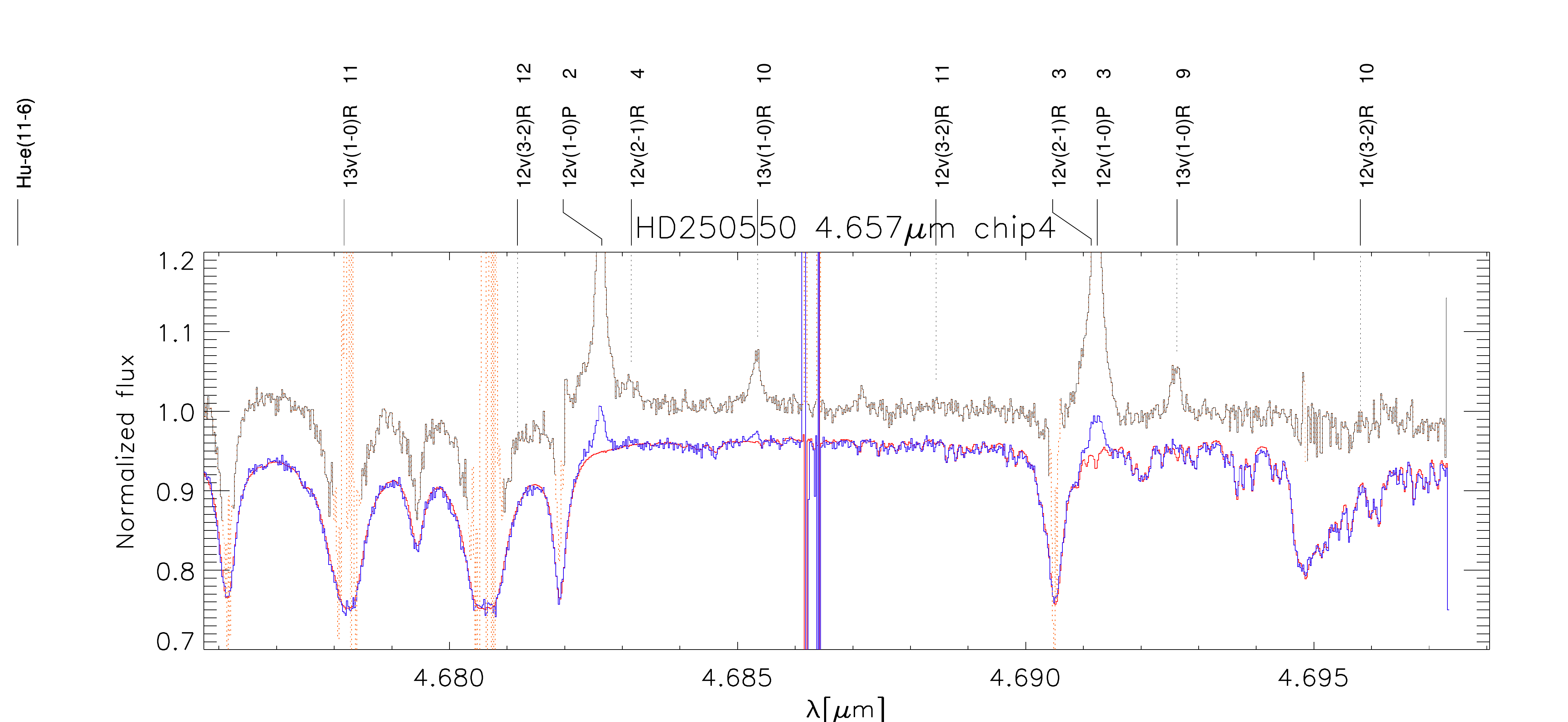}\\
 \end{minipage}
\begin{minipage}[l]{0.4\textwidth}
  \includegraphics[width=\textwidth]{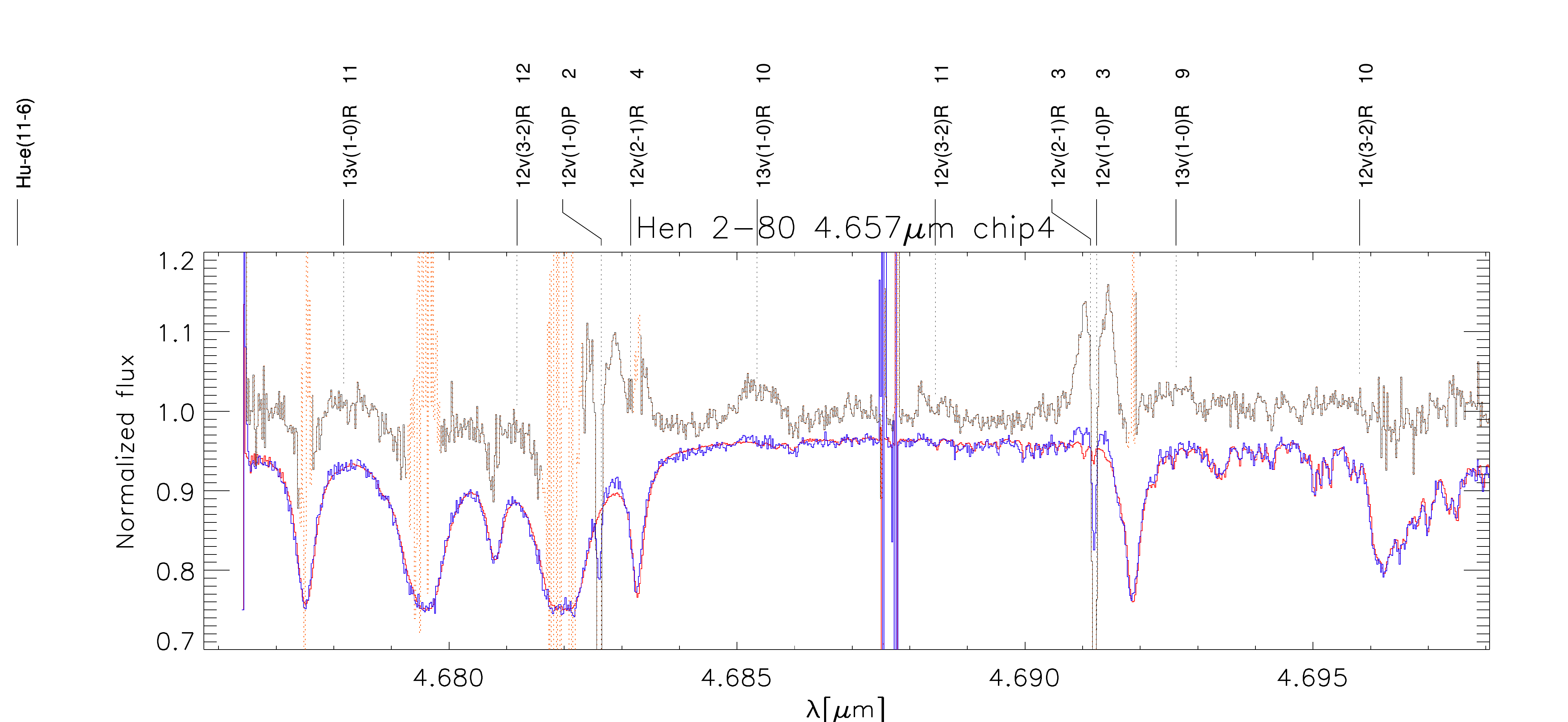}
 \end{minipage}
\begin{minipage}[l]{0.4\textwidth}
  \includegraphics[width=\textwidth]{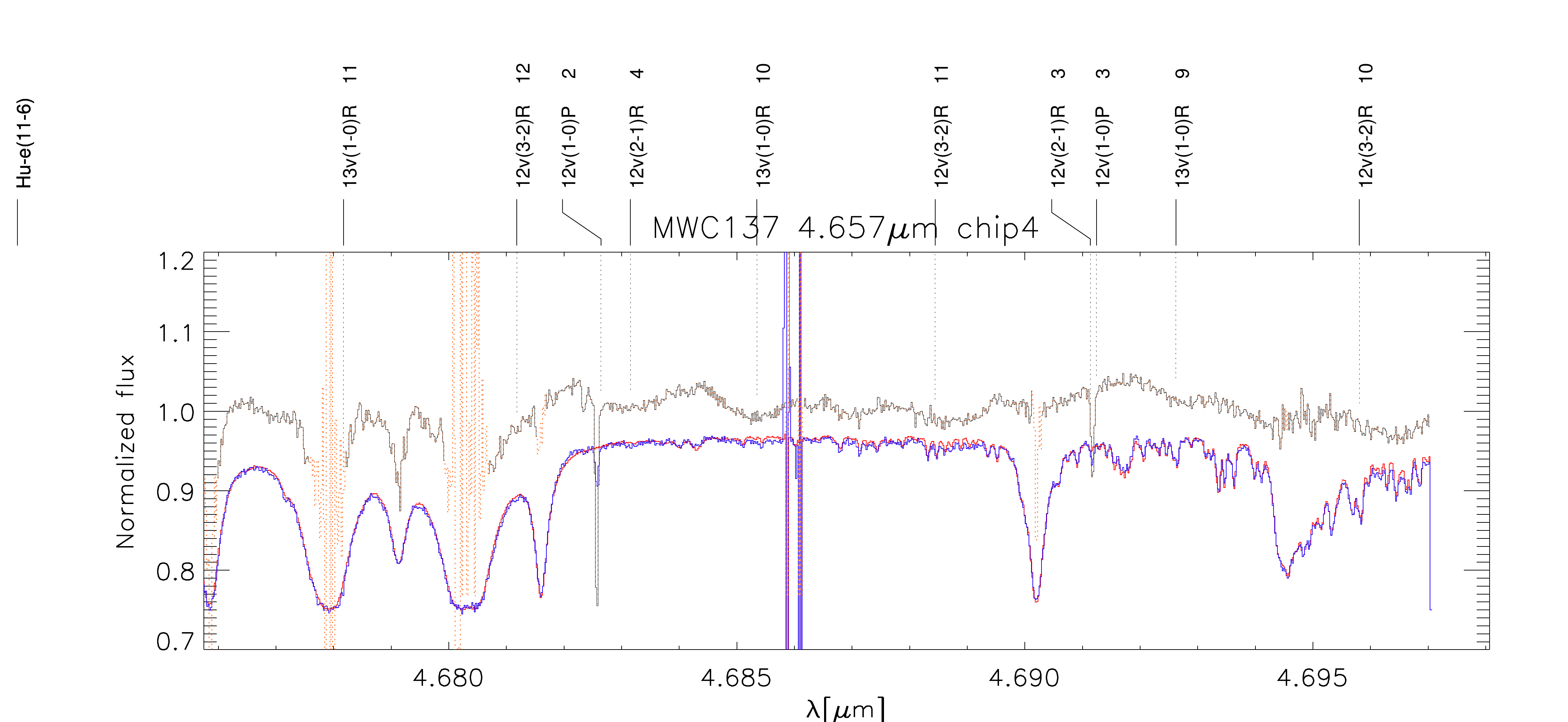}\\
 \end{minipage}
\begin{minipage}[l]{0.4\textwidth}
  \includegraphics[width=\textwidth]{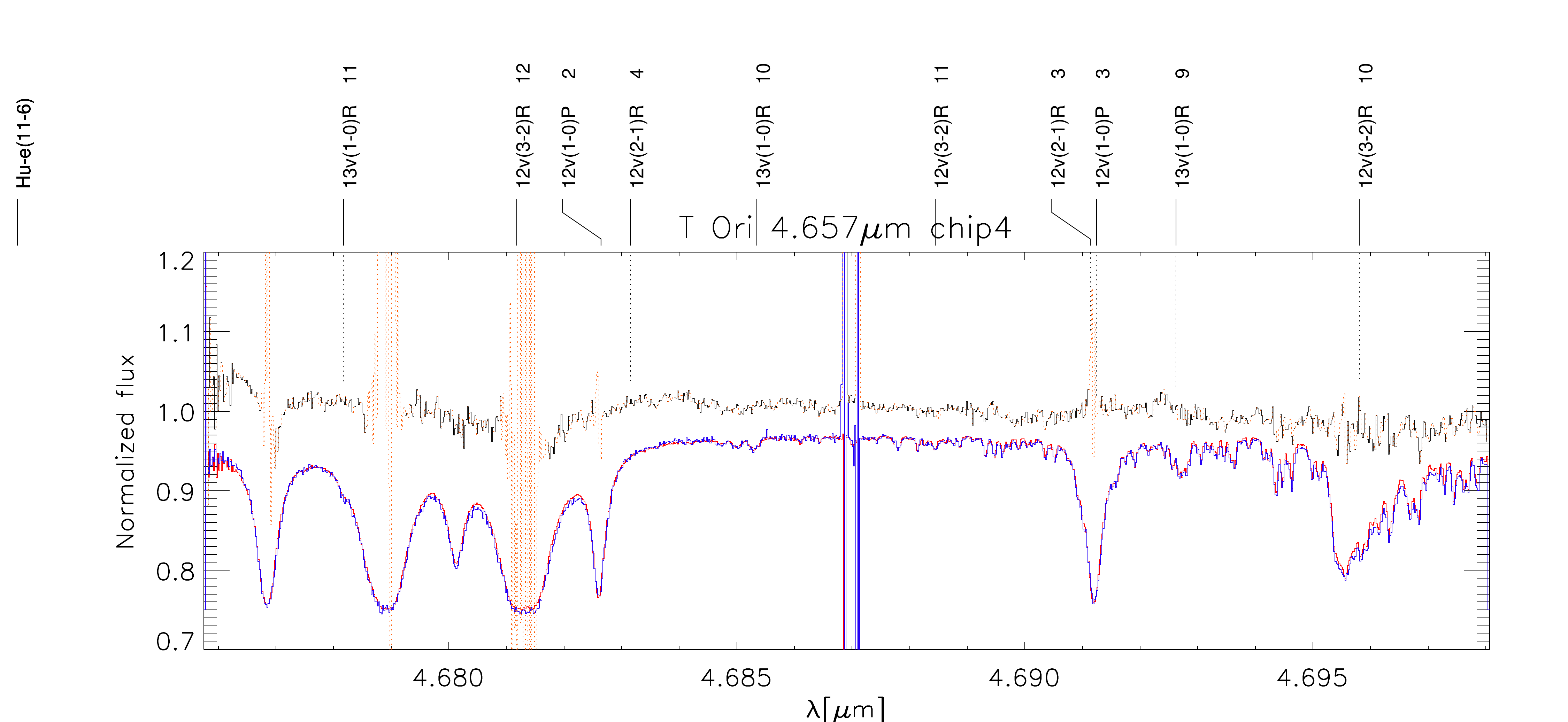}
 \end{minipage}
\begin{minipage}[l]{0.4\textwidth}
  \includegraphics[width=\textwidth]{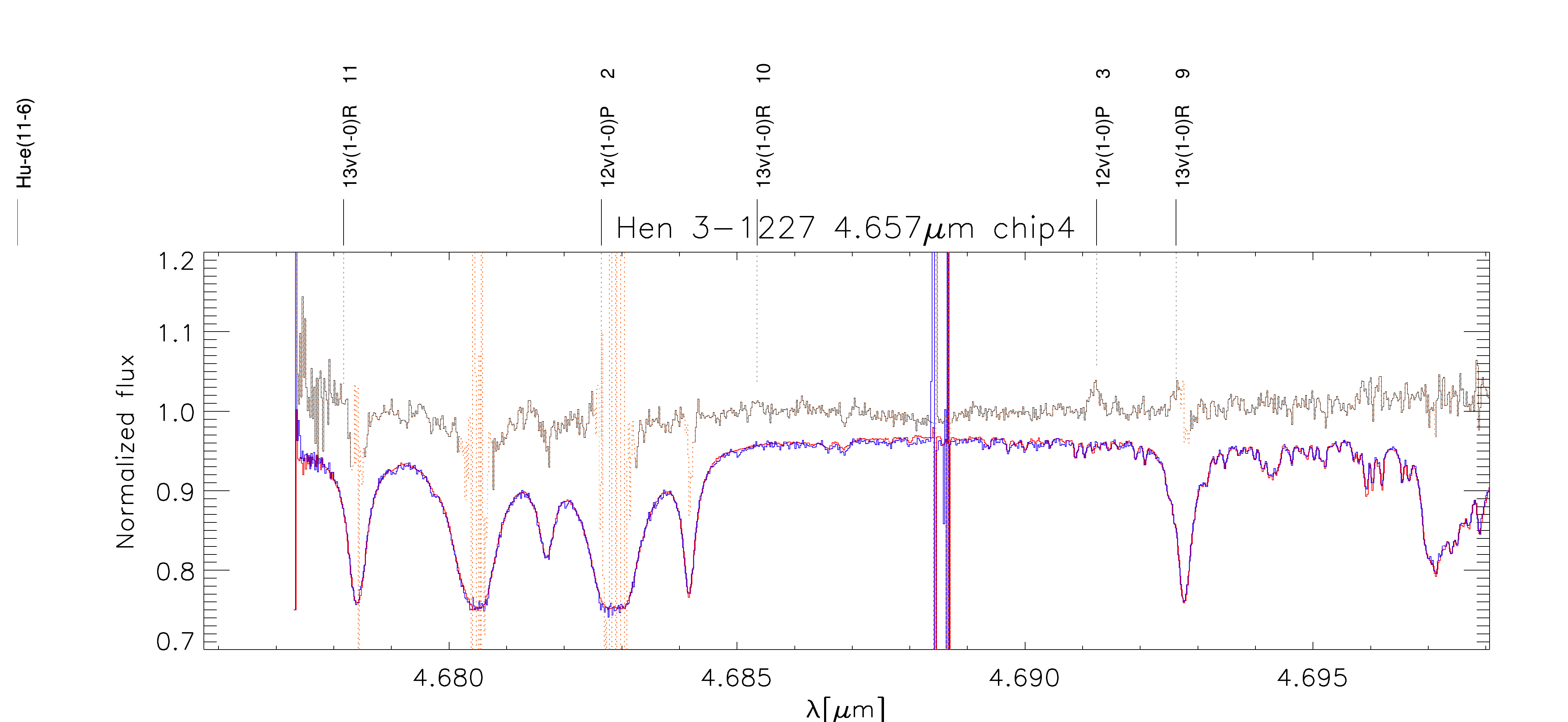}\\
\end{minipage}
\caption{Spectra from all observed sources in our sample (sources names are indicated in the plot titles). These are from chip 3 and chip 4 (indicated in the plot titles) of the spectra taken at wavelength setting 4.657 $\mu$m. The spectra are wavelength calibrated using the H\,{\sc i} lines.}
         \label{fig:spec2}
\end{figure*}

\begin{figure*}[!htbp]
\centering
\begin{minipage}[l]{0.45\textwidth}
   \includegraphics[width=\textwidth]{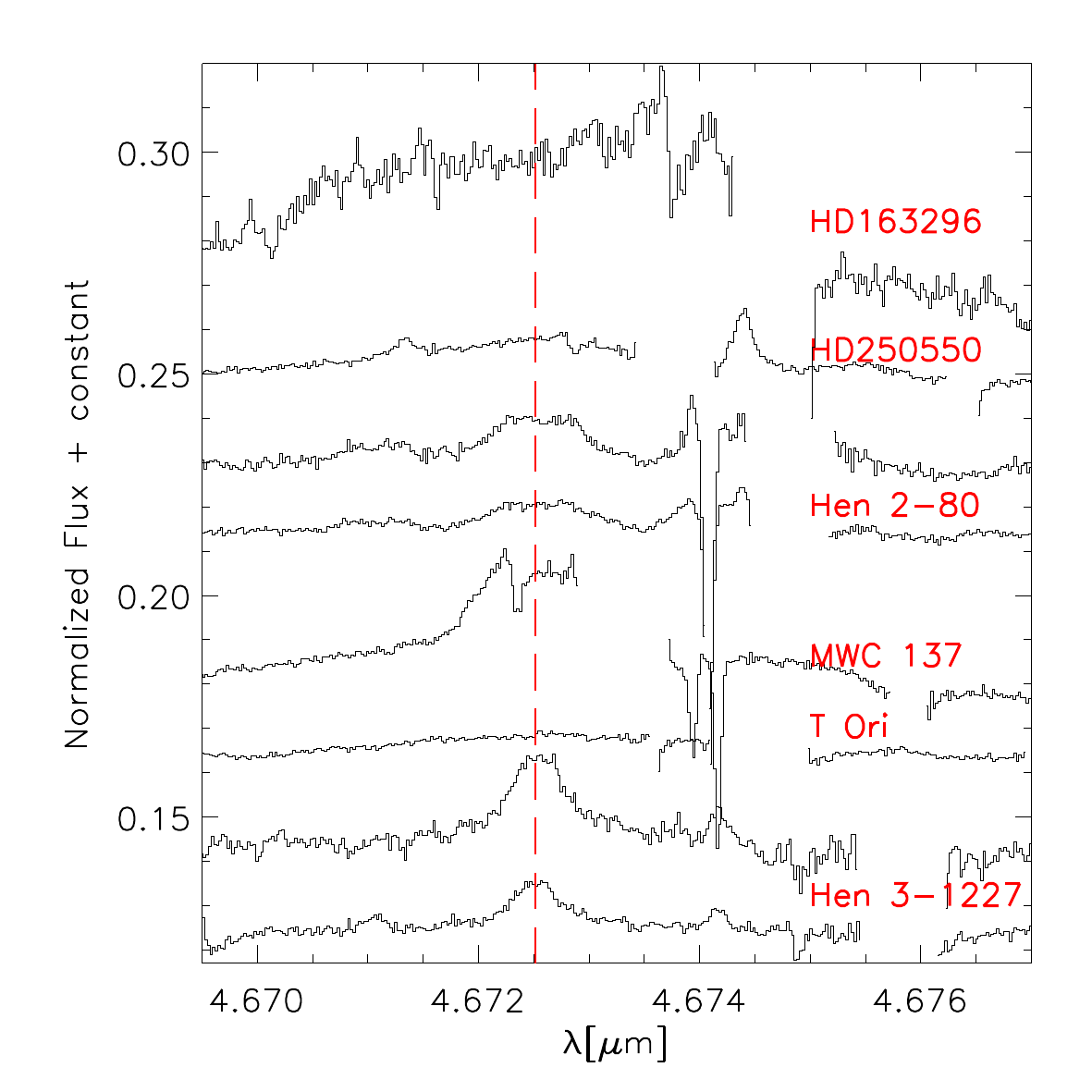}\\
\end{minipage}
\begin{minipage}[l]{0.45\textwidth}
   \includegraphics[width=\textwidth]{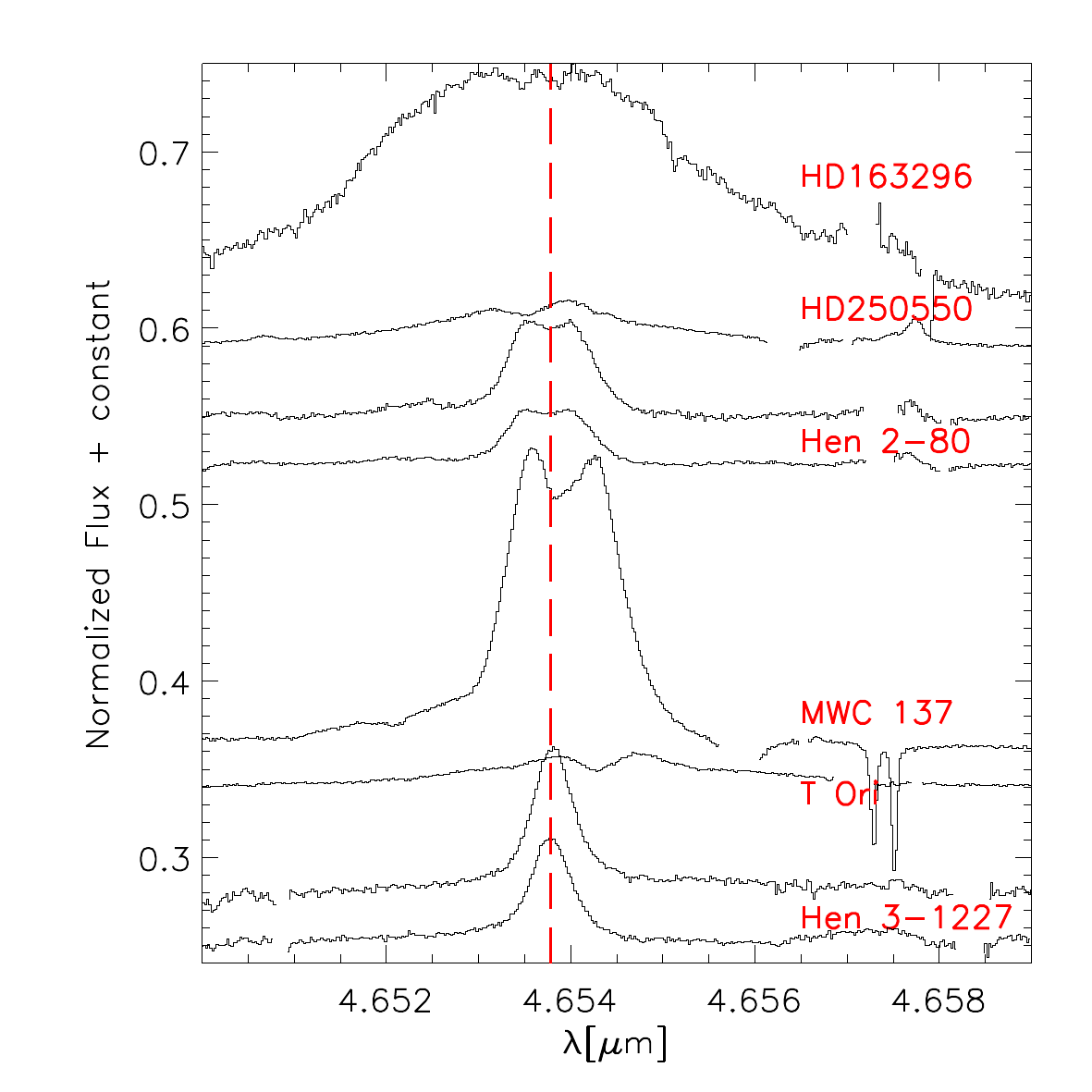}
\end{minipage}
\caption{The Humphreys $\epsilon$ 11-6 lines (left) and the Pfund $\beta$  7-5 lines (right) observed from our sample of stars. To avoid residual telluric features, caused by the saturation of telluric emission lines, we show only those regions of the spectra with transmittance above 35\%. The vertical red dashed line is the location of the line centre. Target names are indicated on the figure.}
         \label{fig:HI}
\end{figure*}

\begin{figure}[!h]
\centering
   \includegraphics[width=.48\textwidth]{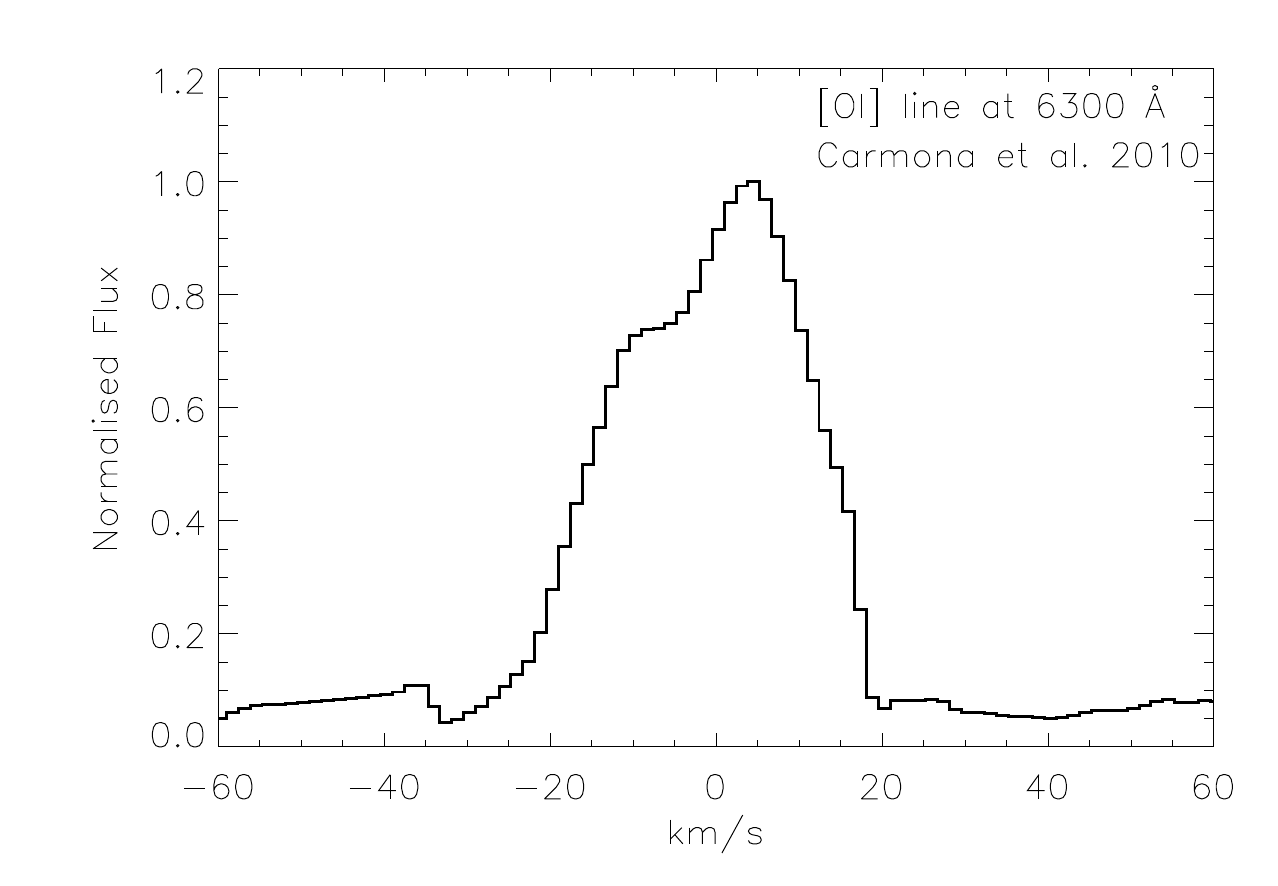}
\caption{{Line profile of the [\ion{O}{i}] line at 6300 \AA\ observed from \mbox{Hen 2-80} by \citet{carmona2010}. The observed spectrum is not corrected for telluric absorption or emission (airglow).}}
         \label{OIline}
\end{figure}

\begin{figure}[!h]
\centering
   \includegraphics[width=.48\textwidth]{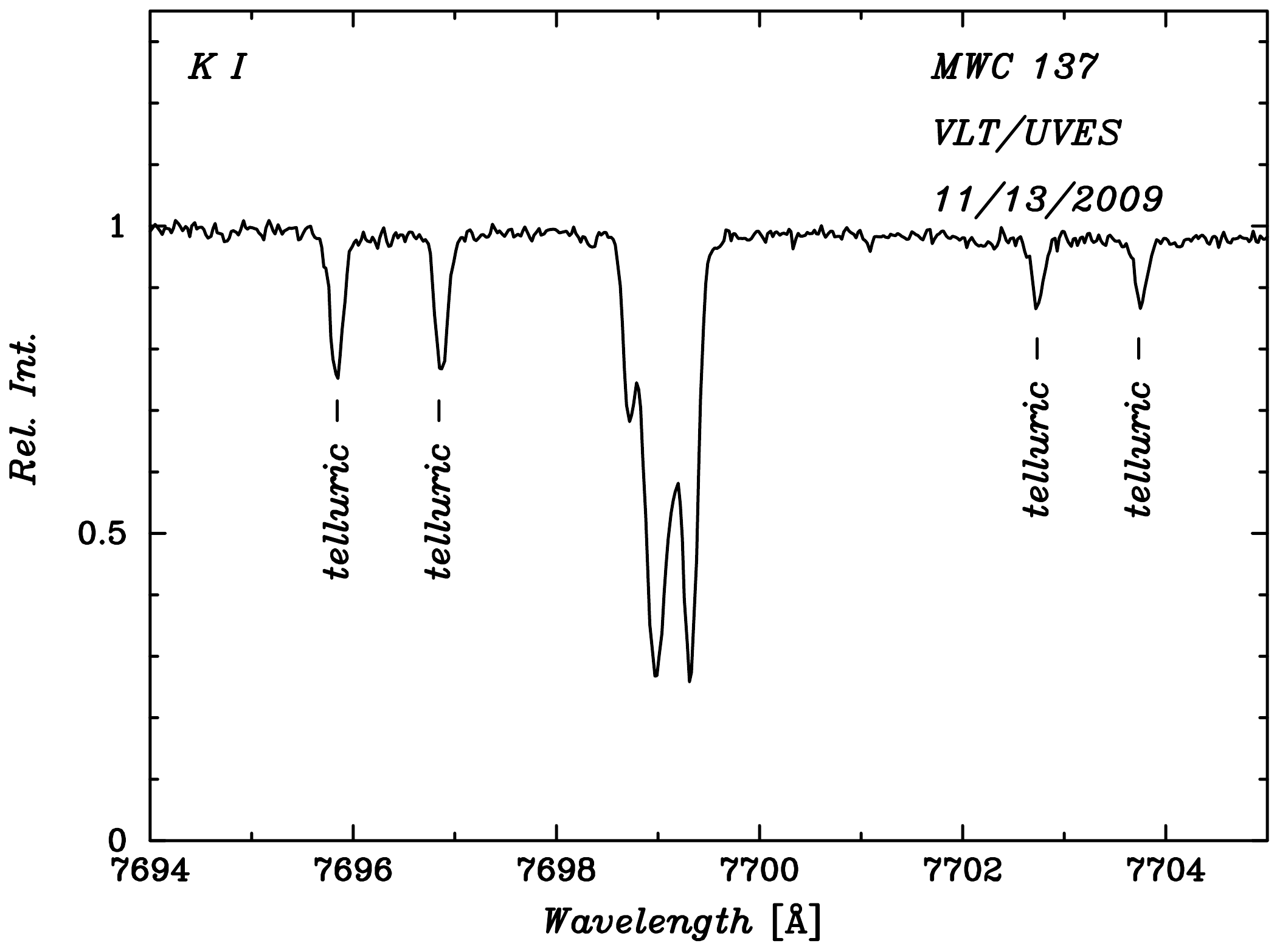}
\caption{{VLT/UVES optical spectrum from MWC~137 (extracted from the ESO data archive) showing multiple absorption components in the \ion{K}{i} 7699 $\AA$ line.}}
         \label{mwc137_ki}
\end{figure}

Figures \ref{fig:spec1} and \ref{fig:spec2} show examples of the near infrared CRIRES spectra from one of the six wavelength settings from all six sources. The final telluric corrected spectra are shown (black) together with the original uncorrected spectra (blue) and the telluric standard spectra (red). CO ro-vibrational and \ion{H}{I} line identifications are indicated on the plots. Low transmission regions in the final corrected spectra are plotted as dotted red lines and are ignored during line extraction. Fig. \ref{fig:HI} shows the \ion{H}{I} lines present in our spectra and Table \ref{table:HI} shows the collected FWHM and velocity shifts of these lines.
{In Fig. \ref{OIline}, we show the line profile of the [OI] line at 6300 \AA\ observed from Hen~2-80 by \citet{carmona2010}, and in Fig. \ref{mwc137_ki} we show a VLT/UVES spectrum from MWC~137 with a detection of the \ion{K}{i} 7699 \AA\ line.}

\begin{table}[!htbp]
\small
\caption{HI line data and velocity shifts. }             
\label{table:HI}      
\centering                          
\begin{tabular}{c c c c c c c c}        
\hline
Pf$\beta$ 7-5& { FWHM} [km/s] & $\lambda_{\rm centre}$ [nm] &$\Delta$v [km/s]  \\   
\hline
{HD~163296}  &333$\pm$21	 &4653.16&-40$\pm$7 		\\   
{HD~250550}  &149$\pm$16	 &4654.11& 21$\pm$6	 \\   
{Hen~2-80 05}&63$\pm$3& 4653.16&-40$\pm$1	\\   
{Hen~2-80 06}& 61$\pm$3& 4653.15&-41$\pm$1	\\   
{MWC137}  & 88$\pm$5 & 4654.84& +68$\pm$2	\\   
{T~Ori}   & 209$\pm$19	 & 4654.10&+20$\pm$4&	\\   
{Hen~3-1227 05} &40$\pm$5 & 4652.22&-100$\pm$2	\\   
{Hen~3-1227 06}  & 38$\pm$3 & 4652.20&-102$\pm$1 \\   
{Lab}  & 	 & 4653.78&	\\   
\hline                           
Hu $\epsilon$ 11-6&{ FWHM} [km/s] & $\lambda_{\rm centre}$ [nm] &$\Delta $v [km/s] \\   
\hline
{HD~163296}  &241$\pm$104&4672.85&-42$\pm$34\\   
{HD~250550}&261$\pm$43&4673.21&45$\pm$9\\   
{Hen~2-80 05}&65$\pm$5&4671.92&-38$\pm$2\\   
{Hen~2-80 06}&68$\pm$6&4671.95&-36$\pm$2\\   
{MWC137} &75$\pm$18&4673.36&+51$\pm$7\\   
{T~Ori}  &164$\pm$22&	4672.17&-22$\pm$7	\\   
{Hen~3-1227 05}&51$\pm$8&4670.95&-100$\pm$3\\   
{Hen~3-1227 06}&41$\pm$9&4670.93&-101$\pm$4 \\   
{Lab}  &	 &	4672.51&	\\   
\hline                           
\end{tabular}
\tablefoot{The  FWHM, the line centres, and the velocity shift are found by fitting the profiles to a Gaussian. The error estimates are the three sigma errors from the fitting of the Gaussian. Large error bars, when present, are due to the HI profiles being poorly approximated by a Gaussian profile.}
\end{table}

\clearpage

\section{Extracted line fluxes and profiles} \label{sec:res}
\begin{sidewaysfigure*}[!h]
\centering
\begin{minipage}[l]{0.23\textwidth}
   \includegraphics[width=\textwidth]{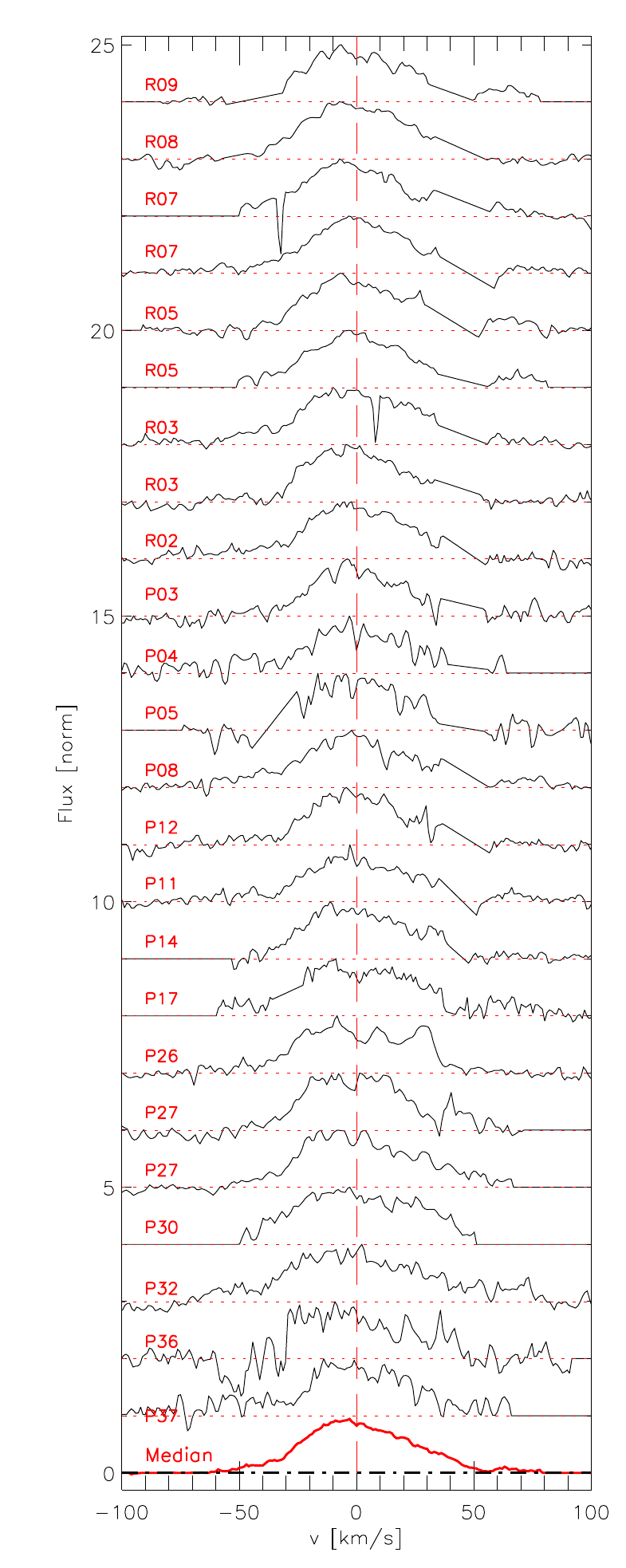}
 \end{minipage}
\begin{minipage}[l]{0.23\textwidth}
   \includegraphics[width=\textwidth]{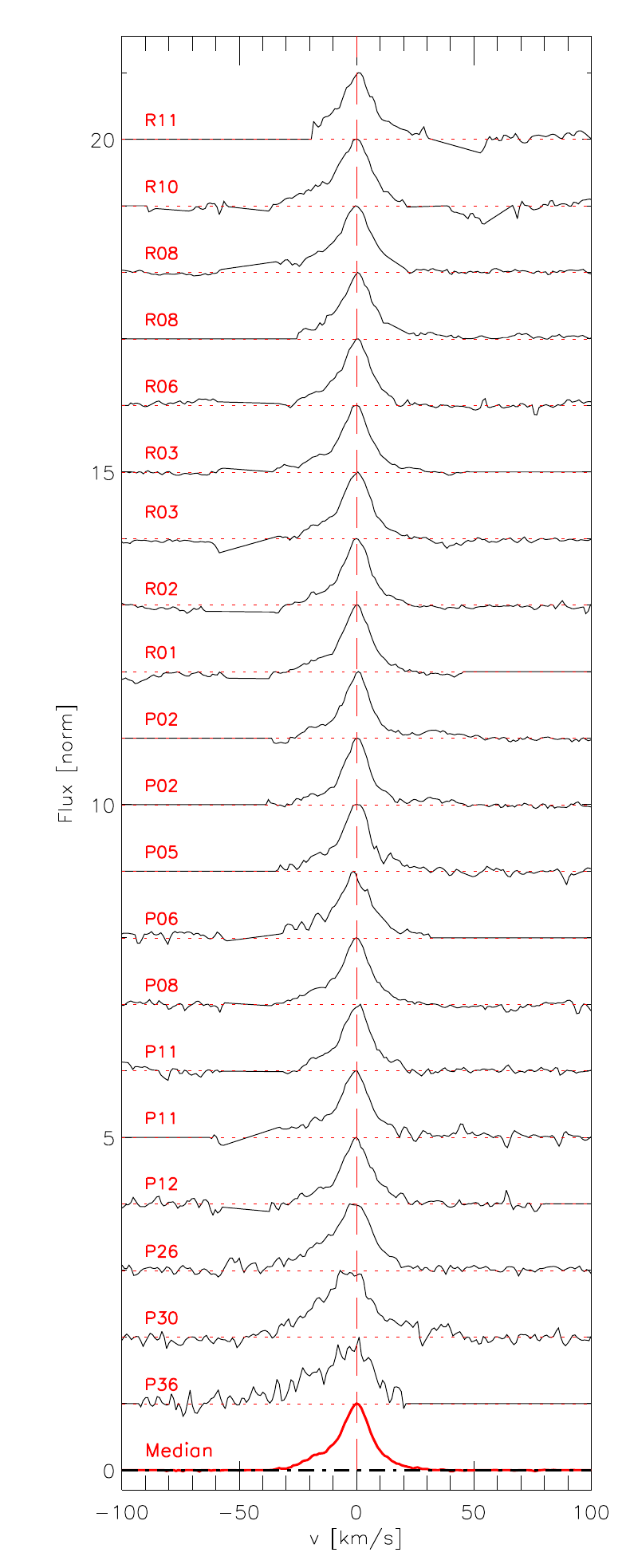}
 \end{minipage}
\begin{minipage}[l]{0.23\textwidth}
   \includegraphics[width=\textwidth]{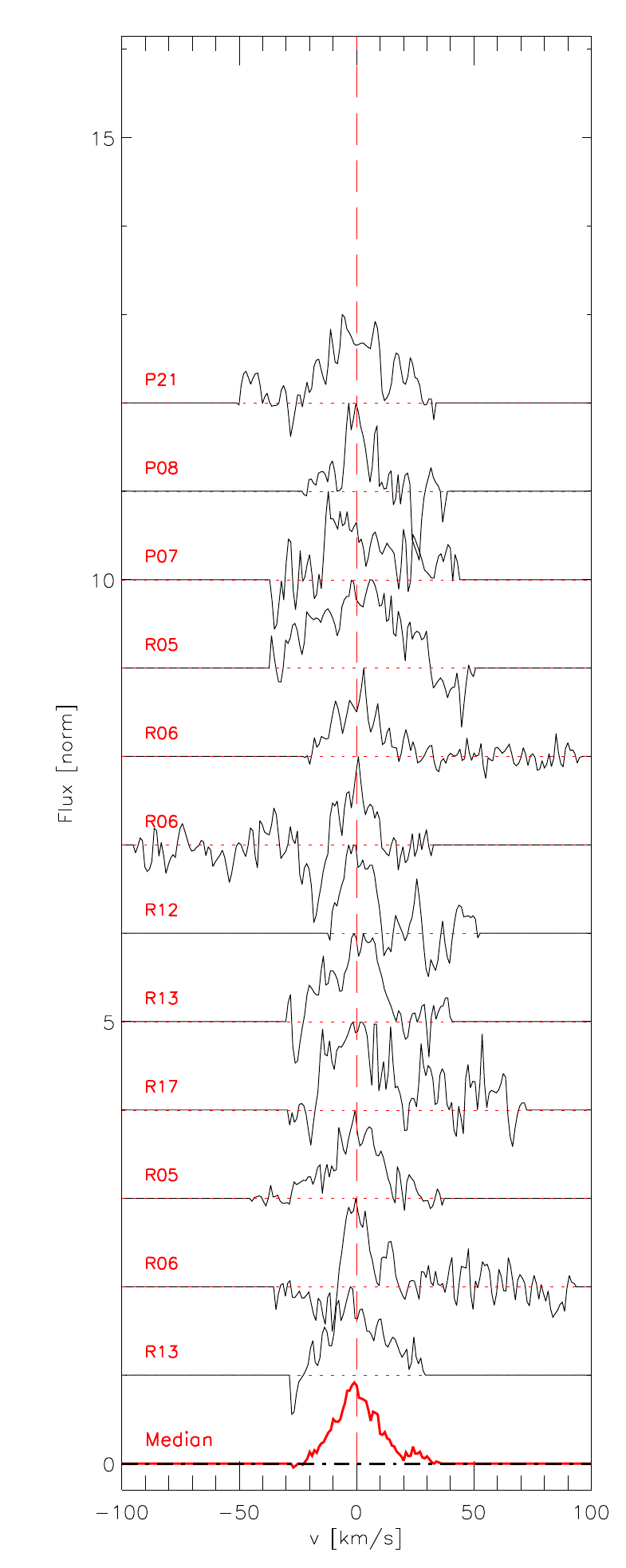}
  \end{minipage}
\begin{minipage}[l]{0.23\textwidth}
   \includegraphics[width=\textwidth]{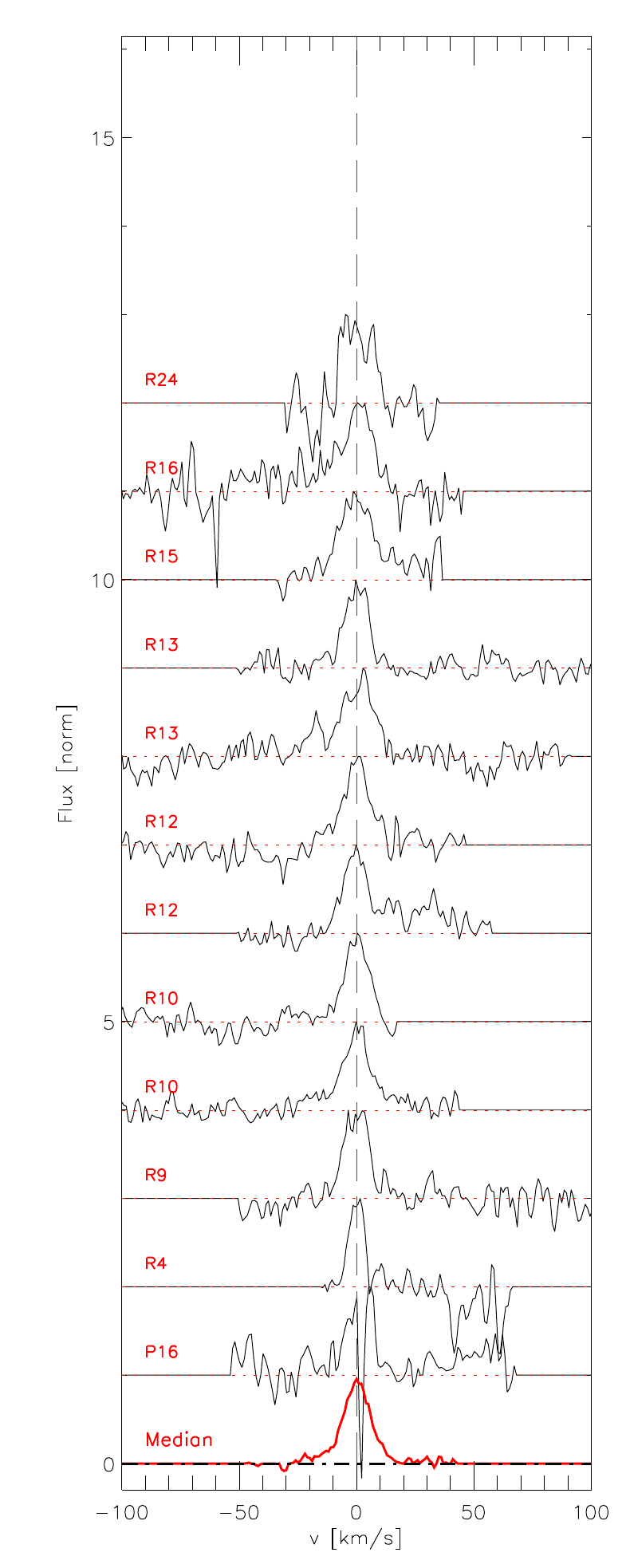}
  \end{minipage}
\caption{from left to right \element[][12]{CO} $v$=1-0 lines observed from HD~163296 and \element[][12]{CO} $v$=1-0 lines, \element[][12]{CO} $v$=2-1 lines and \element[][13]{CO} $v$=1-0 lines from HD~250550.}
         \label{hd16_profile}
\end{sidewaysfigure*}

\begin{sidewaysfigure*}[!h]
\centering
\begin{minipage}[l]{0.23\textwidth}
   \includegraphics[width=\textwidth]{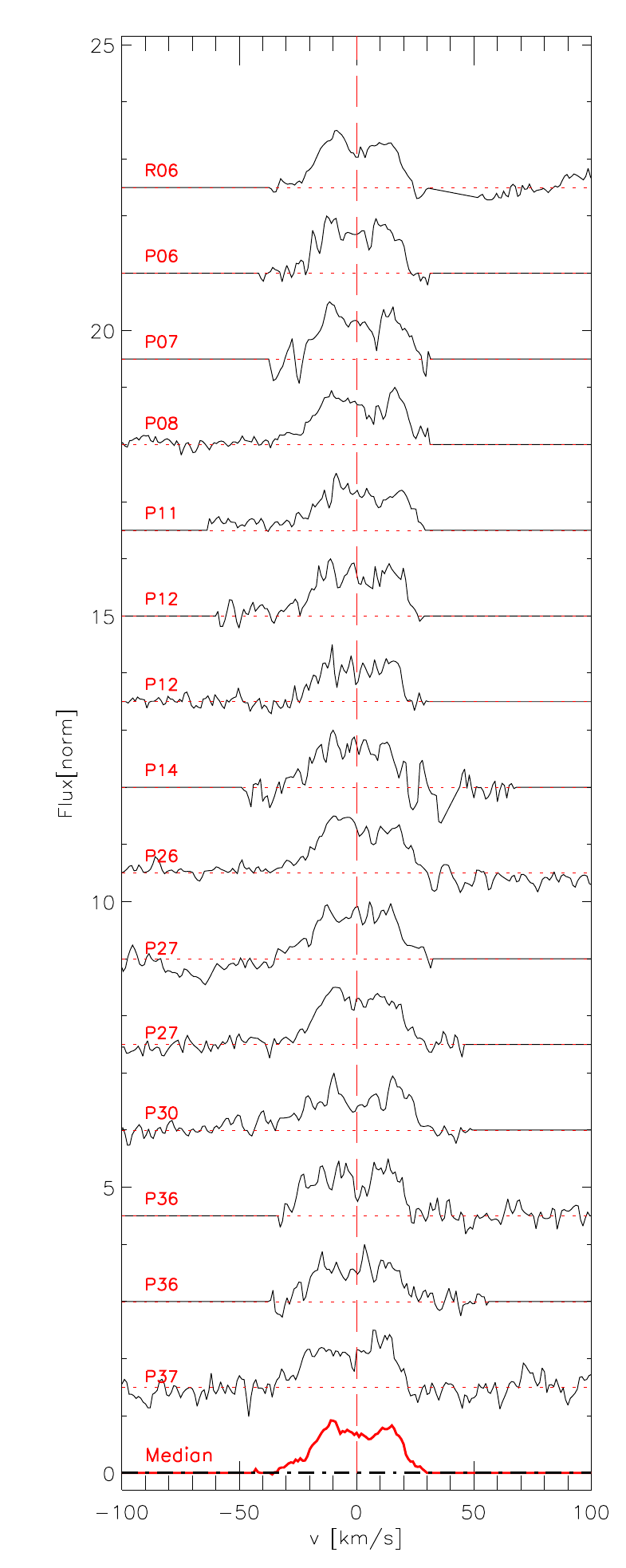}
\end{minipage}
\begin{minipage}[l]{0.23\textwidth}
   \includegraphics[width=\textwidth]{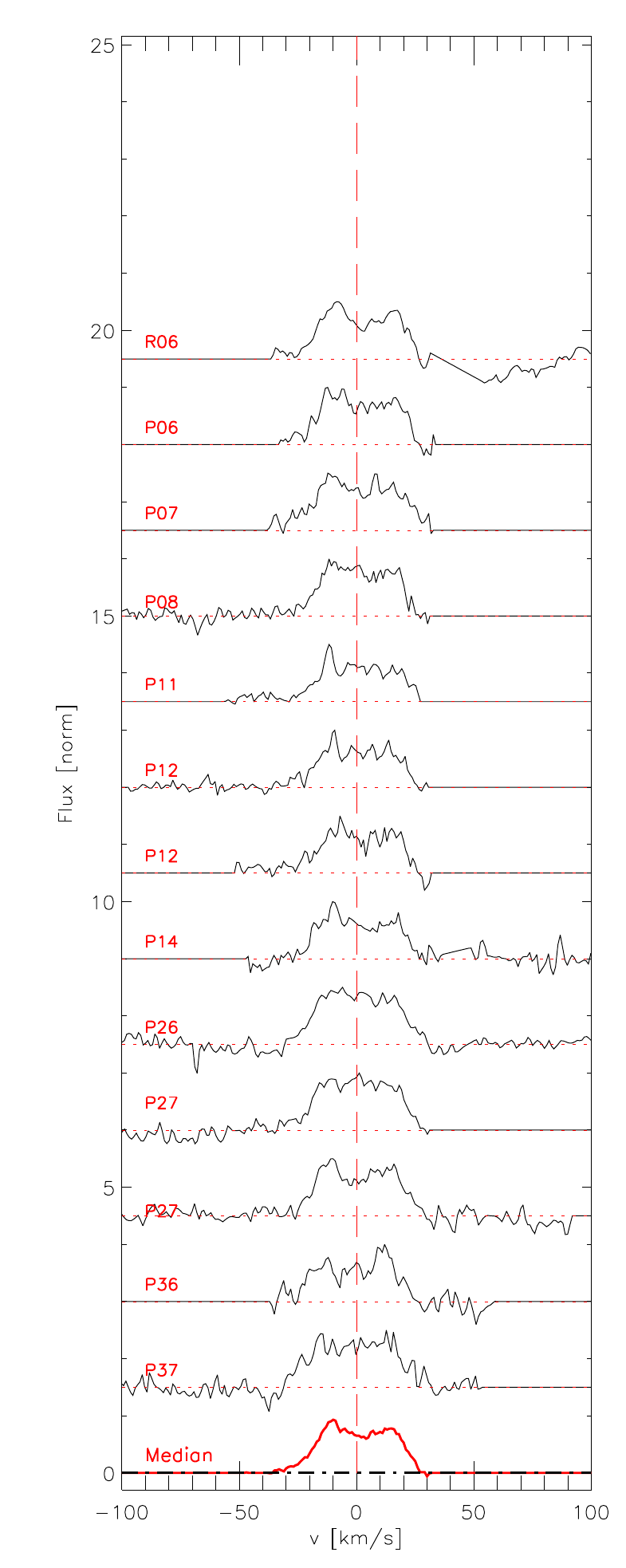}
\end{minipage}
\begin{minipage}[l]{0.23\textwidth}
   \includegraphics[width=\textwidth]{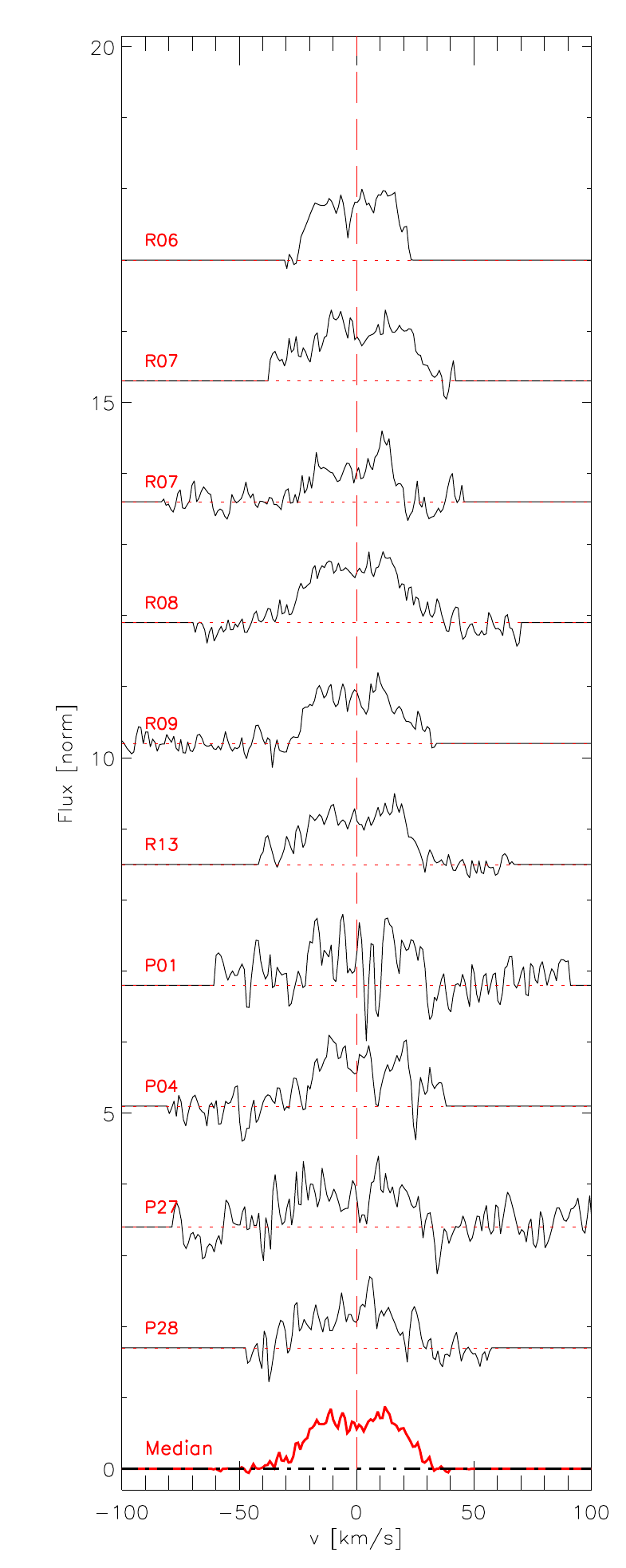}
\end{minipage}
\begin{minipage}[l]{0.23\textwidth}
   \includegraphics[width=\textwidth]{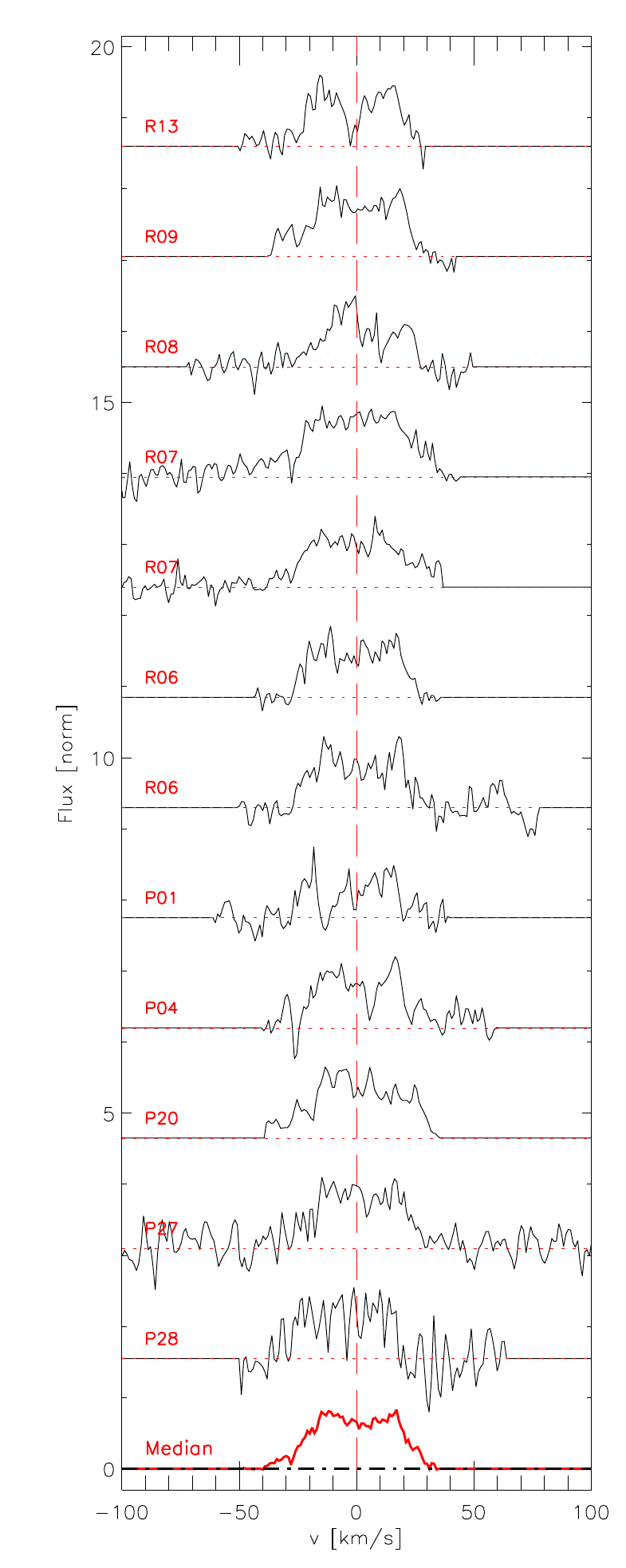}
\end{minipage}
\caption{From left to right\element[][12]{CO} $v$=1-0 lines observed from Hen~2-80 on March 5, 2012 and on March 6, 2012, and \element[][12]{CO} $v$=2-1 lines observed from Hen~2-80 on March 5, 2012 and March 6, 2012.}
         \label{hen280v1_profile}
\end{sidewaysfigure*}

\begin{sidewaysfigure*}[!h]
\centering
\begin{minipage}[l]{0.23\textwidth}
   \includegraphics[width=\textwidth]{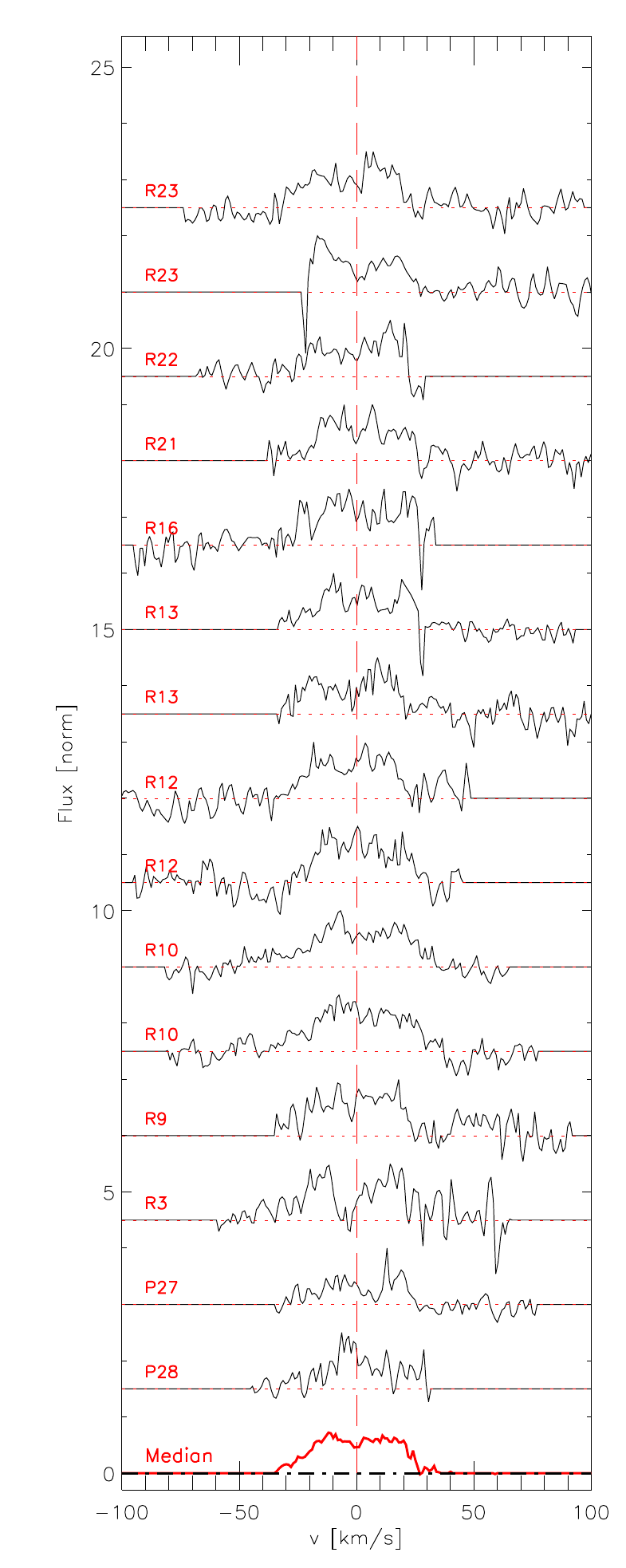}
\end{minipage}
\begin{minipage}[l]{0.23\textwidth}
   \includegraphics[width=\textwidth]{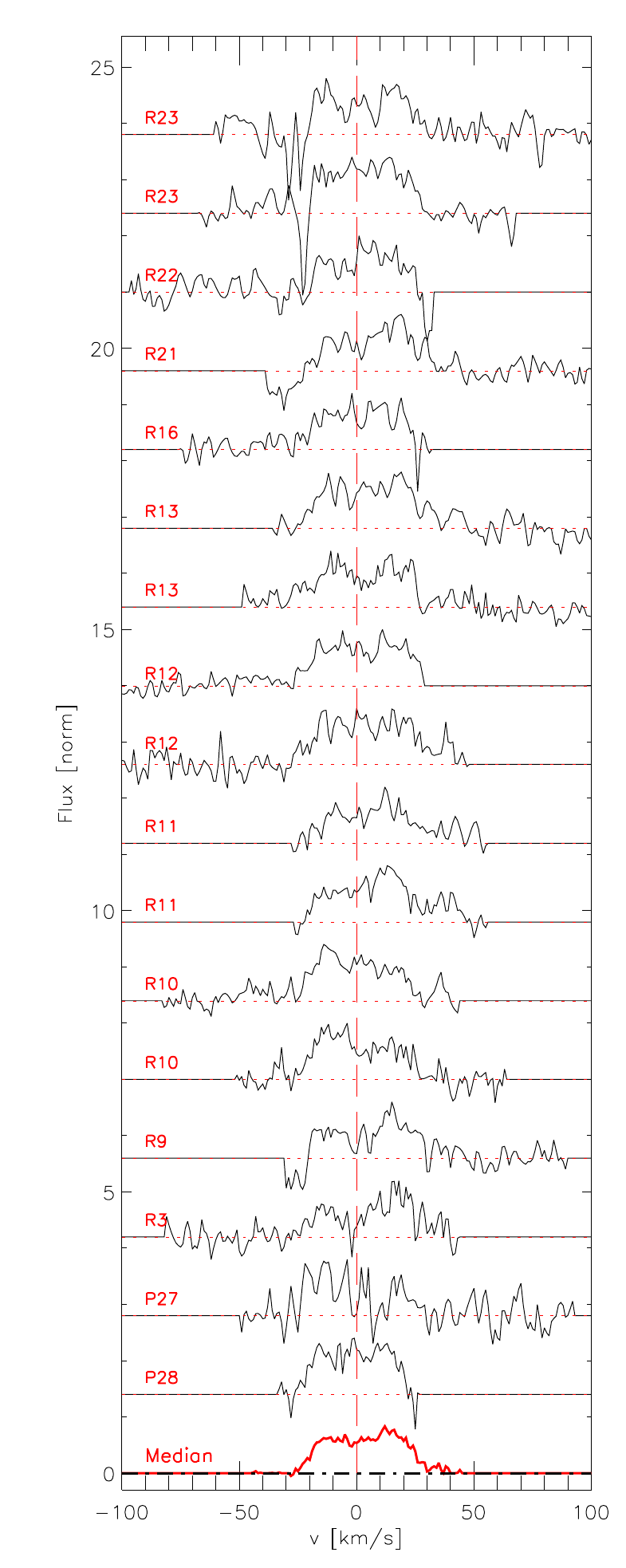}
\end{minipage}
\begin{minipage}[l]{0.23\textwidth}
   \includegraphics[width=\textwidth]{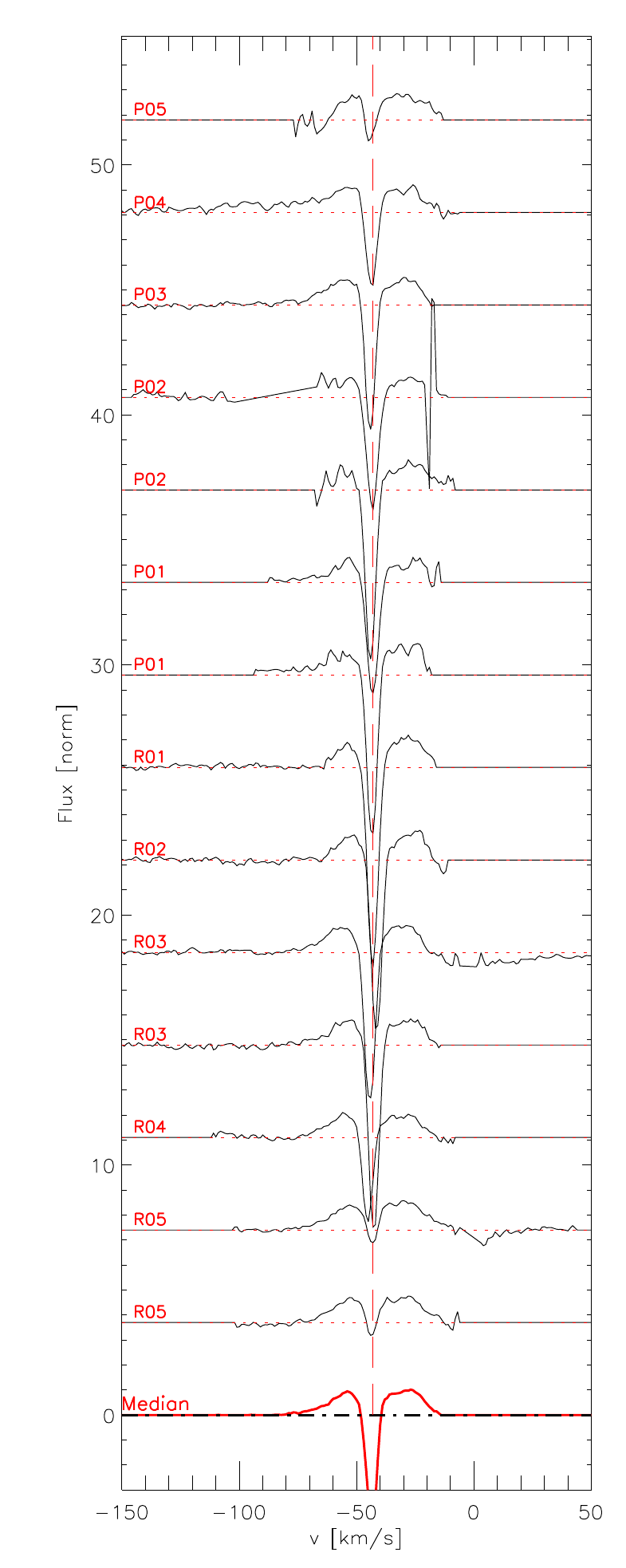}
\end{minipage}
 \caption{From left to right, \element[][13]{CO} $v$=1-0 lines observed from Hen~2-80 on March 5, 2012 and March 6, 2012 and \element[][12]{CO} absorption lines superposed on the emission lines observed from Hen~2-80 on March 5, 2012.}
         \label{hen280v13_profile}
\end{sidewaysfigure*}

\begin{sidewaysfigure*}[!h]
\centering
\begin{minipage}[l]{0.23\textwidth}
   \includegraphics[width=\textwidth]{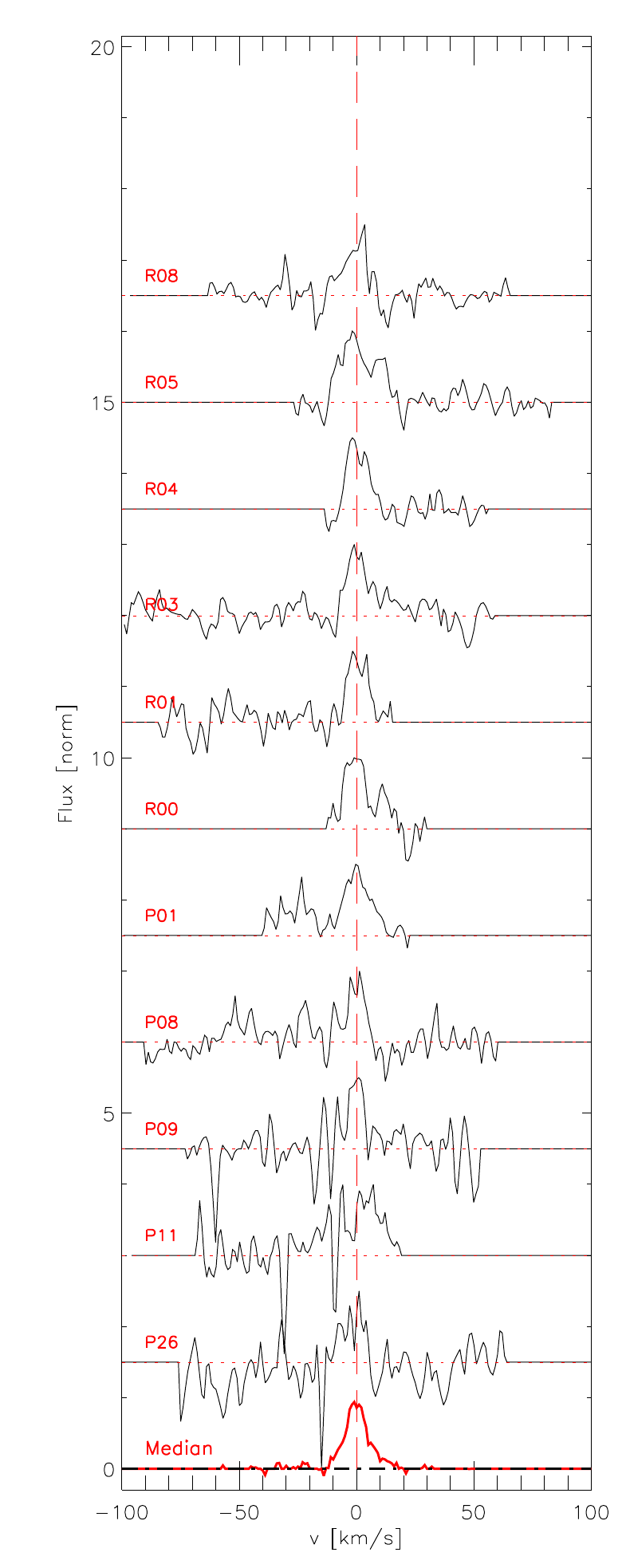}
\end{minipage}
\begin{minipage}[l]{0.23\textwidth}
   \includegraphics[width=\textwidth]{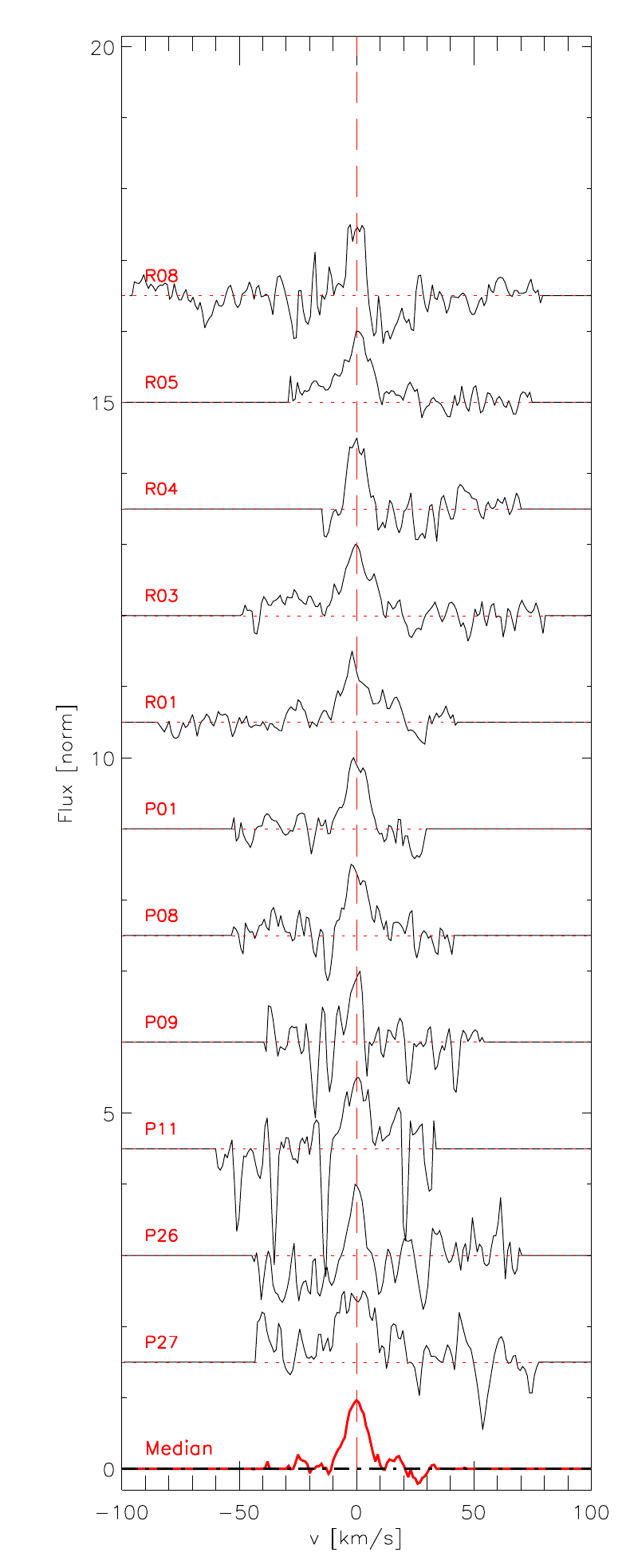}
\end{minipage}
\begin{minipage}[l]{0.23\textwidth}
  \includegraphics[width=\textwidth]{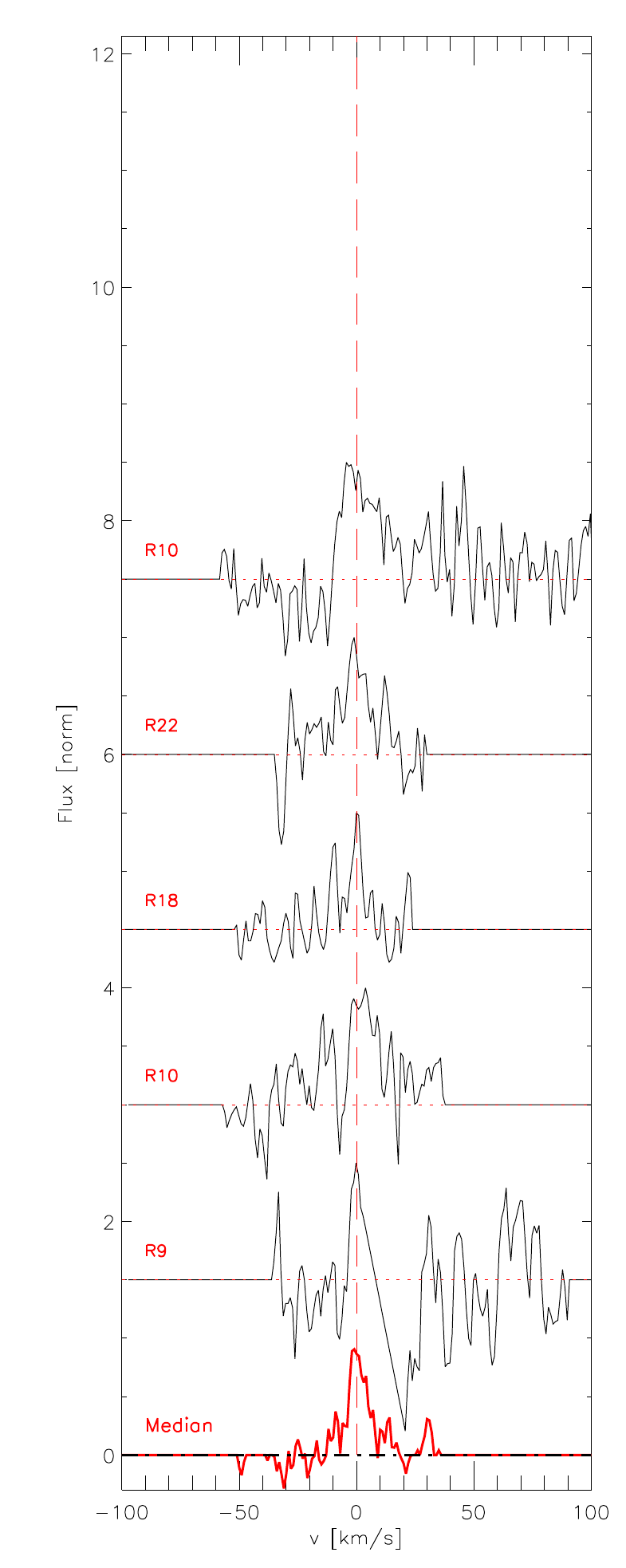}
\end{minipage}
\begin{minipage}[l]{0.23\textwidth}
   \includegraphics[width=\textwidth]{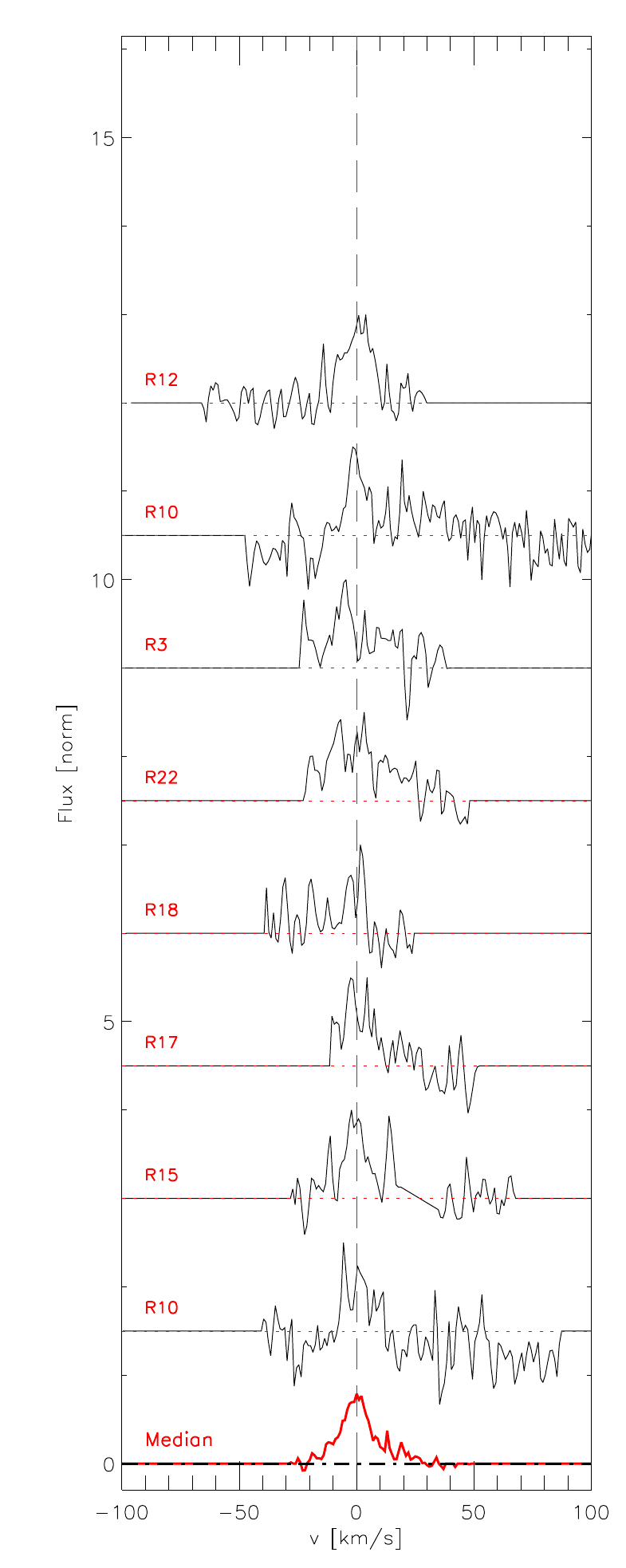}
\end{minipage}
\caption{From left to right, \element[][12]{CO} $v$=1-0 lines observed from Hen~3-1227 on March 5, 2012 and March 6, 2012, and \element[][13]{CO} $v$=1-0 lines observed from Hen~3-1227 on March 5, 2012 and March 6, 2012.}
         \label{hen3v1_profile}
\end{sidewaysfigure*}

The individual CO ro-vibrational lines collected from the four CO detected sources and selected for further analysis are listed in Table \ref{table:COlines} and shown in Figures \ref{hd16_profile} to \ref{hen3v1_profile} together with a median of all transitions at the bottom of each plot. In Fig. \ref{hen280v13_profile}, we show also the low $J$, $v$=1-0 lines collected from Hen~2-80 with CO ro-vibrational absorption superposed. These were not selected for the further analysis.
For the selected lines, line fluxes where calculated and are presented in Tables \ref{tab:hd16} to \ref{tab:hen3}. The line selection approach is described in Sect. \ref{line_selection}. 

\begin{table}[!htbp]
\caption{CO lines selected for further analysis.}             
\label{table:COlines}      
\centering                          
\begin{tabular}{l l  }        
\hline            
 HD~163296 \\
\hline
v=1-0&P37,P36,P32,P30,P27,P26,P17,P14,P12,P11,\\
	&P8,P5,P4,P3,R2,R3,R5,R6,R7,R8,R9\\
\hline            
 HD~250550 \\
\hline
v=1-0&P36,P30,P26,P12,P11,P8,P6,P5,P3,P2,\\
	&R1,R2,R3,R6,R8,R10,R11\\
v=2-1&R17,R13,R12,R6,R5,P7,P8,P21	\\
\element[][13]{CO} $v$=1-0& P16,R4,R9,R10,R12,R13,R15,R16,R24\\
\hline            
Hen~2-80 \\
\hline
v=1-0&P37,P36,P30,P27,P26,P14,P12,P11,P8,P7,\\
	&P6,R6\\
v=2-1&P28,P27,P20,P4,P1,R6,R7,R8,R9,R13\\
\element[][13]{CO} $v$=1-0& P28,P27,R3,R9,R10,R11,R12,R13,R16,\\
	&R21,R22,R23\\
\hline                           
 Hen~3-1227 \\
\hline
v=1-0&P27,P26,P11,P9,P8,P1,R0,R1,R3,R4,\\
	&R5,R8\\
\element[][13]{CO} $v$=1-0&R3,R9,R10,R12,R15,R17,R18,R22\\
\hline            
\end{tabular}
\end{table}

\begin{table}[!h]
\caption{HD~163296 \element[][12]{CO} line fluxes.}             
\label{tab:hd16}      
\centering                          
\begin{tabular}{l llll  }        
            
line	&$\lambda_{\rm line}$[nm] &  Cont.[$\frac{\rm{erg}}{\rm{cm}^2\rm{s}}$] &    $F_{\rm line}$[$\frac{\rm{erg}}{\rm{cm}^2\rm{s}}$]  & Error[$\frac{\rm{erg}}{\rm{cm}^2\rm{s}}$]  \\ [1ex]
\hline
  \multicolumn{2}{l}{\element[][12]{CO} $v$=1-0}\\
 \hline
R09 &   4587.64 &  2.31e-12 &  1.4e-13 &  2e-14\\
R08 &   4594.99 &  2.28e-12 &  2.2e-13 &  2e-14\\
R07 &   4602.44 &  2.29e-12 &  1.9e-13 &  2e-14\\
R07 &   4602.44 &  2.29e-12 &  1.9e-13 &  2e-14\\
R06 &   4610.00 &  2.29e-12 &  1.3e-13 &  2e-14\\
R05 &   4617.66 &  2.28e-12 &  1.4e-13 &  2e-14\\
R05 &   4617.66 &  2.26e-12 &  1.6e-13 &  2e-14\\
R03 &   4633.28 &  2.28e-12 &  1.5e-13 &  2e-14\\
R03 &   4633.28 &  2.28e-12 &  1.1e-13 &  2e-14\\
R02 &   4641.24 &  2.29e-12 &  1.4e-13 &  2e-14\\
P03 &   4691.24 &  2.23e-12 &  1.4e-13 &  3e-14\\
P04 &   4699.95 &  2.20e-12 &  1.7e-13 &  5e-14\\
P05 &   4708.77 &  2.20e-12 &  1.8e-13 &  4e-14\\
P08 &   4735.87 &  2.16e-12 &  1.7e-13 &  3e-14\\
P11 &   4763.99 &  2.15e-12 &  1.8e-13 &  2e-14\\
P12 &   4773.58 &  2.13e-12 &  2.1e-13 &  3e-14\\
P14 &   4793.12 &  2.12e-12 &  2.1e-13 &  4e-14\\
P17 &   4823.31 &  2.07e-12 &  2.1e-13 &  4e-14\\
P26 &   4920.41 &  1.97e-12 &  1.7e-13 &  2e-14\\
P27 &   4931.82 &  1.94e-12 &  1.5e-13 &  3e-14\\
P27 &   4931.82 &  1.95e-12 &  1.1e-13 &  2e-14\\
P30 &   4966.84 &  1.89e-12 &  1.3e-13 &  3e-14\\
P32 &   4990.85 &  1.89e-12 &  1.5e-13 &  3e-14\\
P36 &   5040.48 &  1.89e-12 &  7.0e-14 &  2e-14\\
P37 &   5053.24 &  1.90e-12 &  7.0e-14 &  2e-14\\
\hline                           
\end{tabular}
\end{table}

\begin{table}[!h]
\caption{HD~250550 \element[][12]{CO} and \element[][13]{CO} line fluxes.}             
\label{tab:hd25_12}      
\centering                          
\begin{tabular}{l llll  }        
            
line	&$\lambda_{\rm line}$[nm] &  Cont.[$\frac{\rm{erg}}{\rm{cm}^2\rm{s}}$] &    $F_{\rm line}$[$\frac{\rm{erg}}{\rm{cm}^2\rm{s}}$]  & Error[$\frac{\rm{erg}}{\rm{cm}^2\rm{s}}$] \\ [1ex]
\hline
  \multicolumn{2}{l}{\element[][12]{CO} $v$=1-0}\\
 \hline
R11 &   4573.23 &  3.62e-13 &  5e-14 &  1e-14\\
R10 &   4580.38 &  3.69e-13 &  5e-14 &  1e-14\\
R08 &   4594.99 &  3.66e-13 &  5.7e-14 &  5e-15\\
R08 &   4594.99 &  3.64e-13 &  4.3e-14 &  5e-15\\
R06 &   4610.00 &  3.65e-13 &  4.8e-14 &  6e-15\\
R03 &   4633.28 &  3.65e-13 &  3.6e-14 &  6e-15\\
R03 &   4633.28 &  3.65e-13 &  4.8e-14 &  5e-15\\
R02 &   4641.24 &  3.70e-13 &  3.4e-14 &  5e-15\\
R01 &   4649.31 &  3.85e-13 &  3.0e-14 &  6e-15\\
P02 &   4682.64 &  3.69e-13 &  3.0e-14 &  3e-15\\
P02 &   4682.64 &  3.69e-13 &  3.2e-14 &  4e-15\\
P03 &   4691.24 &  3.63e-13 &  4.1e-14 &  5e-15\\
P05 &   4708.77 &  3.54e-13 &  5.9e-14 &  8e-15\\
P06 &   4717.69 &  3.52e-13 &  5.0e-14 &  1e-14\\
P08 &   4735.87 &  3.55e-13 &  5.0e-14 &  7e-15\\
P11 &   4763.99 &  3.47e-13 &  6.1e-14 &  8e-15\\
P11 &   4763.99 &  3.50e-13 &  5.7e-14 &  6e-15\\
P12 &   4773.58 &  3.45e-13 &  6.4e-14 &  7e-15\\
P26 &   4920.41 &  3.18e-13 &  3.4e-14 &  4e-15\\
P30 &   4966.84 &  3.04e-13 &  3.2e-14 &  4e-15\\
P36 &   5040.48 &  2.91e-13 &  1.6e-14 &  5e-15\\
 \hline                           
\end{tabular}
\end{table}

\begin{table}[!h]
\caption{HD~250550 \element[][12]{CO} $v$=2-1 and \element[][13]{CO} line fluxes.}             
\label{tab:hd25_12}      
\centering                          
\begin{tabular}{l llll  }        
            
line	&$\lambda_{\rm line}$[nm] &  Cont.[$\frac{\rm{erg}}{\rm{cm}^2\rm{s}}$] &    $F_{\rm line}$[$\frac{\rm{erg}}{\rm{cm}^2\rm{s}}$]  & Error[$\frac{\rm{erg}}{\rm{cm}^2\rm{s}}$] \\ [1ex]

\hline
  \multicolumn{2}{l}{\element[][12]{CO} $v$=2-1}\\
 \hline
R17 &   4588.75 &  3.65e-13 &  6e-15 &  3e-15\\
R13 &   4615.98 &  3.65e-13 &  5e-15 &  2e-15\\
R13 &   4615.98 &  3.67e-13 &  5e-15 &  2e-15\\
R12 &   4623.03 &  3.63e-13 &  3e-15 &  2e-15\\
R06 &   4667.50 &  3.79e-13 &  5e-15 &  2e-15\\
R06 &   4667.50 &  3.81e-13 &  5e-15 &  2e-15\\
R06 &   4667.50 &  3.83e-13 &  6e-15 &  2e-15\\
R05 &   4675.28 &  3.83e-13 &  8e-15 &  1e-15\\
R05 &   4675.28 &  3.84e-13 &  9e-15 &  4e-15\\
P07 &   4786.08 &  3.44e-13 &  9e-15 &  6e-15\\
P08 &   4795.37 &  3.49e-13 &  5e-15 &  5e-15\\
P21 &   4926.94 &  3.14e-13 &  7e-15 &  3e-15\\
  \hline                           
  \multicolumn{2}{l}{\element[][13]{CO} $v$=1-0}\\
 \hline
R24 &   4593.51 &  3.63e-13 &  3e-15 &  2e-15\\
R16 &   4643.73 &  3.71e-13 &  7e-15 &  3e-15\\
R15 &   4650.43 &  3.98e-13 &  7e-15 &  2e-15\\
R13 &   4664.11 &  3.77e-13 &  8e-15 &  2e-15\\
R13 &   4664.11 &  3.71e-13 &  1e-14 &  3e-15\\
R12 &   4671.09 &  4.11e-13 &  9e-15 &  3e-15\\
R12 &   4671.09 &  4.04e-13 &  8e-15 &  2e-15\\
R10 &   4685.35 &  3.66e-13 &  10e-15 &  2e-15\\
R10 &   4685.35 &  3.66e-13 &  7e-15 &  2e-15\\
 R9 &   4692.62 &  3.61e-13 &  9e-15 &  2e-15\\
 R4 &   4730.48 &  3.56e-13 &  6e-15 &  4e-15\\
P16 &   4917.83 &  3.16e-13 &  7e-15 &  3e-15\\
 \hline                           
\end{tabular}
\end{table}

\begin{table}[!h]
\caption{Hen~2-80 \element[][12]{CO} line fluxes from March 5, 2012.}             
\label{tab:hen2_12v1}      
\centering                          
\begin{tabular}{l llll  }        
            
line	&$\lambda_{\rm line}$[nm] &  Cont.[$\frac{\rm{erg}}{\rm{cm}^2\rm{s}}$] &    $F_{\rm line}$[$\frac{\rm{erg}}{\rm{cm}^2\rm{s}}$]  & Error[$\frac{\rm{erg}}{\rm{cm}^2\rm{s}}$] \\ [1ex]
\hline
  \multicolumn{2}{l}{\element[][12]{CO} $v$=1-0}\\
 \hline
R06 &   4610.00 &  3.58e-13 &  2.9e-14 &  5e-15\\
P06 &   4717.69 &  3.55e-13 &  2.1e-14 &  9e-15\\
P07 &   4726.73 &  3.53e-13 &  2.4e-14 &  6e-15\\
P08 &   4735.87 &  3.49e-13 &  3.4e-14 &  5e-15\\
P11 &   4763.99 &  3.49e-13 &  3.0e-14 &  7e-15\\
P12 &   4773.58 &  3.59e-13 &  2.3e-14 &  7e-15\\
P12 &   4773.58 &  3.50e-13 &  2.5e-14 &  8e-15\\
P14 &   4793.12 &  3.58e-13 &  2.5e-14 &  8e-15\\
P26 &   4920.41 &  3.48e-13 &  4.0e-14 &  6e-15\\
P27 &   4931.82 &  3.37e-13 &  3.0e-14 &  7e-15\\
P27 &   4931.82 &  3.37e-13 &  2.6e-14 &  6e-15\\
P30 &   4966.84 &  3.16e-13 &  2.4e-14 &  7e-15\\
P36 &   5040.48 &  3.20e-13 &  1.9e-14 &  7e-15\\
P36 &   5040.48 &  3.18e-13 &  2.0e-14 &  6e-15\\
P37 &   5053.24 &  3.21e-13 &  2.9e-14 &  6e-15\\
 \hline
  \multicolumn{2}{l}{\element[][12]{CO} $v$=2-1}\\
 \hline
R13 &   4615.98 &  3.54e-13 &  1.0e-14 &  3e-15\\
R09 &   4644.81 &  3.53e-13 &  1.3e-14 &  4e-15\\
R08 &   4652.27 &  3.72e-13 &  1.0e-14 &  4e-15\\
R07 &   4659.83 &  3.56e-13 &  1.6e-14 &  4e-15\\
R07 &   4659.83 &  3.55e-13 &  1.7e-14 &  5e-15\\
R06 &   4667.50 &  3.54e-13 &  1.2e-14 &  4e-15\\
P01 &   4732.65 &  3.48e-13 &  1.2e-14 &  6e-15\\
P04 &   4758.86 &  3.44e-13 &  1.7e-14 &  8e-15\\
P27 &   4994.72 &  3.15e-13 &  1.6e-14 &  6e-15\\
P28 &   5006.47 &  3.12e-13 &  1.5e-14 &  8e-15\\
\hline                           
\end{tabular}
\end{table}

\begin{table}[!h]
\caption{Hen~2-80 \element[][12]{CO} line fluxes from March 6, 2012.}             
\label{tab:hen2_12v1}      
\centering                          
\begin{tabular}{l llll  }        
            
line	&$\lambda_{\rm line}$[nm] &  Cont.[$\frac{\rm{erg}}{\rm{cm}^2\rm{s}}$] &    $F_{\rm line}$[$\frac{\rm{erg}}{\rm{cm}^2\rm{s}}$]  & Error[$\frac{\rm{erg}}{\rm{cm}^2\rm{s}}$] \\ [1ex]
\hline
  \multicolumn{2}{l}{\element[][12]{CO} $v$=1-0}\\
 \hline
R06 &   4610.00 &  3.12e-13 &  2.6e-14 &  8.e-15\\
P06 &   4717.69 &  3.17e-13 &  1.7e-14 &  7e-15\\
P07 &   4726.73 &  3.12e-13 &  2.6e-14 &  5e-15\\
P08 &   4735.87 &  3.14e-13 &  2.9e-14 &  5e-15\\
P11 &   4763.99 &  3.39e-13 &  3.0e-14 &  6e-15\\
P12 &   4773.58 &  3.16e-13 &  2.3e-14 &  5e-15\\
P12 &   4773.58 &  3.38e-13 &  3.0e-14 &  5e-15\\
P14 &   4793.12 &  3.40e-13 &  2.9e-14 &  7e-15\\
P26 &   4920.41 &  3.32e-13 &  3.7e-14 &  4e-15\\
P27 &   4931.82 &  3.28e-13 &  3.1e-14 &  5e-15\\
P27 &   4931.82 &  3.23e-13 &  3.2e-14 &  5e-15\\
P30 &   4966.84 &  3.31e-13 &  2.9e-14 &  1e-14\\
P36 &   5040.48 &  3.39e-13 &  1.7e-14 &  5e-15\\
P37 &   5053.24 &  3.37e-13 &  2.5e-14 &  5e-15\\
\hline                           
  \multicolumn{2}{l}{\element[][12]{CO} $v$=2-1}\\
 \hline
 R13 &   4615.98 &  3.13e-13 &  7e-15 &  3e-15\\
R09 &   4644.81 &  3.11e-13 &  9e-15 &  4e-15\\
R08 &   4652.27 &  3.30e-13 &  7e-15 &  2e-15\\
R07 &   4659.83 &  3.16e-13 &  1.6e-14 &  4e-15\\
R07 &   4659.83 &  3.18e-13 &  1.7e-14 &  4e-15\\
R06 &   4667.50 &  3.18e-13 &  9e-15 &  4e-15\\
R06 &   4667.50 &  3.21e-13 &  7e-15 &  2e-15\\
P01 &   4732.65 &  3.11e-13 &  7e-15 &  5e-15\\
P04 &   4758.86 &  3.07e-13 &  1.3e-14 &  6e-15\\
P20 &   4916.09 &  3.30e-13 &  1.1e-14 &  3e-15\\
P27 &   4994.72 &  3.23e-13 &  1.7e-14 &  4e-15\\
P28 &   5006.47 &  3.24e-13 &  1.3e-14 &  6e-15\\
\hline                           
\end{tabular}
\end{table}

\begin{table}[!h]
\caption{Hen~2-80 \element[][13]{CO} line fluxes.}             
\label{tab:hen2_13}      
\centering                          
\begin{tabular}{l llll  }        
line	&$\lambda_{\rm line}$[nm] &  Cont.[$\frac{\rm{erg}}{\rm{cm}^2\rm{s}}$] &    $F_{\rm line}$[$\frac{\rm{erg}}{\rm{cm}^2\rm{s}}$]  & Error[$\frac{\rm{erg}}{\rm{cm}^2\rm{s}}$] \\ [1ex]
\hline
  \multicolumn{2}{l}{March 5, 2012}\\
  \multicolumn{2}{l}{\element[][13]{CO} $v$=1-0}\\
 \hline
R23 &   4599.47 &  3.56e-13 &  1.2e-14 &  3e-15\\
R23 &   4599.47 &  3.58e-13 &  1e-14 &  1e-14\\
R22 &   4605.51 &  3.59e-13 &  9e-15 &  4e-15\\
R21 &   4611.65 &  3.54e-13 &  1.2e-14 &  3e-15\\
R16 &   4643.73 &  3.54e-13 &  1.1e-14 &  3e-15\\
R13 &   4664.11 &  3.52e-13 &  1.0e-14 &  3e-15\\
R13 &   4664.11 &  3.52e-13 &  1.0e-14 &  4e-15\\
R12 &   4671.09 &  3.57e-13 &  9e-15 &  3e-15\\
R12 &   4671.09 &  3.58e-13 &  1.3e-14 &  3e-15\\
R10 &   4685.35 &  3.47e-13 &  1.4e-14 &  3e-15\\
R10 &   4685.35 &  3.43e-13 &  1.3e-14 &  3e-15\\
R9 &   4692.62 &  3.50e-13 &  1.0e-14 &  3e-15\\
R3 &   4738.34 &  3.44e-13 &  1.3e-14 &  6e-15\\
P27 &   5035.82 &  3.16e-13 &  1.6e-14 &  9e-15\\
P28 &   5047.28 &  3.15e-13 &  1.1e-14 &  6e-15\\
\hline
  \multicolumn{2}{l}{March 6, 2012}\\
  \multicolumn{2}{l}{\element[][13]{CO} $v$=1-0}\\
 \hline
R23 &   4599.47 &  3.08e-13 &  1.1e-14 &  3e-15\\
R23 &   4599.47 &  3.09e-13 &  1.1e-14 &  4e-15\\
R22 &   4605.51 &  3.10e-13 &  7.8e-15 &  3e-15\\
R21 &   4611.65 &  3.09e-13 &  9.0e-15 &  3e-15\\
R16 &   4643.73 &  3.12e-13 &  9.3e-15 &  3e-15\\
R13 &   4664.11 &  3.15e-13 &  1.1e-14 &  3e-15\\
R13 &   4664.11 &  3.18e-13 &  1.0e-14 &  3e-15\\
R12 &   4671.09 &  3.19e-13 &  1.1e-14 &  3e-15\\
R12 &   4671.09 &  3.17e-13 &  1.1e-14 &  2e-15\\
R11 &   4678.17 &  3.11e-13 &  1.0e-14 &  3e-15\\
R11 &   4678.17 &  3.11e-13 &  8.5e-15 &  4e-15\\
R10 &   4685.35 &  3.14e-13 &  8.6e-15 &  3e-15\\
R10 &   4685.35 &  3.13e-13 &  1.1e-14 &  3e-15\\
 R9 &   4692.62 &  3.12e-13 &  8e-15 &  3e-15\\
 R3 &   4738.34 &  3.12e-13 &  9e-15 &  4e-15\\
P27 &   5035.82 &  3.36e-13 &  1.6e-14 &  6e-15\\
P28 &   5047.28 &  3.36e-13 &  1.0e-14 &  4e-15\\
\hline                           
\end{tabular}
\end{table}

\begin{table}[!h]
\caption{Hen~3-1227 \element[][12]{CO} and \element[][13]{CO} $v$=1-0 line fluxes.}             
\label{tab:hen3}      
\centering                          
\begin{tabular}{l llll  }        
line	&$\lambda_{\rm line}$[nm] &  Cont.[$\frac{\rm{erg}}{\rm{cm}^2\rm{s}}$] &    $F_{\rm line}$[$\frac{\rm{erg}}{\rm{cm}^2\rm{s}}$]  & Error[$\frac{\rm{erg}}{\rm{cm}^2\rm{s}}$] \\ [1ex]
\hline
  \multicolumn{2}{l}{March 5, 2012}\\
   \multicolumn{2}{l}{\element[][12]{CO} $v$=1-0}\\
 \hline
 R08 &   4594.99 &  7.85e-13 &  1.2e-14 &  3e-15\\
R05 &   4617.66 &  7.90e-13 &  1.0e-14 &  2e-15\\
R04 &   4625.42 &  7.91e-13 &  6e-15 &  2e-15\\
R03 &   4633.28 &  7.94e-13 &  1.1e-14 &  3e-15\\
R01 &   4649.31 &  8.01e-13 &  7e-15 &  2e-15\\
R00 &   4657.49 &  8.11e-13 &  6e-15 &  3e-15\\
P01 &   4674.15 &  8.22e-13 &  9e-15 &  3e-15\\
P08 &   4735.87 &  7.94e-13 &  9e-15 &  3e-15\\
P09 &   4745.13 &  7.94e-13 &  1.5e-14 &  7e-15\\
P11 &   4763.99 &  7.87e-13 &  1.3e-14 &  6e-15\\
P26 &   4920.41 &  7.75e-13 &  9e-15 &  4e-15\\
\hline
   \multicolumn{2}{l}{\element[][13]{CO} $v$=1-0}\\
 \hline
 R22 &   4605.51 &  7.89e-13 &  5e-15 &  3e-15\\
R18 &   4630.62 &  7.92e-13 &  5e-15 &  2e-15\\
R10 &   4685.35 &  7.97e-13 &  7e-15 &  2e-15\\
R10 &   4685.35 &  8.01e-13 &  1.0e-14 &  3e-15\\
 R9 &   4692.62 &  8.02e-13 &  1.1e-14 &  4e-15\\
\hline
  \multicolumn{2}{l}{March 6, 2012}\\
  \multicolumn{2}{l}{\element[][12]{CO} $v$=1-0}\\
 \hline
R08 &   4594.99 &  8.17e-13 &  1.3e-14 &  3e-15\\
R05 &   4617.66 &  8.20e-13 &  1.1e-14 &  2e-15\\
R04 &   4625.42 &  8.26e-13 &  6e-15 &  2e-15\\
R03 &   4633.28 &  8.27e-13 &  1.2e-14 &  3e-15\\
R01 &   4649.31 &  8.41e-13 &  1.2e-14 &  3e-15\\
P01 &   4674.15 &  8.56e-13 &  6e-15 &  3e-15\\
P08 &   4735.87 &  8.62e-13 &  9e-15 &  3e-15\\
P09 &   4745.13 &  8.54e-13 &  1.9e-14 &  1e-14\\
P11 &   4763.99 &  8.52e-13 &  1.3e-14 &  1e-14\\
P26 &   4920.41 &  8.56e-13 &  8e-15 &  4e-15\\
P27 &   4931.82 &  8.55e-13 &  1.1e-14 &  3e-15\\
\hline
  \multicolumn{2}{l}{ \element[][13]{CO} $v$=1-0}\\
 \hline
R22 &   4605.51 &  8.16e-13 &  6e-15 &  3e-15\\
R18 &   4630.62 &  8.26e-13 &  3e-15 &  2e-15\\
R17 &   4637.12 &  8.30e-13 &  5e-15 &  1e-15\\
R15 &   4650.43 &  8.44e-13 &  1.0e-14 &  3e-15\\
R12 &   4671.09 &  8.53e-13 &  8e-15 &  2e-15\\
R10 &   4685.35 &  8.50e-13 &  8e-15 &  3e-15\\
R10 &   4685.35 &  8.54e-13 &  1.0e-14 &  3e-15\\
 R3 &   4738.34 &  8.57e-13 &  7e-15 &  5e-15\\
\hline                           
\end{tabular}
\end{table}

\subsection{HD~163296}
We include \element[][12]{CO} $v$=1-0 emission lines from $J$=2 up to $J$=36 in our study. We detect very broad emission lines indicating that the CO ro-vibrational lines are emitted fairly close to the star, consistent with the previous trend of {group I} discs emitting at smaller radii. We find single peaked emission lines for low and mid $J$ values while flat topped asymmetric profiles with shoulders seem to be present at higher $J$. The scatter in the  FWHM measurements, are caused by low S/N and very wide lines with broad wings hard to separate from the continuum. This scatter is reflected well in the associated error bars, based on the standard deviation of the nearby continuum. {The line shapes that we observe are not compatible with the earlier line detections of \citet{blake2004} and \citet{salyk2011} using NIRSPEC/Keck data ($R\sim$25000). At low $J$, \citet{blake2004} find single peaked profiles of similar shape as our observed profiles but much wider than ours. For high $J$ lines, \citet{salyk2011} find double peaked and again much wider profiles than ours (for $J$>25,  FWHM=83 km/s compared to our  FWHM=59 km/s)}. If we can exclude the instrumental setup as the cause for these profile differences (the earlier data were collected using NIRSPEC/Keck as opposed to CRIRES/VLT), this could be connected to the reported variations in NIR brightness \citep{sitko2008,ellerbroek2014}.
An in depth study of our detected CO lines from this source, including a thorough comparison to previous observations and to a model, will be the topic of a forthcoming paper.

\subsection{HD~250550} \label{sec:hd25}
We include \element[][12]{CO} $v$=1-0 emission lines from $J$=1 up to $J$=35, $v$=2-1 emission lines from $J$=6 up to $J$=20, \element[][13]{CO} $v$=1-0 emission lines from $J$=5 up to $J$=25 in our study. The $v$=1-0 lines are single peaked with a blue shifted shoulder that seems to increase for higher $J$. The noticed correlation between the FWHM (of the main component) and $J$ value (increasing FWHM, confirmed with the Kendall's Tau test performed in Sect. \ref{sec:co_lines}) points to a CO emitting region where lines of higher excitation are coming from regions closer and closer to the star.

The lower $J$ line median for HD~250550 shows a narrower Gaussian profile (FWHM=15km/s), with a shoulder at -17 km/s, while the higher $J$ line median shows a wider main component (FWHM=20km/s), where the blue component has grown and merged with the line wing of the main component. 

The line shape of the \element[][12]{CO} v(1-0)R30 line is consistent with that observed previously by \citet{brittain2007}.
The average  FWHM and single peaked line shape of our detected lines is consistent with the  FWHM and line shape of the ro-vibrational OH line at 2.9345 $\mu$m reported by \citet{fedele2011}. The single peaked shape could be caused by low inclination of the disc and based on their observations of the OH line \citet{fedele2011} conclude that a model with an inclination of 10$\degree$ gives the best fit. The velocity resolution for CRIRES is 3 km/s, so double peaks separated by $\le$3 km/s could be observed as single peaks.

\subsection{Hen~2-80}
We include \element[][12]{CO} $v$=1-0 emission lines from $J$=5 up to $J$=36, $v$=2-1 emission lines from $J$=0 to $J$=27, \element[][13]{CO} $v$=1-0 emission lines from $J$=4 to $J$=27.
These lines show broad, double peaked profiles and the fact that the width and line shape stay similar through all $J$ values (see Figures \ref{fwhm_all} and \ref{fig_median}) indicates a similar emitting region for all lines. One of the causes for a similar emitting region could be the presence of a gap which would be consistent with the classification, from the $L_{NIR}/L_{IR}$ versus [12]-[60] colour, of the source as group I. The width of the lines, however, is not consistent with this classification. The peak separations for the observed lines are also fairly large ($\sim$22-25 km/s), indicating that the CO emitting region is not very extended and/or that the disc has an inclination far from face on.

We also detect \element[][12]{CO} $v$=1-0  $J$<6 lines but these coincide with CO ro-vibrational absorption lines (Fig. \ref{hen280v13_profile}). These absorption lines are perfectly centred on the emission lines and their origin could likely be in the reported surrounding nebulosity \citep{carmona2010}. 

\subsection{Hen~3-1227}
We include \element[][12]{CO} $v$=1-0 emission lines from $J$=0 to $J$=26 and \element[][13]{CO} $v$=1-0 emission lines from $J$=4 to $J$=23. All lines are weak, narrow, and single peaked. This could be due to a low inclination angle of the disc. However, we cannot confirm a protoplanetary disc nature for this object (see Sect. \ref{sec:sample}). 
The lack of full $J$ coverage, together with the scatter (due to low S/N) in the FWHM values makes it hard to determine whether the  FWHM is constant or rising with $J$ (Fig. \ref{fwhm_all}). However, the FWHM values are not inconsistent with a constant  FWHM versus $J$ behaviour.

\subsection{MWC~137}
We see broad and blended emission lines over the entire spectrum. The crowded spectrum does not arise from a poorly chosen standard star spectrum for the telluric correction, since the same spectrum was used for the correction of T~Ori, a source that shows a mostly flat spectrum. We have clear detections of the \ion{H}{i} recombination lines and of CO absorption $v$=1-0 from lower $J$ levels ($J_{\rm up}$=0,1,2,3,4). For the CO absorption lines, the lowest $J$ ($J_{\rm up}$=0,1) transitions are seen as narrow double-peaked or multicomponent lines with separation of 12 km/s. {A high-resolution VLT/UVES optical spectrum of MWC~137 extracted from the ESO data archive (see Fig. \ref{mwc137_ki} in the appendix),} also shows multiple absorption components in the \ion{K}{i} 7699 $\AA$ line, with the same velocity separation as seen in the CO absorption lines. Therefore, we interpret the CO absorption seen in our spectra of MWC~137 as being due to the presence of cool molecular material in multiple absorbing foreground clouds.
{The broad blended emission features present throughout the spectrum could be blended CO ro-vibrational lines, since no other molecule has been reported to produce such strong and numerous emission in this particular wavelength region. 
Similar spectra has been observed from other discs, e.g. RW~Aur by \citet{najita2003}, who show that blended very broad CO ro-vibrational lines can explain the observed spectrum, and \mbox{HD 101412} by \citet{plas2014}, where first overtone CO emission was also seen, similar to MWC~137 \citep{oksala2013}.}

\subsection{T~Ori}
For T~Ori no CO emission lines were found in the spectra (Figures \ref{fig:spec1} and \ref{fig:spec2}). The spectrum shows a somewhat flat and sometimes 'bumpy' continuum. In the previous study \citep{plas2014}, CO fundamental lines were detected from a similar source, HD~190073 (group II classification and spectral type A2IVpe). The line fluxes for the $v$=1-0 lines from this source range from 1.4$\pm$0.6$\cdot10^{-14}$ to 3.3$\pm$1.3$\cdot10^{-14}$~erg/cm$^2$/s \citep{plas2014}. Scaled to the distance of T~Ori (i.e. 1.7 times further away), we would expect typical flux levels below $\sim 10^{-14}$~erg/cm$^2$/s. This is of the same order as some of the weakest lines (v=2-1 or \element[][13]{CO}) we detect in sources such as HD~250550 and Hen~3-1227. However, the detected lines in these two sources were narrow and sharply peaked and were therefore easier to separate from the continuum. For T~Ori, being a group II source, we expect emission lines to be wider (emitted closer to the star) and therefore more difficult to separate from the continuum in the case of weak lines. Thus, the observed flat spectrum we see from T~Ori is consistent with the expectation of broad and weak emission lines with line fluxes below the detection limit.
 
A similar explanation is also valid for the $v$=2-1 and for the \element[][13]{CO} lines from HD~163296. These lines, though still visible in the spectra, are too broad and weak to be quantified. The $v$=1-0 lines, detected from HD~163296, are also very wide but strong enough ($10^{-13}$ erg/cm$^2$/s) that they can be separated from the continuum.

\end{appendix}

\end{document}